\documentclass[usenatbib]{mn2e}
\usepackage{graphicx,natbib,amssymb,amsmath}
\usepackage{txfonts, stfloats}
\usepackage{color}

\newcommand{\Mpc}{{\rm Mpc}}

\newcommand{\expf}[1]{{{\rm e}^{#1}}}

\newcommand{\Neff}{N_{\rm eff}}

\newcommand{\keV}{{\rm keV}}
\newcommand{\MeV}{{\rm MeV}}
\newcommand{\GeV}{{\rm GeV}}
\newcommand{\cm}{{\rm cm}}

\newcommand{\TCMB}{T_{\rm CMB}}

\newcommand{\Tgin}{{\Tg^{\rm in}}}

\newcommand{\zh}{{z_{\rm h}}}
\newcommand{\zmu}{{z_{\mu}}}

\newcommand{\nbb}{{n^{\rm pl}}}

\newcommand{\muc}{\mu_{\rm c}}

\newcommand{\taudot}{\dot{\tau}}

\newcommand{\xc}{x_{\rm c}}

\newcommand{\id}{{\,\rm d}}

\newcommand{\beq}{\begin{equation}}   %

\newcommand{\eeq}{\end{equation}}   %

\newcommand{\beqa}{\begin{eqnarray}}   %

\newcommand{\eeqa}{\end{eqnarray}}   %

\newcommand{\beal}{\begin{align}}

\newcommand{\enal}{\end{align}}

\newcommand{\bspl}{\begin{split}}

\newcommand{\espl}{\end{split}}

\newcommand{\bsub}{\begin{subequations}}

\newcommand{\esub}{\end{subequations}}

\newcommand{\bmulti}{\begin{multline}}   %

\newcommand{\beqm}{\begin{mathletters}}   %

\newcommand{\eeqm}{\end{mathletters}}   %

\newcommand{\me}{m_{\rm e}}

\newcommand{\Ne}{N_{\rm e}}

\newcommand{\Te}{T_{\rm e}}

\newcommand{\Tg}{T_{\gamma}}

\newcommand{\The}{\theta_{\rm e}}

\newcommand{\Thg}{\theta_{\gamma}}

\newcommand{\sigT}{\sigma_{\rm T}}

\newcommand{\nBE}{n_{\rm BE}}

\newcommand{\pot}[2]{#1 \times 10^{#2}}

\title[Large energy release]{Thermalization of large energy release in the early Universe}

\begin{document}

\author[Chluba, Ravenni \& Acharya]{
Jens Chluba$^1$\thanks{E-mail:jens.chluba@manchester.ac.uk},
Andrea Ravenni$^1$\thanks{E-mail:andrea.ravenni@manchester.ac.uk}
and Sandeep Kumar Acharya$^{2,1}$\thanks{E-mail:sandeepkumar@theory.tifr.res.in}
\\
$^1$Jodrell Bank Centre for Astrophysics, School of Physics and Astronomy, The University of Manchester, Manchester M13 9PL, U.K.
\\
$^2$Department of Theoretical Physics, Tata Institute of Fundamental Research, Mumbai 400005, India
}

\date{\vspace{-3mm}{Accepted 2020 July 14. Received 2020 May 22}}

\maketitle

\begin{abstract}
Spectral distortions of the cosmic microwave background (CMB) provide a unique tool for learning about the early phases of cosmic history, reaching deep into the primordial Universe. At redshifts $z\lesssim 10^6$, thermalization processes become inefficient and existing limits from {\it COBE/FIRAS} imply that no more than $\Delta \rho/\rho\lesssim \pot{6}{-5}$ ($95\%$ c.l.) of energy could have been injected into the CMB. However, at higher redshifts, when thermalization is efficient, the constraint weakens and $\Delta \rho/\rho\simeq 0.01-0.1$ could in principle have occurred. Existing computations for the evolution of distortions commonly assume $\Delta \rho/\rho \ll 1$ and thus become inaccurate in this case. Similarly, relativistic temperature corrections become relevant for large energy release, but have previously not been modeled as carefully. Here we study the evolution of distortions and the thermalization process after {\it single large energy release} at  $z\gtrsim 10^5$. We show that for large distortions the thermalization efficiency is significantly reduced and that the distortion visibility is sizeable to much earlier times. This tightens spectral distortions constraints on low-mass primordial black holes with masses $M_{\rm PBH}\lesssim\pot{2}{11}\,{\rm g}$. Similarly, distortion limits on the amplitude of the small-scale curvature power spectrum at wavenumbers $k\gtrsim 10^4\,\Mpc^{-1}$ and short-lived decaying particles with lifetimes $t_X\lesssim 10^7\,{\rm s}$ are tightened, however, these still require a more detailed time-dependent treatment. We also briefly discuss the constraints from measurements of the effective number of relativistic degrees of freedom and light element abundances and how these complement spectral distortion limits.
\end{abstract}


\section{Introduction}
The cosmic microwave background (CMB) has delivered a wealth of information about the Universe we live in, clearly estabilishing $\Lambda$CDM as the preferred cosmological model \citep{WMAP_params, Planck2015params}. One of the important pillars of this model was forged with the discovery of the CMB \citep{Penzias1965}, yet another with the proof of the CMB's blackbody nature \citep{Mather1994, Fixsen1996}. However, the path towards precision cosmology was successfully paved by studies of the CMB temperature and polarization anisotropies, and we are now looking at a bright future for observations of CMB polarization signals and secondary anisotropies, with many experiments coming online or being planned \citep{SOWP2018, Suzuki2018, Delabrouille2018, Hanany2019PICO, Jacques2019Voyage, Basu2019Voyage}.

One of the next frontiers in CMB research is the measurements of CMB {\it spectral distortions} from the early Universe \citep{Chluba2019BAAS, Chluba2019Voyage}. Energy release in the early Universe caused by both standard and non-standard processes can change the thermodynamic equilibrium between matter and radiation, imprinting various departures of the CMB energy spectrum from that of a blackbody, thereby providing a probe of the thermal history deep into the pre-recombination era  \citep{Zeldovich1969, Sunyaev1970mu, Danese1982, Burigana1991, Hu1993}.
The measurements of {\it COBE/FIRAS} in the early 90s still stand as the most robust and broadly applicable limit on early energy release, implying that no more than $\Delta\rho/\rho \lesssim \pot{6}{-5}$ \citep[95\% c.l.;][]{Fixsen1996, Fixsen2009} was injected into the CMB at redshifts $z\lesssim 10^6$. However, we are now experimentally ready for the next level of precision, and innovative spectrometer designs may open a new window to the early Universe in the decades to come \citep{Kogut2011PIXIE, PRISM2013WPII, Kogut2016SPIE, Kogut2019BAAS, Jacques2019Voyage}. 

The great potential of spectral distortions lies in their ability to constrain a wide range of standard and non-standard processes \citep{Sunyaev2009, Chluba2011therm, Sunyaev2013, Tashiro2014, deZotti2015, Chluba2016, Lucca2020}. This can complement other cosmological probes in particular by shedding light on processes occurring at redshifts $z\gtrsim 10^3$, in principle giving us the opportunity to distinguish various scenarios of distortions through precise CMB spectroscopy \citep[e.g.,][]{Chluba2013fore, Chluba2013PCA}.

The evolution of spectral distortions for various energy release scenarios can be described in detail using {\tt CosmoTherm} \citep{Chluba2011therm}. This thermalization code already includes state-of-the-art descriptions of the various thermalization mechanisms, also accounting for leading order temperature relativistic corrections. However, like most treatments of the thermalization problem it assumes that the distortions are {\it small} at any stage of their evolution. 
At redshifts $z\lesssim 10^6$, when the photon production by double Compton (DC) and Bremsstrahlung (BR) slows, this indeed is a requirements to ensure that the constraints from {\it COBE/FIRAS} are not violated as the distortion visibility\footnote{This defines the amount of energy that is still seen as a distortion today, and will be defined carefully in Sect.~\ref{sec:Jbb_section}.} approaches unity. However, earlier, a significant amount of energy, reaching $\Delta \rho/\rho\simeq 0.01-0.1$, can in principle still be ingested. In this case, the thermalization problem becomes {\it non-linear}, and the numerical treatment has to be modified to obtain reliable CMB distortion limits. 

In previous works \citep{Burigana1991, Hu1993}, this situation was already studied numerically for a few examples using the Kompaneets equation \citep{Kompa56} to describe Compton scattering, the Lightman approximation \citep{Lightman1981} for the DC emissivity, and various simple approximations for the BR process. This showed that the thermalization of large distortions is indeed slower, such that the limits on early energy release ought be tighter. However, due to the lack of simple approximations and efficient numerical treatments, current distortion constraints on various energy release scenarios are commonly obtained using the {\it small distortion limit}. For {\it COBE/FIRAS}, this is expected to become inaccurate at\footnote{This redshift can be estimated using the small-distortion visibility function and asking when the possible $\Delta \rho/\rho$ exceeds $\simeq 0.01$ at a given distortion sensitivity (see Fig.~\ref{fig:dlnrho_limits}).} $z\gtrsim \pot{3}{6}$, which is well inside the $\mu$-distortion era.

To obtain accurate constraints, the problem cannot be simply augmented without including additional effects. For large energy release, the temperature of the electrons can become high (i.e., $\The = k\Te/\me c^2 \gtrsim  0.01$ or a few keV), such that relativistic corrections become important. Even at $z\simeq 10^7$, the temperature of the cosmic plasma only reaches $\The\simeq \pot{4.6}{-10}(1+z)\simeq \pot{5}{-3}$ for a standard thermal history, such that relativistic corrections usually remain fairly small \citep{Chluba2005, Chluba2014}. However, the Compton equilibrium temperature of electrons in a strongly distorted CMB spectrum can be much higher \citep{Sazonov2001}.
This affects the DC and BR emissivity of the plasma, which can now be accurately described using {\tt DCpack} \citep{DCpack2020} and {\tt BRpack} \citep{BRpack2020}. In addition, the Comptonization process is no longer described using the simple Kompaneets equation \citep{Sazonov2000}, and a Compton kernel approach is more appropriate. With {\tt CSpack}, this can today be done quasi-exactly \citep{CSpack2019}, thus in principle allowing us to eliminate all approximations to the main thermalization calculations.

In this paper, we will take an important step into this direction, considering the evolution of spectral distortions after {\it single energy injection} in the $\mu$-distortion era. We will provide the general formulation of the problem and then explain various simplifications valid in the $\mu$-era for single-injection scenarios (Sect.~\ref{sec:formulation}). We mainly study the problem numerically, but provide supporting analytical estimates where possible. This allows us to demonstrate that the distortion visibility is greatly enhanced for energy release in excess of $\Delta \rho/\rho\simeq 0.01$, which affects the distortion constraints at early times (see Fig.~\ref{fig:J_bb}). 

We will highlight the relevance of various effects and point out the most important differences between the small and large distortion regimes (Sect.~\ref{sec:QS_sols_gamma}). Our analysis implies that a treatment of {\it continuous energy release} scenarios requires a more careful consideration also including modifications to the Hubble expansion rate as well as detailed time-dependent effects, which depend on the energy release history. Indeed, we argue that even for small distortion scenarios, time-dependent corrections cannot be independently added to the computations, such that distortion visibility approaches \citep[e.g.,][]{Chluba2005, Khatri2012b, Chluba2014} become inaccurate. A detailed treatment of continuous energy release scenarios is, however, left to a forthcoming paper. 

We use our calculations to update CMB distortion constraints on primordial black holes (PBH), slightly improving previous distortion limits on BH with masses $M_{\rm PBH}\lesssim\pot{2}{11}\,{\rm g}$ that evaporate through Hawking radiation in a very bursty, quasi-instantaneous manner. The constraints from {\it COBE/FIRAS} could be improved with a {\it PIXIE}-like spectrometer \citep{Kogut2011PIXIE, Kogut2019BAAS}, further extending our reach into even earlier epochs.
We also briefly discuss the expected constraints on the small-scale power spectrum at wavenumber $k\gtrsim 10^4\,\Mpc^{-1}$ \citep{Sunyaev1970diss, Daly1991, Hu1994, Chluba2012, Chluba2012inflaton}, and decaying particles with lifetimes $t_X\lesssim 10^7\,{\rm s}$ \citep[e.g.,][]{Sarkar1984, Hu1993b, Kawasaki2005}, but leave a more detailed treatment to the future.
We furthermore contrast CMB distortions constraints to those obtained from measurements of light elements \citep[e.g.,][]{Kawasaki2005} and the effective number of relativistic degrees of freedom, $\Neff$ \citep[e.g.,][]{SS2008}, however, the focus of this paper is to revise CMB distortion limits. 

\vspace{-5mm}
\section{Formulation of the problem}
\label{sec:formulation}
Early on, until some $\simeq 10^8-10^9$ seconds after the big bang, thermalization processes are extremely rapid such that after a very short relaxation time following a disturbance away from the initial full equilibrium state (e.g., by heating the matter or direct photon injection), the photon distribution evolves along a sequence of {\it quasi-stationary stages}\footnote{This just broadly means that the shape and amplitude of the distortion evolve very slowly compared to Comptonization timescales.} \citep{Sunyaev1970mu, Danese1982}. 
The degrees of freedom in the photon-baryon system in the early phase are indeed quite limited, and we can resort to a {\it macroscopic} description of the problem. 

We shall assume that the electron and baryon distribution functions are given by a relativistic Maxwell-Boltzmann distributions, which are all characterized by the electron temperature, $\Te$, and the fixed comoving number densities of particles (defined by $\Ne$, $N_{\rm p}$ etc at various redshifts $z$). We also assume that, due to the exceedingly large number of photons over baryons, the heat capacity of the baryon-electron system, henceforth referred to collectively as {\it baryons}, is negligible for most practical purposes. In this case, any injection of energy ultimately is stored by the photon field. Density perturbations of the medium will be neglected too and only the uniform background evolution has to be considered. We will for now also assume only one single injection occurs, even if many of the equations are kept general.

For the photons, the initial spectrum shall be described by that of a blackbody at a temperature $\Tgin=\Te^{\rm in}$. This fixes the initial photon energy and number densities according to the blackbody relations, $\rho_\gamma\propto \Tg^4$ and $N_\gamma \propto \Tg^3$. At redshift, $\zh$, we shall inject some energy and photons into the medium. Injected photons also carry energy and directly appear as a change in the photon energy density, while injected heat may first only modify the matter temperature before equilibrating with the photon fluid (moving mirrors would directly 'heat' photons but usually heat transfer detours via the matter inside the Universe). The total injected energy (heat and the energy that the injected photons carry) increases the energy density of the photon field to $\rho_\gamma'=\rho_\gamma(1+\epsilon_\rho)$. This defines the final temperature\footnote{A small fraction of energy is drained to the baryons, which we neglect here but return to later.} of the new equilibrium blackbody 
$\Tg^{\rm eq}=\Tgin\,(1+\epsilon_\rho)^{1/4}$.
It is important to recognize that this temperature will only be reached if the thermalization process completes, which in practice depends on the injection redshift $\zh$. The final temperature is furthermore independent of the injected number of photons, which just affect how far away from $\Tg^{\rm eq}$ one will start \citep{Chluba2014}. In full equilibrium we will eventually have $\Te^{\rm eq}=\Tg^{\rm eq}$.

The evolution towards the final equilibrium (if ever reached) has several features. Again, we assume that quasi-stationary evolution occurs after a very short relaxation time, with the evolution of the photon-baryon system depending mainly on the momentary photon energy and particle number densities. The electron temperature is driven close to the Compton equilibrium temperature, $\Te^{\rm eq}$, which depends on the spectrum of the photon distribution as described below. The small corrections accounting for the energy taken to produce/absorb photons by BR and DC emission or the slow adiabatic expansion of the Universe can be easily added under quasi-stationary conditions \citep{Chluba2011therm}.

Before full equilibrium is reached, speaking of the 'photon temperature' is not very meaningful as the distorted spectrum is defined by two 'macroscopic' parameters, allowing to separately fix the momentary photon energy and number densities. Any photon distribution may be described by a Bose-Einstein spectrum
\begin{align}
\label{eq:BE_n}
\nBE(x)
&= \frac{1}{\expf{x+\mu(x)}-1},
\end{align}
with $x=h\nu/k\Te$ and {\it frequency-dependent chemical potential} $\mu(x)$. We directly chose the electron temperature as a scale in the problem for reasons that will become evident below. However, generally we could have chosen any other convenient reference. 

While $\mu(x)$ depends on frequency, by assuming quasi-stationary evolution the only real degree of freedom is the overall normalization. The {\it shape} of the distortion is fully determined by Comptonization and the photon production processes\footnote{For continuous energy release also the time-dependence of the injection mechanism becomes another parameter, as we explain below.} as described below and used in the early works on the problem \citep{Sunyaev1970mu, Danese1982}. It is set by 'microscopic' degrees of freedom, determining how the plasma reacts to heat and photon injection. For simplicity we can thus think of the spectral evolution as an evolution of the chemical potential amplitude or equivalently the number density of photons. However, the {\it shape} depends both on the particles densities and the overall level of departure from full equilibrium as we will further explain.

\vspace{-3mm}
\subsection{Photon Boltzmann equation and its moments}
\label{sec:Boltzman}
Including both Compton scattering (CS), photon production processes (DC and BR) and a photon source term (e.g., from a decaying particle), the photon Boltzmann equation in the expanding Universe reads \citep[e.g.,][]{Hu1993, Chluba2011therm}
\begin{align}
\label{eq:Dn_evol}
\left.\frac{\partial \, n}{\partial \tau}\right|_x - x\left.\frac{\partial \, n}{\partial x}\right|_\tau\,
\left[\frac{H}{\taudot}+\frac{\partial_\tau \Te}{\Te}\right]
&=\left.\frac{\text{d}n}{\text{d}\tau}\right|_{\rm CS}+\left.\frac{\text{d}n}{\text{d}\tau}\right|_{\rm em}+\left.\frac{\text{d}n}{\text{d}\tau}\right|_{\rm S}.
\end{align}
The collision terms will be specified in Sect.~\ref{sec:QS_sols_gamma}.
For convenience, we use the Thomson scattering optical depth, $\id \tau=\dot\tau \id t=c\Ne\sigT\id t$, as time-coordinate.  Assuming that the photon distribution evolves under exact quasi-stationary conditions, the left hand side of this equation vanishes\footnote{Naturally $\partial_\tau n\approx 0$ and $\frac{H}{\taudot}+\frac{\partial_\tau \Te}{\Te}=\partial_\tau \ln a \Te \approx 0$.}. Compton scattering alone leads to a Bose-Einstein spectrum with constant chemical potential, but photon adjusting terms restore $\mu=0$ at low frequencies. At a fixed energy density, this also modifies the high-frequency spectrum, affecting the heat capacity of the photon field \citep{Chluba2014}. These features will be studied in more detail below. 

To make progress, we compute the first two moments of the Boltzmann equation, corresponding to photon number and energy densities. This leads to
\begin{align}
\label{eq:Dn_evol_moments}
\frac{\id \ln a^3 N_\gamma}{\id \tau} 
&=\left.\frac{\text{d}\ln a^3 N_\gamma}{\text{d}\tau}\right|_{\rm em}+\left.\frac{\text{d}\ln a^3 N_\gamma}{\text{d}\tau}\right|_{\rm S}
\\
\nonumber
\frac{\id\ln a^4 \rho_\gamma}{\id \tau} 
&=\left.\frac{\text{d}\ln a^4  \rho_\gamma}{\text{d}\tau}\right|_{\rm CS}
+\left.\frac{\text{d}\ln a^4 \rho_\gamma}{\text{d}\tau}\right|_{\rm em}+\left.\frac{\text{d}\ln a^4 \rho_\gamma}{\text{d}\tau}\right|_{\rm S},
\end{align}
where we used the fact that Compton scattering conserves photon number. The terms on the right hand side are all caused by collisions, which can be thought of in comoving coordinates. Notice that on the left hand side no explicit dependence on $\Te$ appears, as the corresponding terms disappear when taking the moments. This simply is a reflection of the fact that without collision each of the quantities on the left hand side are conserved. Even for the standard thermal history and in the absence of perturbations in the medium, this case is never reached due to the adiabatic cooling of the baryons, which continuously extracts a tiny amount of energy from the photons \citep{Chluba2005, Chluba2011therm}.

\subsection{Evolution of the baryons}
\label{sec:heat_cap_baryons}
We assume that all baryons follow relativistic Maxwell-Boltzmann distributions at a common temperature. In this case. the energy density of each species ${\rm i}=\{{\rm e}, {\rm H}^+, {\rm He}^{++}\}$ is given by \citep[compare, e.g.,][]{Chluba2005}
\begin{align}
\label{eq:rho_matter}
\rho_{\rm i}&=m_{\rm i} c^2 N_{\rm i} + \frac{3}{2} k \Te\,N_{\rm i} \, F\left(\frac{k \Te}{m_{\rm i} c^2}\right)
\\ \nonumber
F(x)&=\left[2+\frac{2K_1(1/x)-2K_2(1/x)}{3xK_2(1/x)}\,\right]
\approx 1+\frac{5}{4}x-\frac{5}{4}x^2+\frac{45}{64}x^3
\end{align}
where $K_m(y)$ denotes the modified Bessel function of kind $m$. Since the masses of hydrogen ${\rm H}^+$ and ${\rm He}^{++}$ are much larger than the electron mass, at a fixed temperature one can neglect relativistic corrections to the kinetic energy for those species ($F\approx 1$), however, they are easy to include.

We are now interested in phases right after Big Bang Nucleosynthesis (BBN) but well before the recombination era. We then have $\id \ln a^3 N_{\rm i}/\id \tau = 0$ for each species. For the evolution of the total baryon energy density ($\rho_{\rm b}=\sum \rho_{\rm i}$ and $N_{\rm b}=\sum N_{\rm i}$), we then obtain \citep[e.g., see][for derivations]{Hu1995PhD, Chluba2005}
\begin{align}
\label{eq:evol_matter}
\frac{\id a^3 \rho_{\rm b}}{a^3 \!\id \tau}+3\frac{H}{\taudot} P_{\rm b}
&= \frac{3}{2} k N_{\rm b} \frac{\id \left[\Te F_{\rm b}(\Te)\right]}{\id \tau}+3\,k N_{\rm b}\,\frac{H}{\taudot}\,\Te
\nonumber\\
&=\left.\frac{\id a^3 \rho_{\rm b}}{a^3 \!\id \tau}\right|_{\rm h} + \left.\frac{\id a^3 \rho_{\rm b}}{a^3 \!\id \tau}\right|_{\gamma}
\nonumber
\\
F_{\rm b}(\Te)&\approx 1 + \frac{\Ne}{N_{\rm b}}\left[F\left(\frac{k \Te}{m_{\rm e} c^2}\right)-1\right],
\end{align}
which assumes that Coulomb interactions ensure Maxwellians for all species at $\Te\equiv T_{\rm i}$ but can also cause net heating by external energy injection into the system, determined by $\id a^3 \rho_{\rm b}/a^3 \!\id \tau\big|_{\rm h}$. The interaction with the photons, $\propto \id a^3 \rho_{\rm b}/a^3 \!\id \tau\big|_{\gamma}$, are mostly mediated through the electrons by the Compton scattering and DC/BR emission terms.
We next define the baryon heat capacity
\begin{align}
\label{eq:heat_cap_matter}
C^*_V&=\left.\frac{\partial \rho_{\rm b}}{\partial \Te}\right|_{V,N}
= \frac{3}{2} k N_{\rm b} \left[F_{\rm b}(\Te)+\Te \partial_{\Te} F_{\rm b}(\Te) \right]
\nonumber
\\
&\approx \frac{3}{2} k \left[ N_{\rm b}  + \frac{5}{2}\The\,\Ne \left(1-\frac{3}{2}\The +\frac{9}{8}\The^2\right)\right]
= \frac{3}{2} k N_{\rm b} \Big(1   + \lambda \Big),
\end{align}
where $\lambda=\frac{5}{2}\The\,\frac{\Ne}{N_{\rm b}} \left(1-\frac{3}{2}\The +\frac{9}{8}\The^2\right)$ accounts for leading order relativistic temperature corrections caused by the electrons \citep{Chluba2005}. With $H/\taudot=\id_\tau \ln a$, we can then write the temperature evolution equation in the form
\begin{align}
\label{eq:Te_evol_def}
\frac{\id \ln a \Te }{\id \tau}
&= \left[1- \frac{3\,k N_{\rm b}}{C^*_V}\right] \frac{\id \ln a}{\id \tau}
\nonumber\\
&\qquad + \frac{\rho_{\rm b}}{C^*_V \Te}\left[\left.\frac{\id \ln a^3 \rho_{\rm b}}{\id \tau}\right|_{\rm h} 
+ \left.\frac{\id \ln a^3 \rho_{\rm b}}{\!\id \tau}\right|_{\gamma}\right].
\end{align}
The coefficient in front of $\id \ln a/\id \tau$ can be further simplified to
\begin{align}
\label{eq:coeff}
1- \frac{3\,k N_{\rm b}}{C^*_V}
&\approx -\frac{1  - \frac{5}{2}\The\,\frac{\Ne}{N_{\rm b}} \left(1-\frac{3}{2}\The +\frac{9}{8}\The^2\right)}
{1  + \frac{5}{2}\The\,\frac{\Ne}{N_{\rm b}} \left(1-\frac{3}{2}\The +\frac{9}{8}\The^2\right)}
=-1+\frac{2\lambda}{1+\lambda}.
\end{align}
The second term in Eq.~\eqref{eq:Te_evol_def} is due to heating of the baryons by collision with other matter particles and the last is due to interactions with the photon, i.e., photon emission and Compton scattering.

\subsection{Thermodynamics for constant chemical potential}
\label{sec:BE_solutions}
Let us first entirely omit the effects of DC and BR emission. In this case, kinetic equilibrium between the electrons and photons is reached once a constant chemical potential distortion is established, $\mu=\muc$, simply as a consequence of Compton scattering \citep{Sunyaev1970SPEC}. Following the energy injection, the electron temperature can in principle depart significantly from the final equilibrium temperature, $\Tg^{\rm eq}=\Tgin\,(1+\epsilon_\rho)^{1/4}$, but after a short relaxation time the new equilibrium is determined by \citep[e.g.,][]{Burigana1991, Chluba2015GreensII}
\bsub
\label{eq:BE_n_sol}
\begin{align}
\label{eq:BE_n_sol_a}
\left(\frac{\Tgin}{\Te}\right)^{3}\big[1+\epsilon_N\big]&\equiv \left(\frac{\Tg^{\rm eq}}{\Te}\right)^{3}\frac{1+\epsilon_N}{(1+\epsilon_\rho)^{3/4}}=\mathcal{G}_2(\Te)
\\
\label{eq:BE_n_sol_b}
\left(\frac{\Tgin}{\Te}\right)^{4}\big[1+\epsilon_\rho\big]&\equiv \left(\frac{\Tg^{\rm eq}}{\Te}\right)^{4}=\mathcal{G}_3(\Te),
\\
\mathcal{G}_k(T)&=\frac{1}{G^{\rm pl}_k}\int \frac{x^k\id x}{\expf{x+\muc(T)}-1},
\end{align}
\esub
with $\epsilon_N=\Delta N/N$ and $\epsilon_\rho=\Delta \rho/\rho$ characterizing the total disturbance away from the initial equilibrium blackbody at temperature $\Tgin$. The integrals $\mathcal{G}_k(T)$ can be expressed using poly-logarithms. Here, we also used the blackbody integrals, $G^{\rm pl}_k=\int x^k/(\expf{x}-1)\id x$, yielding $G^{\rm pl}_2\approx 2.404$ and $G^{\rm pl}_3\approx 6.4939$. We furthermore expressed the conditions in terms of the final (to be reached) equilibrium blackbody temperature $\Tg^{\rm eq}=\Tgin(1+\epsilon_\rho)^{1/4}$.

Assuming that $\Tgin$, $\epsilon_\rho$ and $\epsilon_N$ are given we can then solve for $\Te$ and $\muc$. Alternatively, we can also fix $\Tgin$, $\Te$ and $\epsilon_N$ or any other combination of three parameters to determine the final state as convenient. For $\muc\neq 0$ this means $\Te\neq\Tg^{\rm eq}$, and without also adjusting the photon number, $\Te^{\rm eq}=\Tg^{\rm eq}$ cannot be fulfilled. This is where DC and BR emission come into play, as we explain below.

\subsubsection{Small distortions ($\muc\ll 1$)}
While the conditions, Eq.~\eqref{eq:BE_n_sol}, for general $\epsilon_N$ and $\epsilon_\rho$ are easily solved numerically, in the limit, $\epsilon_\rho\ll 1$ and $\epsilon_N\ll 1$, this leads to the well-known solution \citep{Sunyaev1970mu, Burigana1991, Hu1993, Chluba2015GreensII}
\bsub
\label{eq:BE_n_sol_small_mu}
\begin{align}
\frac{\Tgin}{\Te}&\approx 1 - 0.4561 \muc- \frac{\epsilon_N}{3}\approx 1+0.5185\epsilon_N-0.6389\,\epsilon_\rho
\\
\muc&\approx 1.401 \epsilon_\rho-1.8675\epsilon_N =1.401 \left(\epsilon_\rho-\frac{4}{3}\epsilon_N\right).
\end{align}
\esub
In this work, we are interested in cases with significant $\muc$ and thus we generally use Eq.~\eqref{eq:BE_n_sol} to compute the boundary conditions.

From Eq.~\eqref{eq:BE_n_sol_small_mu}, we can still read off the main behaviour of the solution. Heating the medium increases both the chemical potential and electron temperature. Adding photons at fixed chemical potential further increases the electron temperature, while at fixed $\epsilon_\rho$ the electron temperature and chemical potential decrease together. Comparing to the final equilibrium temperature 
$\Tg^{\rm eq}/\Tgin\approx 1+\epsilon_\rho/4$ with $\Te/\Tgin\approx 1+0.6389\,\epsilon_\rho-0.5185\epsilon_N$ one can find that for $\epsilon_N\gtrsim(3/4)\,\epsilon_\rho$ the initial electron temperature is $\Te<\Tg^{\rm eq}$. In this case, a negative chemical potential is formed \citep[e.g.,][]{Chluba2015GreensII} and the average energy of the photons
\begin{align}
\left<E_\gamma\right> &=k\Te \frac{\int \frac{x^3\id x}{\expf{x+\muc}-1}}{\int \frac{x^2\id x}{\expf{x+\muc}-1}}
\approx 2.701 k\Te(1+0.2578\muc)
\nonumber
\\
&=2.701k\Tgin(1+\epsilon_\rho-\epsilon_N)
\nonumber
\end{align}
is lower than $2.701 k\Tg^{\rm eq}$ of the final equilibrium blackbody. Note that $\epsilon_N$ can never exceed $\epsilon_\rho$ in physical systems. It is also clear that a {\it constant negative chemical potential} is physically inconsistent and that at low frequencies one has $\mu\rightarrow 0$ due to DC and BR emission.

\vspace{-0mm}
\subsubsection{Large distortions ($\muc\gg 1$)}
Assuming large energy release or photon injection, we can write $n\approx \expf{-x-\muc}$. In this case, the integrals in Eq.~\eqref{eq:BE_n_sol} can be carried out analytically yielding
\bsub
\label{eq:BE_n_sol_large}
\begin{align}
\label{eq:BE_n_sol_large_a}
\left(\frac{\Tgin}{\Te}\right)^{3}\big[1+\epsilon_N\big]&
\equiv \left(\frac{\Tg^{\rm eq}}{\Te}\right)^{3}\frac{1+\epsilon_N}{(1+\epsilon_\rho)^{3/4}}\approx \frac{2}{G^{\rm pl}_2}\,\expf{-\muc}
\\
\label{eq:BE_n_sol_large_b}
\left(\frac{\Tgin}{\Te}\right)^{4}\big[1+\epsilon_\rho\big]&
\equiv \left(\frac{\Tg^{\rm eq}}{\Te}\right)^{4}\approx \frac{6}{G^{\rm pl}_3}\,\expf{-\muc}.
\end{align}
\esub
Hence, the chemical potential grows as $\muc\approx 4\ln(0.9804\,\Te/\Tg^{\rm eq})$ with the electron temperature. In contrast, from Eq.~\eqref{eq:BE_n_sol_b} for small distortions we have $\muc\approx 3.602(\Te-\Tg^{\rm eq})/\Tg^{\rm eq}$.
These relation are useful for estimates of the expected range of electron temperatures.

\vspace{-0mm}
\subsubsection{Heat capacity of the photon-baryon fluid for $\mu(x)={\rm const}$}
\label{sec:heat_cap_muc}
By assuming that photon production processes are negligible, we can obtain explicit expressions for the heat capacity of the photon-baryon fluid. The energy density of the system is given by the sum of the photon and matter energy densities. The heat capacity at constant number of photons and constant volume is then given by
\begin{align}
\label{eq:heat_cap}
\frac{\partial\rho_{\rm tot}}{\partial T}\Bigg|_{N,V}&\approx\frac{3}{2}\,k\,N_{\rm b}+\frac{\partial\rho_{\gamma}}{\partial T}\Bigg|_{N,V},
\end{align}
where we used the non-relativistic expression for the thermal energy of the gas particles. The temperature appearing in the above equation is the electron temperature. The photon number density is given by $N_\gamma(T)=N_\gamma^{\rm pl}(T)\,\mathcal{G}_2(T)$, while the energy density is $\rho_\gamma(T)=\rho_\gamma^{\rm pl}(T)\,\mathcal{G}_3(T)$. Here, $N_\gamma^{\rm pl}(T)\propto T^3$ and $\rho_\gamma^{\rm pl}(T)\propto T^4$ are the corresponding blackbody relations. 
The derivative of the photon energy density with respect to $T$ then is given by
\begin{align}
\label{eq:heat_cap_rho}
\frac{\partial\rho_{\gamma}}{\partial T}\Bigg|_{N,V}=\frac{4\rho_{\gamma}}{T}-3kN_{\gamma}T \,\frac{\partial\muc}{\partial T}\Bigg|_{N,V}.
\end{align}
Usually, the chemical potential derivative is set to zero, as photons are quickly replenished (e.g., by the walls of the photon cavity).
However, here we assume a constant photon number. Thus, in our case, $\muc$ can be eliminated using 
\begin{align}
\label{eq:heat_cap_N}
0&=\frac{\partial N_{\gamma}}{\partial T}\Bigg|_{N,V}=\frac{3N_{\gamma}}{T}-2N_{\gamma} \,\frac{G_1^{\rm pl}\,\mathcal{G}_1(T)}{G_2^{\rm pl}\,\mathcal{G}_2(T)}\,\frac{\partial\muc}{\partial T}\Bigg|_{N,V}
\nonumber\\
\Rightarrow & \quad \frac{\partial\muc}{\partial T}\Bigg|_{N,V}=\frac{3}{2 T}
\,\frac{G_2^{\rm pl}\,\mathcal{G}_2(T)}{G_1^{\rm pl}\,\mathcal{G}_1(T)}.
\end{align}
Altogether this then yields
\begin{align}
\label{eq:heat_cap_rho_fin}
\frac{\partial\rho_{\gamma}}{\partial T}\Bigg|_{N,V}=\frac{4\rho_{\gamma}}{T}-\frac{9}{2}\,k\,N_{\gamma}
\,\frac{G_2^{\rm pl}}{G_1^{\rm pl}}\,\frac{\mathcal{G}_2(T)}{\mathcal{G}_1(T)}.
\end{align}
with $G_2^{\rm pl}/G_1^{\rm pl}\approx 1.4615$. Even if we initially assume $\muc=0$, this expression shows that a second term appears in addition to the expected first contribution. {\it How should we interpret this term?} By adding heat to the matter, photons up-scatter and thus reduce the effective increase of the matter temperature. However, since we assume that the number of photons is fixed, at this higher temperature, the heat capacity of the photon fluid is reduced in comparison with the heat capacity of a blackbody at the electron temperature. This reduction is described by the second term. The required energy to increase the matter temperature by some amount is thus significantly lower than it would be without this modification.
For an initial blackbody spectrum we obtain
\begin{align}
\label{eq:heat_cap_rho_Planck}
\frac{\partial\rho^{\rm pl}_{\gamma}}{\partial T}\Bigg|_{N,V}=
\left[4\frac{G_3^{\rm pl}}{G_2^{\rm pl}}-\frac{9}{2}
\,\frac{G_2^{\rm pl}}{G_1^{\rm pl}}\right]\,k\,N^{\rm pl}_{\gamma}(T)\approx 4.2279\,k\,N^{\rm pl}_{\gamma}(T).
\end{align}
Without the correction this would be $\partial_T\rho^{\rm pl}_{\gamma}\big|_{N,V}\approx 10.805\,k\,N^{\rm pl}_{\gamma}(T)$, so more than a factor of 2 larger.

It is interesting to ask when this aspect is relevant. Even when we include DC and BR photon production, we still find that the basic picture is not changed that much. The integrals $\mathcal{G}_k(T)$ are merely modified a little, unless thermalization processes are so rapid that full equilibrium is basically restored instantaneously. In the early Universe, this is only possible during the first $\simeq 10^6\,{\rm s}$, corresponding to up to about a month after the big bang. The baryon density is $N_{\rm b} \gtrsim 10^{11}\,\cm^{-3}$ at those times, but afterwards, even until today, some $\simeq 14$ billion years later, the spectrum would still be distorted \citep{Burigana1991, Hu1993}, and thus the heat capacity reduced, violating the simple adiabatic case. On the other hand, thinking about a blackbody in a cavity with reflecting walls, one can likely assume adiabatic conditions at any state given that the thermalization timescales are those of electrons in a solid state.

\subsection{Generalization to include photon production}
\label{sec:BE_solutions_photon_production}
In the early Universe, photon production by DC and BR are not negligible and the momentary spectrum will no longer be described by a Bose-Einstein spectrum with {\it constant} chemical potential. However, generalizing to a frequency-dependent chemical potential $\mu(x)$ allows to describe the new quasi-stationary state. We are again not interested in the precise evolution of the system during the very short relaxation time that is needed to reach the quasi-stationary state. Let us first only consider heating with no external photon injection. Right after the energy release, the electron temperature will increase above $\Tg^{\rm eq}$. Then up-scattering of photons will reduce the temperature with the quasi-stationary state given by Eq.~\eqref{eq:BE_n_sol} were we neglect DC and BR emission. However, the DC and BR photon production during the relaxation process will modify the resulting quasi-stationary state in two ways: i) the Compton equilibrium temperature will decrease slightly and ii) the number of photons is no longer just defined by $\epsilon_N=0$. Indeed because of ii), without following the exact relaxation of the system to the quasi-stationary state we cannot precisely perform the mapping from $\Tgin$, $\epsilon_N$ and $\epsilon_\rho$ to the quasi-stationary state of the spectrum shortly after.

{\it How can we then make progress?} By simply ignoring the exact physics of the relaxation process and imposing (or computing) the conditions Eq.~\eqref{eq:BE_n_sol} after the relaxation is finished.
This perspective defines a two-step scheme to describing the evolution of spectral distortions in the early Universe. We first need to determine a solution for the {\it shape} of the spectral distortion by means of the frequency-dependent chemical potential, $\mu(x)$, and then determine the slow quasi-stationary evolution of the photon number by thermalization processes afterwards using the solution for $\mu(x)$. In the second step, one can think of the evolution of $N_\gamma$ as being linked to the high-frequency amplitude of the chemical potential. This essentially imposes suitable normalization conditions on the average chemical potential \citep{Chluba2014}. In contrast to the evolution of small distortions, we will generally find that the shape also depends significantly on the total amplitude of the distortion, thus leading to non-linear effects that determine the precise trajectory that the photon distribution will take along the sequence of quasi-stationary states. Note, however, that this simplified picture only applies to single energy release, as we detail below.

\subsubsection{Generalized evolution equations for $\Te$ and $\mu_\rho$}
\label{sec:BE_solutions_photon_production_evol_eq}
We can now obtain the required evolution equation for the photon number and energy densities during the quasi-stationary stage, assuming that we already have the solution for $\mu(x)$. It is then easy to show that in the expanding Universe we have 
\bsub
\label{eq:evol_N_rho}
\begin{align}
\frac{\id a^3 N_\gamma}{a^3 \id \tau}&=\frac{3N_{\gamma}}{a\Te}\,\frac{\id a\Te}{\id \tau}-N_\gamma \frac{\int \frac{x^2\expf{x+\mu(x)}\id x}{(\expf{x+\mu(x)}-1)^2} \frac{\id \mu(x)}{\id \tau}}{\int \frac{x^2\id x}{\expf{x+\mu(x)}-1}},
\\
\frac{\id a^4 \rho_\gamma}{a^4 \id \tau}&=\frac{4 \rho_{\gamma}}{a\Te}\,\frac{\id a\Te}{\id \tau}-\rho_\gamma \frac{\int \frac{x^3\expf{x+\mu(x)}\id x}{(\expf{x+\mu(x)}-1)^2} \frac{\id \mu(x)}{\id \tau}}{\int \frac{x^3\id x}{\expf{x+\mu(x)}-1}}.
\end{align}
\esub
Defining the weighted average
\begin{align}
\nonumber
\left<Y(x)\right>_k&
=\frac{\int x^k n(1+n) Y(x)\id x}{\int x^k n \id x}
=\frac{\int \frac{x^k \expf{x+\mu(x)}\id x}{(\expf{x+\mu(x)}-1)^2} Y(x)}{\int \frac{x^k\id x}{\expf{x+\mu(x)}-1}},
\end{align}
this can also be case into the more compact form
\bsub
\label{eq:evol_N_rho_compact}
\begin{align}
\label{eq:evol_N_rho_compact_a}
\frac{\id \ln a^3 N_\gamma}{\id \tau}&=3\frac{\id \ln a\Te}{\id \tau}-\left<\frac{\id \mu}{\id \tau}\right>_2,
\\
\frac{\id \ln a^4 \rho_\gamma}{\id \tau}&=4\frac{\id \ln a\Te}{\id \tau}-\left<\frac{\id \mu}{\id \tau}\right>_3.
\end{align}
\esub
We now introduce $\mu(x, \tau)=\mu_\rho(\tau)\,\xi(x, \tau)$, where $\mu_\rho(\tau)$ is a suitable amplitude of the chemical potential, e.g., always defined at $x=10$. While this choice is arbitrary, at high frequencies we find $\mu(x)\simeq {\rm const}$, further motivating this definition. With this we then obtain the two evolution equations
\bsub
\label{eq:evol_Te_murho_compact}
\begin{align}
\frac{\id \ln a\Te}{\id \tau}
&=
\frac{\mathcal{M}_2\id \ln a^4 \rho_\gamma}{\kappa \id \tau}-\frac{\mathcal{M}_3\id \ln a^3 N_\gamma}{\kappa\id \tau}
\\ \nonumber&\qquad\qquad
+\frac{\mu_\rho}{\kappa}\left[\mathcal{M}_2\left<\frac{\id\xi}{\id \tau}\right>_3
-\mathcal{M}_3\left<\frac{\id\xi}{\id \tau}\right>_2\right],
\\[1mm]
\frac{\id \mu_\rho}{\id \tau}
&=\frac{3\id \ln a^4 \rho_\gamma}{\kappa\id \tau}-\frac{4\id \ln a^3 N_\gamma}{\kappa\id \tau}
\nonumber\\&\qquad\qquad
+\frac{\mu_\rho}{\kappa}\left[3\left<\frac{\id \xi}{\id \tau}\right>_3
-4\left<\frac{\id \xi}{\id \tau}\right>_2\right],
\\[1mm]
\mathcal{M}_k&=\left<\xi(x, \tau)\right>_k, 
\quad
\kappa=4\mathcal{M}_2-3 \mathcal{M}_3.
\end{align}
\esub
For $\mu_\rho\ll 1$, the problem can be linearized and we obtain the formulation given by Eq.~(9) of \citet{Chluba2014}.

The above equations can be further reduced by eliminating the terms $\propto \id \ln a^4 \rho_\gamma/\id \tau$. Using Eq.~\eqref{eq:evol_matter}, we can write
\begin{align}
\nonumber
\frac{\id a^4 \rho_\gamma}{a^4 \id \tau} + \frac{\id a^3 \rho_{\rm b}}{a^3 \id \tau}+3\frac{H}{\taudot} P_{\rm b}
&=\left.\frac{\id a^4 \rho_{\rm b}}{a^4 \!\id \tau}\right|_{\rm S}+\left.\frac{\id a^3 \rho_{\rm b}}{a^3 \!\id \tau}\right|_{\rm h} \equiv \dot{Q}_{\rm ex}
\end{align}
for the photon-baryon system. Here, $\dot{Q}_{\rm ex}$ defines the energy density added externally to the photon-baryon system, thought of as heat and photons. Here, 'external' can mean the decay of a dark matter particle or some scalar field or even PBH evaporation.
Since 
\begin{align}
\nonumber
\frac{\id a^3 \rho_{\rm b}}{a^3 \id \tau}+3\frac{H}{\taudot} P_{\rm b}
&=\frac{3}{2} k N_{\rm b} \Te \left[\Big(1  + \lambda \Big) \frac{\id \ln a\Te}{\id \tau} + \Big(1  - \lambda \Big) \frac{\id \ln a}{\id \tau}\right],
\end{align}
with $C_V^*=\frac{3}{2} k N_{\rm b}(1+\lambda)$ we find
\begin{align}
\nonumber
\frac{\id \ln a^4 \rho_\gamma}{\id \tau} 
&= \frac{\dot{Q}_{\rm ex}}{\rho_\gamma}
-\frac{C_V^*\Te}{\rho_\gamma} 
\left[\frac{\id \ln a\Te}{\id \tau} + \frac{1  - \lambda}{1  + \lambda} \frac{\id \ln a}{\id \tau}\right].
\end{align}
Together with Eq.~\eqref{eq:evol_N_rho_compact_a}, after a few rearrangements, we then have
\begin{align}
\label{eq:evol_murho_final}
\frac{\id \ln a\Te}{\id \tau}
&=
\frac{\mathcal{M}_2}{\kappa^*} \frac{\dot{Q}^*_{\rm ex}}{\rho_\gamma}-\frac{\mathcal{M}_3\id \ln a^3 N_\gamma}{\kappa^*\id \tau}
\nonumber \\ \nonumber&\qquad\qquad
+\frac{\mu_\rho}{\kappa^*}\left[\mathcal{M}_2\left<\frac{\id\xi}{\id \tau}\right>_3
-\mathcal{M}_3\left<\frac{\id\xi}{\id \tau}\right>_2\right]
\nonumber
\\[1mm]
\nonumber
&\equiv \frac{\dot{Q}^{\rm b}_{\rm ex}}{C_V^*\Te}-\frac{1  - \lambda}{1  + \lambda} \frac{\id \ln a}{\id \tau}
-\frac{\rho_\gamma}{C_V^*\Te} \frac{\id \ln a^4 \rho_\gamma}{\id \tau}\Bigg|_{\rm CS+em},
\\[2mm]
\frac{\id \mu_\rho}{\id \tau}
&=\frac{3}{\kappa^*} \frac{\dot{Q}^*_{\rm ex}}{\rho_\gamma}-\frac{4 \alpha \id \ln a^3 N_\gamma}{\kappa^*\id \tau}
+\frac{\mu_\rho}{\kappa^*}\left[3\left<\frac{\id \xi}{\id \tau}\right>_3
-4\alpha \left<\frac{\id \xi}{\id \tau}\right>_2\right],
\nonumber 
\\[1mm]
\frac{\dot{Q}^*_{\rm ex}}{\rho_\gamma}&=\frac{\dot{Q}_{\rm ex}}{\rho_\gamma}-\frac{C_V^*\Te}{\rho_\gamma} 
\frac{1  - \lambda}{1  + \lambda}\frac{\id \ln a}{\id \tau}, \quad \dot{Q}_{\rm ex}=\dot{Q}^{\rm b}_{\rm ex}+\dot{Q}^\gamma_{\rm ex}
\\[1mm]
\nonumber
\kappa^*&=4\alpha\mathcal{M}_2-3 \mathcal{M}_3, 
\quad
\alpha=1+\frac{C_V^*\Te}{4\rho_\gamma},
\end{align}
where $\dot{Q}^{\rm b}_{\rm ex}$ and $\dot{Q}^\gamma_{\rm ex}$ describe the external sources of heating for the baryons and photons (e.g., via photon injection) separately. We also introduced $\left.\id \ln a^4 \rho_\gamma/\id \tau\right|_{\rm CS+em}$ to include both the contributions from Compton scattering and DC/BR emission (see Sect.~\ref{sec:QS_sols_gamma}).

Because $\frac{3}{2} k N_{\rm b}\Te\ll 4\rho_\gamma$, we have $\alpha \approx 1$ and thus $\kappa^*\approx \kappa$, which we will use in our computations below. The second term of $Q^*_{\rm ex}/\rho_\gamma$ is due to the adiabatic cooling of the baryons and causes a negative chemical potential distortion of order $\mu_\rho\simeq -\pot{3}{-9}$ \citep{Chluba2005, Chluba2011therm}, but will also be neglected here. 
The corrections due to changes in the shape of the distortions ($\propto \id \xi/\id \tau$) usually also remain negligible until redshifts below the quasi-stationary phase \citep{Chluba2014}, as we also discuss below.

Equation~\eqref{eq:evol_murho_final} is still general and could be directly solved by adding the photon Boltzmann equations for $\xi(x, \tau)$, describing the evolution of the main macroscopic parameters of the distorted photon field under continuous energy release and with external photon injection.\footnote{Note that in this case, $\id \ln a^3 N_\gamma/\it \tau$ generally has two contributions, one from DC and BR due to the distortion and the other from external sources.} However, in this case time-dependent corrections come into play such that the photon emission process is modified. These can no longer be solved independently, since the shape of the distortion will depend directly on the energy release history. Below, we assume that only one single energy injection occurred. In this case, time-dependent effects can be included approximately, but we leave a more general discussion to future work.

\vspace{-3mm}
\subsection{Computational procedure}
\label{sec:procedure}
We now have a formulation of the full evolution problem in the quasi-stationary phase. 
Given some initial redshift to fix the matter number densities and initial temperature of the plasma, we can obtain a solution for $\mu(x)$ assuming quasi-stationary conditions. Since we need to compute the Comptonization of the photon field, which directly depends on the electron temperature, it is most convenient to define the injection process by providing $\Te$ together with $\epsilon_N$. This avoids having to update the Compton scattering kernel, which greatly accelerates the computation. This then determines the expected value for $\muc$ assuming that DC and BR are negligible. By next demanding that the energy density of the photons does not change one can obtain a quasi-stationary solution for $\mu(x)$. The evolution towards the quasi-stationary condition leads to a change in the number of photons depending on the initial conditions. After the quasi-stationary state is reached one can then compute the values of $\epsilon_N$ and $\epsilon_\rho$ with respect to the initial blackbody that were required to reach the final quasi-stationary state under consideration. Since during the evolution towards the quasi-stationary state we fixed the photon energy density, this means that $\epsilon_N$ has to adjust slightly with respect to the initial condition given at this step. However, this then determines $\mu=\mu(x, t, \epsilon^*_N, \epsilon_\rho)$, where $\epsilon^*_N$ is close to $\epsilon_N$ but with slight modifications from DC and BR emission during the relaxation process. This completes the first step of the calculation.

In the next step, we first compute the rate of change of the photon number density caused given the solution for $\mu$. Crucially, $\mu(x)\rightarrow 0$ at $x\ll 1$, which makes the net emission integral finite but would not be the case for $\mu={\rm const}$ (see Sect.~\ref{sec:DC_terms}). Similarly, we can compute the integrals $\mathcal{M}_k$, which together with Eq.~\eqref{eq:evol_murho_final} allows us to advance the solution for $\mu_\rho$. By calculating tables of $\mu=\mu(x, t, \epsilon_N, \epsilon_\rho)$ for various values of $\epsilon_N$ and $\epsilon_\rho$ we can then pre-tabulate all the required emissivities and integrals to quickly advance the solution from the initial state to a final time when DC and BR emission start freezing. This then completes the second step of the calculation. 

As a last step, by comparing the initial amount of energy in the distortion to the final value we can then obtain the {\it distortion visibility function} for different cases of single injection. Assuming very small energy release implies that $\mu\approx \mu_\rho(t, \epsilon^*_N, \epsilon_\rho)\,\xi(x, t)$ with $\xi(x, t)$ not directly depending on the initial values for $\epsilon_N$ and $\epsilon_\rho$. This case was studied in detail in \citet{Chluba2014} including various corrections to the classical solutions of \citet{Sunyaev1970mu} and \citet{Danese1982}. In this work, we can now add modifications to the thermalization efficiency in the regime of large initial distortions. We will find that this implies the visibility of distortions is significantly higher for large energy release.

\vspace{-3mm}
\subsubsection{Computational procedure for single injection}
\label{sec:procedure_single}
For a single injection of energy and/or photons, the problem further simplifies. We require that the final energy density of the photons equals that of the CMB blackbody with temperature $\Tg=T_0(1+\zh)$ at $\zh$. For a given cosmology this also fixes the number densities of the baryonic component, which affects the BR emissivity, and ensure that the subsequent expansion is described by the standard Hubble parameter, $H(z)$. Since for the computation it is beneficial to use a fix electron temperature, by choosing $\Te$ with Eq~\eqref{eq:BE_n_sol_b} and $\Tgin(1+\epsilon_\rho)^{1/4}=\Tg(\zh)$ this directly fixes $\muc$. By requiring a fixed photon energy density, we can then move towards the full quasi-stationary solution $\mu(x)$ at redshift $\zh$. Since the electron temperature is given by the Compton equilibrium, all that matters is the amplitude of the distortion, $\mu_\rho(\zh)$. Through Eq.~\eqref{eq:BE_n_sol_a}, we can then also fix $\Tgin(1+\epsilon_N)^{1/3}$. Together with $\Tgin(1+\epsilon_\rho)^{1/4}$ (which is also determined) this allows us to map any combination of $\epsilon_N$ and $\epsilon_\rho$ onto this case, by essentially adjusting $\Tgin$. This highlights that for a single injection all that matters is the temperature difference 
\begin{align}
\label{eq:DT_Nrho}
\frac{\Delta T_{N\rho}}{\Tg}&=\frac{\Tgin(1+\epsilon_N)^{1/3}-\Tgin(1+\epsilon_\rho)^{1/4}}{\Tgin(1+\epsilon_\rho)^{1/4}}=\frac{(1+\epsilon_N)^{1/3}}{(1+\epsilon_\rho)^{1/4}}-1,
\end{align}
which determines how far away from the final equilibrium temperature $\Tg$ the photon distribution still is. Several injection scenarios therefore lead to exactly the same initial distortion evolution. Balanced injection scenarios, where just the right amount of photons and energy was added to the plasma start with $\Delta T_{N\rho}/\Tg=0$, whereas scenarios with $\epsilon_N=0$ (i.e., pure energy injection) start with $\Delta T_{N\rho}/\Tg<0$ (i.e., a deficit of photons). In the subsequent evolution this deficit is continuously reduced by DC and BR, which manifests in a reduction of $\mu_{\rho}$. For $\epsilon_N>(1+\epsilon_\rho)^{3/4}-1\approx (3/4) \epsilon_\rho$ the photon field contains more photons than the equilibrium blackbody and thus DC and BR cause a net absorption. 
Overall, we thus only need to compute the quasi-stationary solutions for $\mu(x)$ at various redshifts $\zh$ and for various values of $\Te$, which then allows us to determine the solution $\mu(\zh, \mu_\rho)$ across time.  With these solutions, we can compute all the required rates, e.g., DC and BR photon production rate, to allow us advancing the solution using Eq.~\eqref{eq:evol_murho_final}. The latter are stored in tables for multiple evaluations.

\vspace{-3mm}
\subsection{Continuous injection and cosmology dependence}
\label{sec:procedure_cont_cosmo}
The above procedure breaks down when continuous energy release scenarios are being considered. This statement is even true for small distortion scenarios, once time-dependent corrections are included. {\it How can we see this?}
Neglecting any time-dependent corrections to the shape of the distortion, one can in principle compute the required solutions for the DC and BR emission, and then follow the above procedure. However, time-dependent corrections modify the shape of the distortion and then in turn affect the DC and BR emission rates. For a single energy release, this can again be approximately included as we discuss below, but for continuous energy release, the solution no longer becomes independent of the injection history. This statement applies to both small and large distortions, such that visibility approaches generally are expected to become inaccurate \citep{Chluba2005, Khatri2012b, Chluba2014}. 

One main way around this problem is to compute extended tables for various energy release scenarios. In a similar way, changes of the cosmological parameters can in principle be all covered by using a few mappings that relate redshift or time to matter densities. However, the dimensionality of the problem quickly increases, rendering such an approach moot. 
In addition, for large energy injection, we generally cannot ignore the effect on the expansion rate and evolution of the scale factor. For example, if a massive particle decays into radiation the redshifting laws change and correspondingly the Hubble factor. Even if for single injection this is not an issue (the expansion rate is assumed to match that of the standard case, after the injection), for scenarios with continuous energy release this can become relevant.

As these statements illustrate, whenever going beyond the simplest case of single injection the problem becomes more complex and it is best to solve the full set of evolution equations for the matter temperature and distortion shape. Current thermalization codes are already quite efficient in achieving reasonable runtimes and we plan to investigate more general cases in some future work.

\vspace{-3mm}
\section{Quasi-stationary solutions}
\label{sec:QS_sols_gamma}
In the previous section we mainly focussed on the macroscopic aspects of the distortion evolution. This has to be supplemented by a microphysical solution for the distortion shape. To obtain the quasi-stationary solutions of the problem, we now first specify the various collision terms in the photon evolution equation, Eq.~\eqref{eq:Dn_evol}. We then explain how to obtain numerical solutions of the problem using a Fredholm equation approach. Finally, we give simple analytic approximations that we will compare to below.

\begin{figure}
\centering 
\includegraphics[width=\columnwidth]{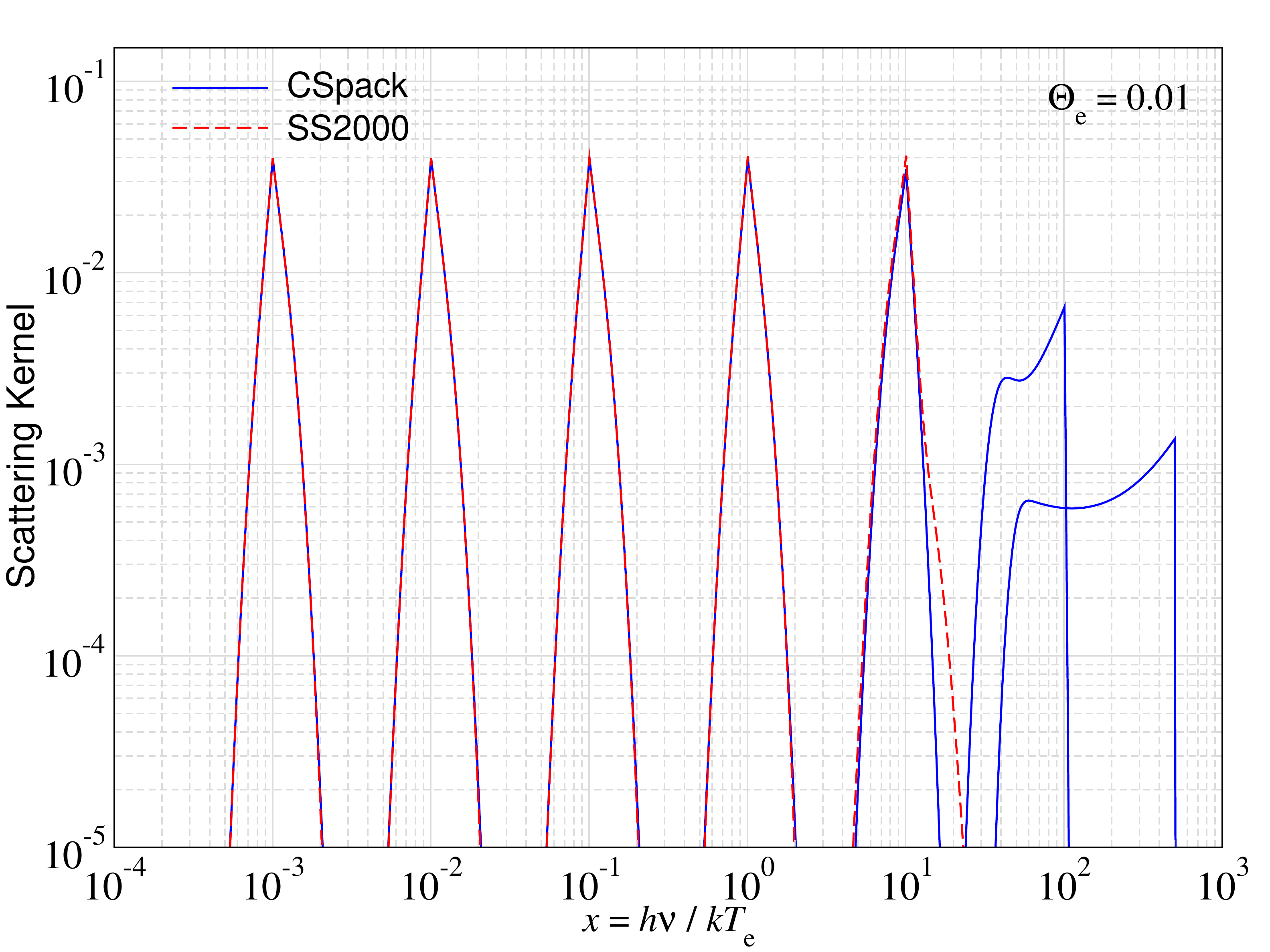}
\\[1mm]
\includegraphics[width=\columnwidth]{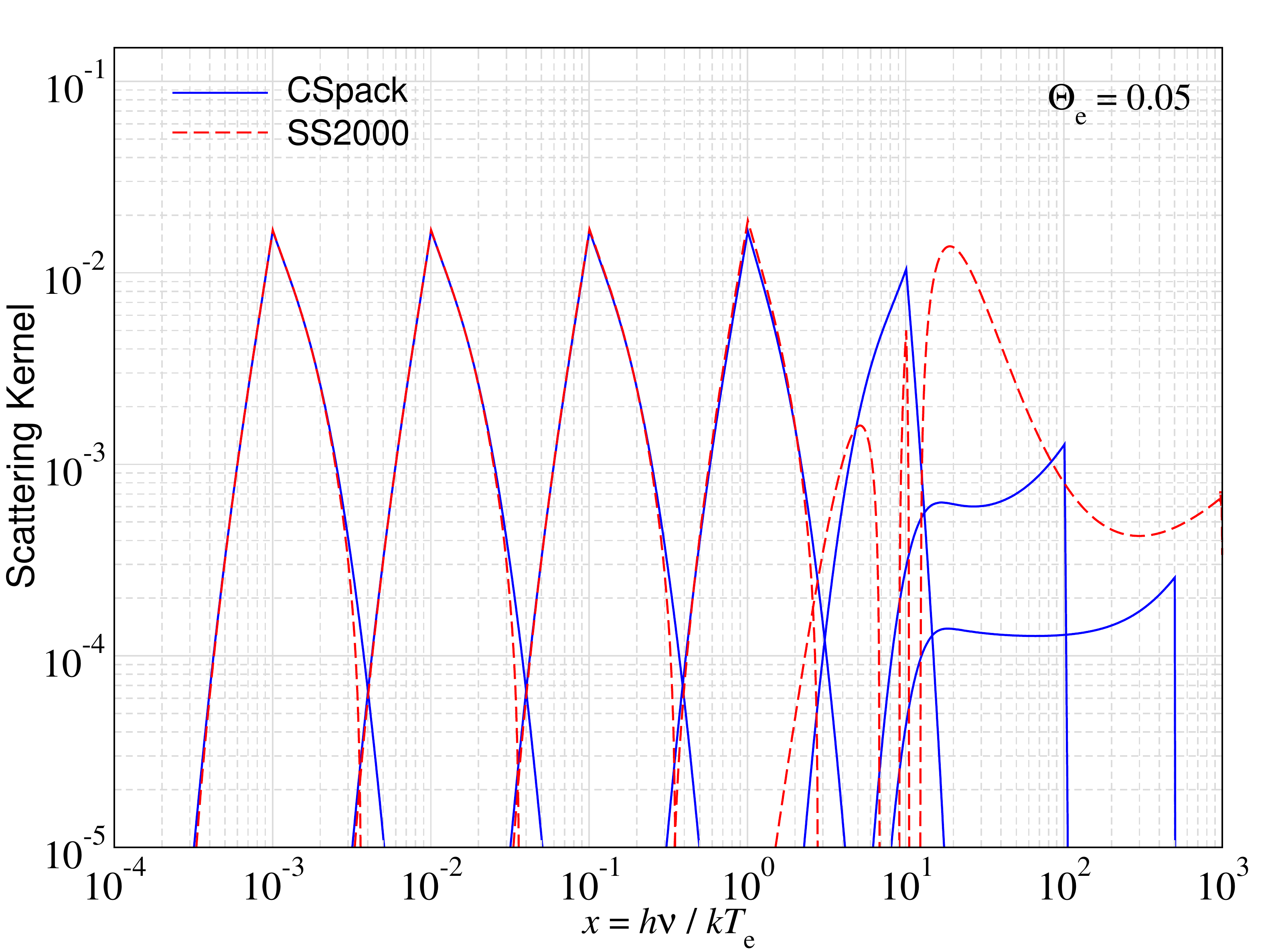}
\caption{Compton kernel, $P(\omega_0 \rightarrow \omega, \The)$ [multiplied by appropriate weight factors], for $\The = 0.01$ (upper panel) and $\The = 0.05$ (lower panel) and various central frequencies $x_0=\omega_0/\The$.  The exact result obtained with {\tt CSpack} is compared with the kernel approximation given by Eq.~(19) of \citet{Sazonov2000}, showing very good agreement up to $x\simeq 10$ for $\The = 0.01$. The cases for $x_0=10^2$ and $500$ are for {\tt CSpack} only, as the approximation fails. For $\The = 0.05$, departures of the approximation from the exact result already become visible at $x\simeq 1$ and are catastrophic at $x=10$.}
\label{fig:Kernel}
\end{figure}

\vspace{-2mm}
\subsection{Compton scattering terms}
\label{sec:CS_terms}
We can describe the redistribution of photons via Compton scattering using the photon collision term \citep[e.g.,][]{Sazonov2000, CSpack2019}
\begin{align}
\label{eq:col_CS}
\left.\frac{\text{d}n}{\text{d}\tau} \right|_{\rm CS}
&= \int P(\omega \rightarrow \omega', \The) \,\Big[\expf{x'-x} \,n'(1+n) - n(1+n') \Big]\,\id\omega'.
\end{align}
Here, $\The=k\Te/\me c^2$, $\omega=h\nu/\me c^2$, and $n=n(x)$ and $n'=n(x')$ described the photon occupation number at the dimensionless frequencies $x=h\nu/k\Te=\omega/\The$ and $x'=h\nu'/k\Te=\omega'/\The$. The Compton scattering kernel, $P(\omega \rightarrow \omega', \The)$, can be efficiently and accurately computed using {\tt CSpack} \citep{CSpack2019}. For some cases, we will also use the expressions of \citet{Sazonov2000}, however, these are valid in a more limited range of photon energies and electron temperatures. The shape of the scattering kernel at several frequencies is shown in Fig.~\ref{fig:Kernel}, illustrating this point. From the figure we can also see that Klein-Nishina corrections start to become important at high frequencies, with an overall suppression of the kernel amplitude. These corrections modify the high-frequency part of the distortion, as we show below.

We now insert a Bose-Einstein spectrum into Eq.~\eqref{eq:col_CS}, but generalize to a {\it frequency-dependent} chemical potential $\muc\rightarrow \mu(x)$. After a few rearrangements, this then yields
\begin{align}
\label{eq:col_CS_mu}
\frac{1}{n}\,\left.\frac{\text{d}n}{\text{d}\tau} \right|_{\rm CS}
&= \int P(\omega \rightarrow \omega', \The) \,(1+n')\,\Big[\expf{\mu-\mu'} -1\Big]\id\omega'
\nonumber\\
&= \int P(\omega \rightarrow \omega', \The) \,\frac{\expf{\mu-\mu'} -1}{1-\expf{-x'-\mu'}}\id\omega',
\end{align}
where $\mu=\mu(x)$ and $\mu'=\mu(x')$. Written in this form it is immediately evident that in equilibrium one has $\mu=\mu'={\rm const}$.
We can also observe the presence of stimulated scattering effects, which depend on the factor $1+n'=1/[1-\expf{-x'-\mu'}]$. For small chemical potential one has $1+n' \approx 1/x'$, which causes photons to up-scatter more slowly than without this term \citep{Chluba2008d}. This term is highly relevant to the main shape of the distortion as we demonstrate below by omitting it in the computation.

\subsection{Photon production terms}
\label{sec:DC_terms}
The production of photons by DC and BR can be described using \citep[e.g.,][]{Chluba2011therm, Chluba2015GreensII}
\begin{align}
\label{eq:col_em}
\left.\frac{\text{d}n}{\text{d}\tau} \right|_{\rm em}
&= \frac{\Lambda(x)\,\expf{-x}}{x^3}\,\Big[1-n\,(\expf{x}-1)\Big]
\end{align}
where the emission coefficient, $\Lambda(x)$, includes both DC and BR. It depends on the temperature and density of particles in the medium but generally varies slowly with frequency. The dominant scaling is this determined by the factor $\expf{-x}/x^3$, which renders the emission term most important at low frequencies.
Inserting a Bose-Einstein spectrum into Eq.~\eqref{eq:col_em}, we find
\begin{align}
\label{eq:col_em_mu}
\frac{1}{n}\,\left.\frac{\text{d}n}{\text{d}\tau} \right|_{\rm em}
&= \frac{\Lambda(x)}{x^3}\,\Big[\expf{\mu}-1\Big].
\end{align}
Here it becomes clear that only for $\mu=0$ do emission and absorption terms vanish. 

We can also directly compute the net photon emission rate that is needed when evolving the chemical potential with Eq.~\eqref{eq:evol_murho_final}. The change in the number of photons is $\propto \int x^2 \id n/\id \tau \id x$. We therefore may write
\begin{align}
\label{eq:DN_N_def}
\left.\frac{\id \ln a^3 N_\gamma}{\id \tau}\right|_{\rm em}
&= \frac{\int \frac{\Lambda(x)}{x}\,n \Big[\expf{\mu}-1\Big]\id x}{\int x^2 n \id x}.
\end{align}
Due to the $1/x$ scaling of the numerator, it is apparent that for constant $\mu$ this integral cannot be evaluated without truncating at the low-frequency side. However, this is hardly surprising, since constant $\mu$ means we did indeed neglect DC and BR in the consideration (i.e., $\Lambda(x)=0$). Once included, $\mu$ vanishes rapidly as $x$ decreases, naturally regularizing the integral. For small distortions, one has \citep{Sunyaev1970mu, Chluba2014}
\begin{align}
\label{eq:DN_N_small}
\left.\frac{\id \ln a^3 N_\gamma}{\id \tau}\right|_{\rm em}
&\approx \frac{\The \xc \mu_\rho}{G_2^{\rm pl}},
\end{align}
where $\xc$ is the critical frequency, which is defined by the competition between DC+BR emission and the Compton process (see below). As a rule of thumb, the thermalization efficiency thus drops with $\xc$, however, details of the active emission region around $\xc$ make the assessment a little more complicated in detail.
We also mention that naturally $\id \ln a^3 N_\gamma/\id \tau=0$ for the Compton process.

\vspace{-1mm}
\subsubsection{Computing the BR emission term}
To compute $\Lambda(x)$ we use {\tt BRpack} \citep{BRpack2020} for the modeling of the BR emission. {\tt BRpack} integrates the differential cross section of \citet{ElwertHaug1969}, which is based on Sommerfeld-Maue eigenfunctions for the electron in a Coulomb potential.
This allows us to accurately capture the differences in the Gaunt factors for hydrogen and helium for non-relativistic to mildly relativistic temperatures. We thereby overcome several limitations of the fitting formulae of \citet{Draine2011Book} and \citet{Itoh2000}, which were previously used in {\tt CosmoTherm}. However, the differences are not as important during most of the quasi-stationary evolution phase, since DC emission dominates over BR at $z\gtrsim \pot{4}{5}$ \citep{Hu1993}, such that we do not go into more detail here.

\vspace{-1mm}
\subsubsection{Computing the DC term}
While {\tt BRpack} produces highly accurate results for the  BR Gaunt factors, the DC process can only be modeled more approximately. Given that in the quasi-stationary evolution phase DC is the most important emission process, we shall compare the results of several approximations below. The classical Lightman-Thorne approximation reads\footnote{In reality a factor $\expf{x}$ should be added here but consistent with the limit for which the approximation is valid we drop it.} \citep{Lightman1981, Thorne1981}
\begin{align}
\label{eq:Lam_LT}
\Lambda_{\rm LT}(x)&=\frac{4\alpha}{3\pi}\,\The^2 \int x^4 n(1+n) \id x, 
\end{align}
which is frequency independent and neglects corrections due to the energy of the incoming photon or moving electron. In thermalization computations for $\mu\ll 1$, the DC emission integral gives $\mathcal{I}_4=\int x^4 n(1+n) \id x \approx \int x^4 \nbb(1+\nbb) \id x=4\pi^4/15\approx 25.976$. Most of the DC emission comes from photons with energy 
\begin{align}
\bar{x}=\frac{\int x^5 n(1+n) \id x}{\int x^4 n(1+n) \id x}\approx 4.79,
\end{align}
where the chemical potential becomes quasi-constant. For cases $\mu\gg 1$, $\bar{x}$ increases slightly approaching $\bar{x}=5$. 

For non-zero $\mu$, one simple augmentation of the Lightman-Thorne approximation is the replacement $\Lambda(x)\rightarrow \Lambda(x)\,\expf{-\bar{\mu}}$ \citep{Chluba2005}, where $\bar{\mu}=\mu(\bar{x})$. This already captures the leading order effects for increasing values of $\mu$. A more accurate approximation assuming constant $\mu=\bar{\mu}$ is 
\begin{align}
\label{eq:mu_Int}
\mathcal{I}_4&=\int x^4 n(1+n) \id x=24\,{\rm Li}_4(\expf{-\bar{\mu}})
=24\,\expf{-\bar{\mu}}\sum_{k=0}^\infty \frac{\expf{-k\bar{\mu}}}{(k+1)^4}
\nonumber\\
&\approx 24\,\expf{-\bar{\mu}}\left[1+\frac{\expf{-\bar{\mu}}}{16}+\frac{\expf{-2\bar{\mu}}}{81}+\frac{\expf{-3\bar{\mu}}}{256}+\ldots\right], 
\end{align}
where ${\rm Li}_n(z)$ denotes the polylogarithm. The sum converges very quickly and usually including $\simeq 10$ terms is sufficient. For $\bar\mu=0$, the series naturally reduces to $\mathcal{I}^{\rm pl}_4\approx4\pi^4/15\approx 25.976$.

\begin{figure}
\centering 
\includegraphics[width=\columnwidth]{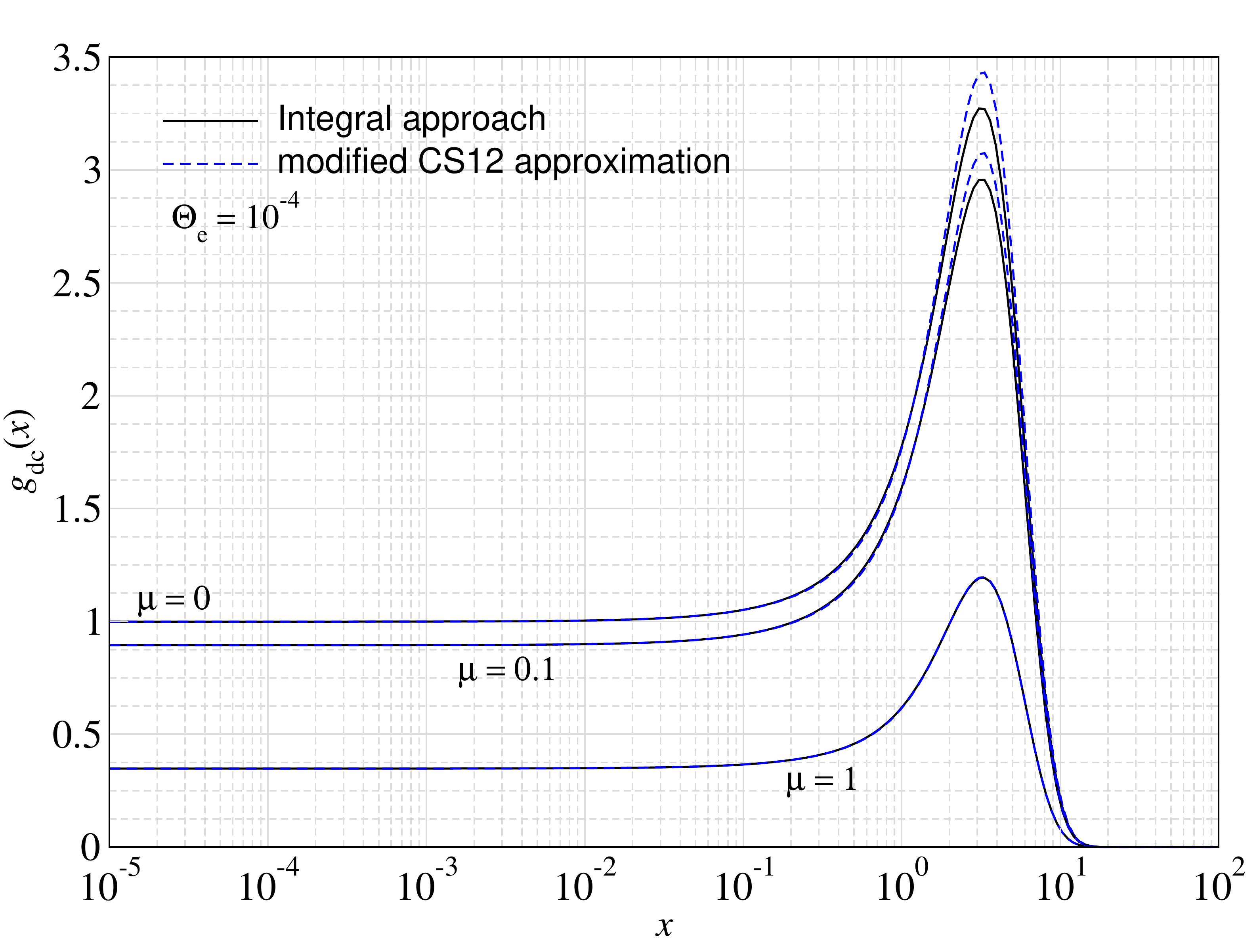}
\\[-0mm]
\includegraphics[width=\columnwidth]{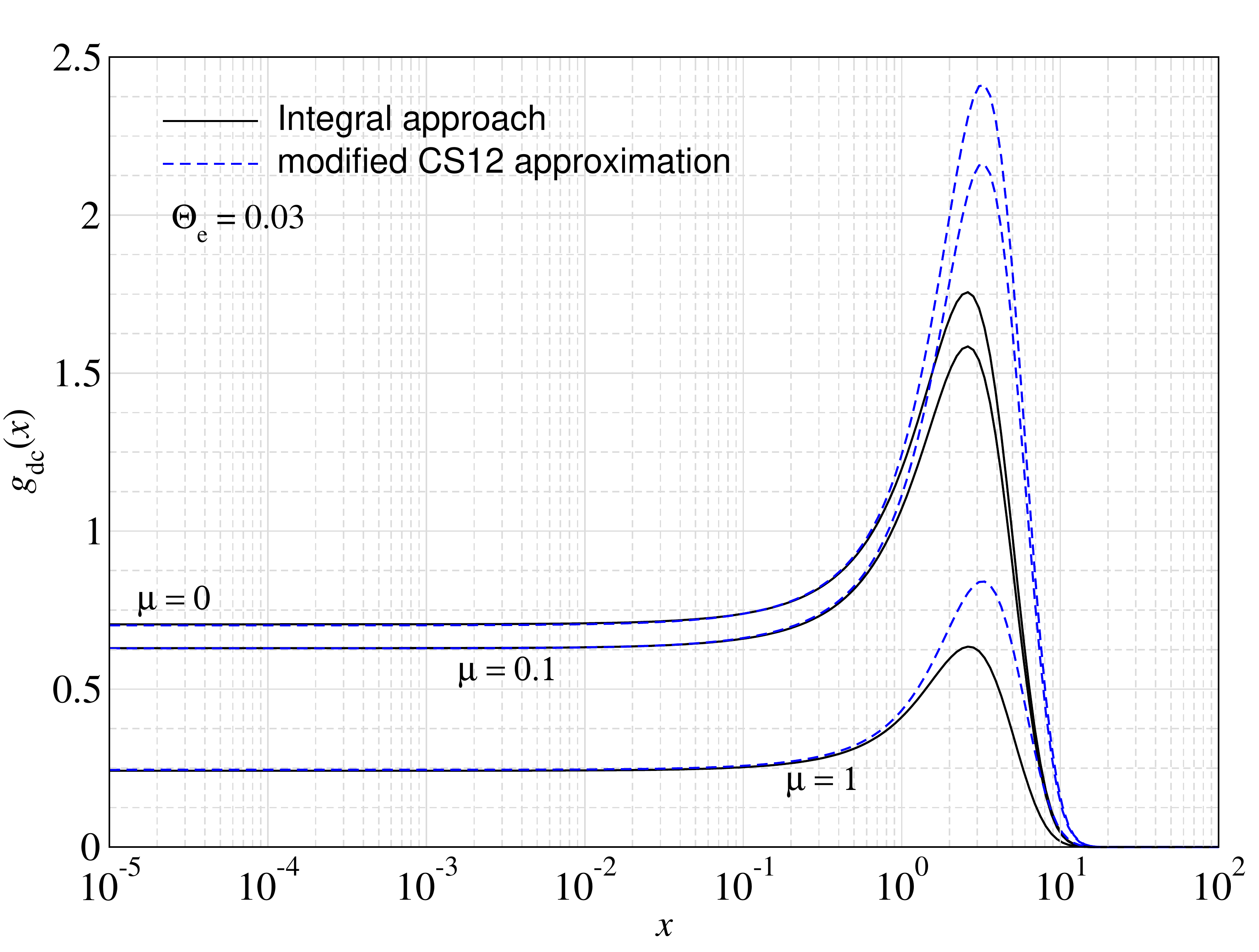}
\caption{DC Gaunt factor for $\The=10^{-4}$ (upper panel) and $\The=0.03$ (lower panel) and various values of constant $\mu$. The dashed lines are the approximation Eq.~\eqref{eq:Lam_CS} modified by the factor $g_{\mu}=\mathcal{I}_4/\mathcal{I}^{\rm pl}_4$ from Eq.~\eqref{eq:mu_Int}, while the solid lines give the results from the integral approach, Eq.~\eqref{eq:Lam_improved}. Overall, the approximation modified CS12 approximation works very well at low frequencies, but shows departures from Eq.~\eqref{eq:Lam_improved} at high frequencies.}
\label{fig:gdc_mu}
\end{figure}

As shown previously \citep{Chluba2005, Chluba2007a}, additional relativistic corrections to the DC emissivity become important at early times. The main corrections have already been included in {\tt CosmoTherm} and individual effects were discussed in \citet{Chluba2014}. Close to equilibrium, the most important effect is a suppression of the DC emissivity with temperature \citep{Chluba2005, Chluba2007a}. This is counteracted by frequency-dependent corrections which account for effects beyond the soft-photon limit, used for the Lightman-Thorne formula. Close to full equilibrium ($\mu\ll 1$), this leads to \citep[][henceforth referred to as CS12 approximation]{Chluba2005, Chluba2007a, Chluba2011therm}
\bsub
\label{eq:Lam_CS}
\begin{align}
\Lambda^{\rm eq}_{\rm rel}(x)&=\frac{4\alpha}{3\pi}\,\The^2\,\mathcal{I}^{\rm pl}_4\,\frac{\expf{x}\,H_{\rm dc}(x)}{1+14.16\The} , 
\\[1mm]
H_{\rm dc}(x)&=\frac{\int_{2x}^\infty {x'}^4 n(x')\big[1+n(x'-x)\big]\,\left[\frac{x}{x'} H_{\rm G}\left(\frac{x}{x'}\right)\right]\id x'}
{\mathcal{I}^{\rm pl}_4}
\nonumber\\
&\approx \expf{-2x}\left[1+\frac{3}{2}x+\frac{29}{24}x^2+\frac{11}{16}x^3+\frac{5}{12}x^4\right]
\end{align}
\esub
with the Gould factor \citep{Gould1984} $H_{\rm G}(w)=[1-3y+3y^2/2-y^3]/y$ for $y=w(1-w)$. At $x\ll 1$, the temperature suppression of the DC emissivity relative to the Lightman-Thorne approximation is captured by the term $\propto [1+14.16\The]^{-1}$, while corrections beyond the soft photon limit give rise to $H_{\rm dc}(x)$. Simply multiplying $\Lambda_{\rm rel}(x)$ by $g_{\mu}=\mathcal{I}_4/\mathcal{I}^{\rm pl}_4$ from Eq.~\eqref{eq:mu_Int} provides an extension to arbitrary chemical potential, which works very well (see Fig.~\ref{fig:gdc_mu}).

However, we can improve the above approximations by using the expressions from \citet{Chluba2007a}. For a fixed electron momentum and incoming photon energy, the correction to the soft photon DC emissivity relative to the Lightman-Thorne approximation is given by\footnote{Notice a typo in $f_3$ that was corrected in \citet{McKinney2017}.} \citep{Chluba2007a, McKinney2017} 
\bsub
\label{eq:G_soft}
\begin{align}
G_{\rm m}(\omega_0, \beta_0)&=\frac{\gamma_0^2(1+\beta^2_0)}{1+\sum_{k=1}^4 f_k(\beta_0) \gamma_0^k\,\omega_0^k}
\\[1mm]
f_1(\beta_0)
&=\;\;\,\frac{1}{1+\beta_0^2}\,\left[\frac{21}{5}+\frac{42}{5}\beta_0^2+\frac{21}{25}\beta^4_0\right]
\\
f_2(\beta_0)
&=\;\;\,\frac{1}{(1+\beta_0^2)^2}\,\left[\frac{84}{25}+\frac{217}{25}\beta_0^2+\frac{1967}{125}\beta^4_0\right]
\\
f_3(\beta_0)
&=-\frac{1}{(1+\beta_0^2)^3}\,\left[\frac{2041}{875}+\frac{1306}{125}\beta_0^2\right]
\\
f_4(\beta_0)
&=\;\;\,\frac{1}{(1+\beta_0^2)^4}\,\frac{9663}{4375}
\end{align}
\esub
with $\beta_0=\varv_0/c$ and Lorentz factor $\gamma_0=1/(1-\beta_0^2)^{1/2}$. This can be thermally-averaged over a relativistic Maxwell-Boltzmann distribution once the electron temperature is fixed, yielding $G_{\rm th}(x, \The)$. Together with the Gould-factor, this then gives
\begin{align}
\label{eq:Lam_improved}
\Lambda_{\rm rel}(x)&=\frac{4\alpha}{3\pi}\,\The^2\,\mathcal{I}^{\rm pl}_4\,g_{\rm dc}(x), 
\\[1mm]\nonumber
g_{\rm dc}(x)&\approx\frac{\int_{2x}^\infty {x'}^4 n(x')\big[1+n(x'-x)\big]\,G_{\rm th}(x', \The)\,\left[\frac{x}{x'} H_{\rm G}\left(\frac{x}{x'}\right)\right]\id x'}{\expf{-x}\mathcal{I}^{\rm pl}_4},
\end{align}
where we introduced the DC Gaunt-factor, $g_{\rm dc}(x)$. To compute the DC Gaunt factor, it is beneficial to pre-tabulate $G_{\rm th}(x, \The)$ once the temperature is set. This eases the numerical integration of the photon distribution in every iteration.

In Fig.~\ref{fig:gdc_mu}, we compare the approximation for the DC Gaunt factor, Eq.~\eqref{eq:Lam_CS}, modified by the factor $g_{\mu}=\mathcal{I}_4/\mathcal{I}^{\rm pl}_4$ from Eq.~\eqref{eq:mu_Int}, with the results from Eq.~\eqref{eq:Lam_improved} for constant chemical potential. At low frequencies the modified CS12 approximation works very well while at high frequencies it is less accurate. The high-frequency part of the DC emission is not as important for the computations of the distortions, such that Eq.~\eqref{eq:Lam_CS}, modified by the suppression factor $g_{\mu}=\mathcal{I}_4/\mathcal{I}^{\rm pl}_4$ from Eq.~\eqref{eq:mu_Int} should suffice for our purposes. All the main figures will be computed with this approximation; however, we find in Sect.~\ref{sec:DC_BR_precision} that the differences are negligible.
We also compared the results for the DC Gaunt factor with those obtained from the more rigorous treatment of {\tt DCpack} \citep{DCpack2020} and found very good agreement.

\vspace{-3mm}
\subsection{Numerical quasi-stationary solution of the problem}
\label{sec:QS_sol_eq}
Starting from Eq.~\eqref{eq:Dn_evol}, we can determine the quasi-stationary solution for the photon distribution with the system
\bsub
\label{eq:QS_final}
\begin{align}
\label{eq:QS_final_a}
0
&\approx  \int P(\omega \rightarrow \omega', \The) \,\frac{\expf{\mu-\mu'} -1}{1-\expf{-x'-\mu'}}\id\omega' 
+\frac{\Lambda(x)}{x^3}\,\Big[\expf{\mu}-1\Big]
\\
0&=\int \frac{x^3\id x}{\expf{x+\muc}-1}\,\frac{1-\expf{\muc-\mu}}{1-\expf{-x-\mu}}.
\label{eq:QS_final_b}
\end{align}
\esub
To discard the trivial solution $\mu=0$, the second equation ensures that the overall solution has the energy density given by the initial condition, Eq.~\eqref{eq:BE_n_sol}. This is equivalent to thinking of the solution as $\mu(x)=\mu_{\rho}\,\xi(x)$, where $\mu_{\rho}$ is a normalization constant that requires the extra equation, while Eq.~\eqref{eq:QS_final_a} mainly fixes the overall shape. Enforcing overall energy conservation generally implies some small amount of photon production with respect to the initial solution $\mu=\muc$ during a short relaxation phase. This is captured using Eq.~\eqref{eq:QS_final} \citep[e.g.,][]{Chluba2015GreensII}.

\subsubsection{Analytic approximation for $\mu\ll 1$}
For small $\muc$ and at low temperature, the problem Eq.~\eqref{eq:QS_final} can be approximately solved analytically. This classical result \citep{Sunyaev1970mu, Danese1982} can be obtained by starting from the Kompaneets equation and then assuming $x\ll 1$. Alternatively, one can directly perform a Fokker-Planck expansion of Eq.~\eqref{eq:QS_final_a} and then take the limit $x\ll 1$. This leads to the differential equation \citep[see][for a step-by-step derivation]{Chluba2014}
\begin{align}
\label{eq:QS_classical}
0\approx x^2 \partial_x^2 \mu +2x \,\partial_x\mu - \frac{\xc^2}{x^2}\,\mu=\partial_x x^2 \partial_x \,\mu - \frac{\xc^2}{x^2}\,\mu
\end{align}
with $\xc\approx \sqrt{\Lambda(\xc)/\The}$. The first terms are due to Compton scattering and stimulated Compton scattering, while the last captures the approximate effect of DC and BR emission. Evidently, this latter term rapidly increases at low frequencies, implying that $\mu(x)\rightarrow 0$ at $x\ll 1$.
The solution to the above equation is 
\begin{align}
\label{eq:QS_classical_sol}
\mu(x)=\muc \expf{-\xc/x},
\end{align}
which captures the main dependence even at large $x$ extremely well, in spite of assuming $x\ll 1$ in the main derivation. 

The critical frequency $\xc$ is determined by the competition between DC/BR and Compton scattering. A simple approximation can be obtained using \citep{Chluba2014}
\bsub
\label{eq:QS_xc}
\begin{align}
\xc&\approx \sqrt{x_{\rm c, BR}^2+x_{\rm c, DC}^2}
\\
x_{\rm c, BR}&\approx \pot{1.23}{-3}\,\left[\frac{1+z}{\pot{2}{6}}\right]^{-0.672}
\\
x_{\rm c, DC}&\approx x^{\rm nr}_{\rm c, DC}\,\left[\frac{1+\frac{1}{4} x^{\rm nr}_{\rm c, DC}}{1+14.16\,\Thg^{\rm eq}(z)}\right]^{0.5}
\\
x^{\rm nr}_{\rm c, DC}&\approx \pot{8.60}{-3}\,\left[\frac{1+z}{\pot{2}{6}}\right]^{0.5},
\label{eq:QS_xc_c}
\end{align}
\esub
which also accounts for leading order relativistic corrections to the DC process and more accurate fits to the BR emissivity. Here, we set $\The\approx \Thg^{\rm eq}(z)=\pot{4.60}{-10}(1+z)$ assuming $\mu\ll 1$. 

\subsubsection{Analytic approximation without stimulated scattering}
For large values of $\mu$, we will see below that stimulated scattering terms at low frequencies will become unimportant. To mimic this effect, another simple approximation can be obtained again starting from the Kompaneets equation, but omitting terms $\propto n^2$: 
\begin{align}
\label{eq:QS_classical_large_mu_K_ansatz}
\left.\frac{\text{d}n}{\text{d}\tau} \right|_{\rm CS}
&\approx \frac{\The}{x^2}\partial_x x^4 \big[\partial_x n +n\big].
\end{align}
Inserting $n=\expf{-x-\mu}$ and only keeping first order terms in $\mu$, together with the emission terms we then obtain
\begin{align}
\label{eq:QS_classical_large_mu}
0\approx x^2 \partial_x^2 \mu +4x \,\partial_x\mu - \frac{\xc^2}{x^3}\,\mu.
\end{align}
Omitting stimulated scattering terms is equivalent to setting the factor $1-\expf{-x'-\mu'}\simeq 1$ in the denominator of kernel in Eq.~\eqref{eq:QS_final_a}, which enhances the relative importance of DC and BR emission terms by a factor of $\nbb \simeq 1/x$. Equation~\eqref{eq:QS_classical_large_mu} has the solution
\begin{align}
\label{eq:QS_classical_sol_large_mu}
\mu(x)=\muc \, \frac{2}{3}\frac{\xc}{x^{3/2}} K_1\left(\frac{2}{3}\frac{\xc}{x^{3/2}}\right),
\end{align}
which again describes the overall behaviour of the solution even at high frequencies quite well. As we will show below, in comparison to Eq.~\eqref{eq:QS_classical_sol}, the transition from low to high-frequency regime becomes more steep, highlighting the importance of stimulated scattering for the shape of the distortion.

\subsubsection{Analytic approximation for large $\mu$}
While the solution in Eq.~\eqref{eq:QS_classical_sol_large_mu} highlights the importance of stimulated terms, it assumes $\mu\ll 1$. We can overcome this aspect and obtain another solution valid for $\mu\gg 1$. Starting from Eq.~\eqref{eq:QS_final_a} and performing a Fokker-Planck expansion of $f_\mu=1-\expf{-\mu}$, we find 
\begin{align}
\label{eq:QS_classical_large_mu_full}
0\approx x^2 \partial_x^2 f_\mu +4x \,\partial_x f_\mu - \frac{{\xc^*}^2}{x^3}\, f_\mu.
\end{align}
We will return to the definition of $\xc^*$ below, but for now assume it is simply constant.
By comparing with Eq.~\eqref{eq:QS_classical_large_mu} and imposing suitable boundary conditions we then have
\begin{align}
\label{eq:QS_classical_sol_large_mu_full}
\mu(x)=-\ln\left[1-(1-\expf{-\muc}) \, \frac{2}{3}\frac{\xc^*}{x^{3/2}} K_1\left(\frac{2}{3}\frac{\xc^*}{x^{3/2}}\right)\right].
\end{align}
For $\muc\ll 1$ this naturally reduces to Eq.~\eqref{eq:QS_classical_sol_large_mu}, however, the shape of the chemical potential differs significantly for large $\muc$ (see Fig.~\ref{fig:large_mu}).

In the derivation above we introduced $\xc^*$, which we determine such that it captures the leading order frequency correction to the DC emissivity. We encounter ${\xc^*}^2=\Lambda_{\rm DC}(\bar{x}, z, \The, \muc)/\The$, but need to determine $\bar{x}$ to allow the evaluation. The well-justified assumption is that at low frequencies $\Lambda_{\rm DC}$ scales very slowly with $x$ and thus the variation of the term $\propto {\xc^*}^2 f_\mu/x^3$ is mainly determined by $f_\mu/x^3$. The maximum of this term, given the solution Eq.~\eqref{eq:QS_classical_sol_large_mu_full}, with $\xi=(2/3)\,\xc^*/x^{3/2}$ is determined by the equation $\xi K_0\left(\xi \right)=2K_1\left(\xi\right)$, which has the solution $\xi = 2.387$, or $\bar{x}=0.4273\,\xc^*$.  We then have
\begin{align}
\label{eq:xc_new_def}
{\xc^*}^2&=\frac{\Lambda_{\rm DC}(\bar{x}, z, \The, \muc)}{\The}
\approx \frac{4\alpha}{3\pi}\,\The\,\mathcal{I}^{\rm pl}_4\,g_{\mu}\,\frac{(1+\bar{x}/2)}{1+14.16\,\The},
\end{align}
where $g_{\mu}=\mathcal{I}_4/\mathcal{I}^{\rm pl}_4\approx 0.924\,\expf{-\muc}$ and $\The\approx 1.02\,\expf{\muc/4}\,\Thg^{\rm eq}$. This means
\begin{align}
\label{eq:critical}
\xc^*\approx 0.971\,x^{\rm nr}_{\rm c, DC}(z)\,\expf{-3\muc/8}\,\left[\frac{1+0.207 [x^{\rm nr}_{\rm c, DC}(z)]^{2/3}\,\expf{-3\muc/8}}{1+14.44\,\Thg^{\rm eq}(z)\,\expf{\muc/4}}\right]^{0.5}
\end{align}
with $x^{\rm nr}_{\rm c, DC}(z)$ given by Eq.~\eqref{eq:QS_xc_c}. This expression implies that overall the critical frequency reduces when $\muc$ increases. 

\vspace{-0mm}
\subsubsection{Iterative method based on the Fredholm equation}
One simple approach for solving Eq.~\eqref{eq:QS_final} exactly is to map it into an approximate {\it homogeneous Fredholm equation} of second kind. Introducing $f_\mu=1-\expf{-\mu}$, we may write
\begin{align}
\label{eq:QS_final_FH}
\frac{\Lambda(x)}{x^3}\,f_\mu
=  \int K(x,x',f_{\mu'}) \,[f_{\mu'}-f_{\mu}]\id\omega'.
\end{align}
The kernel of this integral equation,
\begin{align}
K(x,x',f_{\mu'})=P(\omega \rightarrow \omega', \The) \,(1+n')=\frac{P(\omega \rightarrow \omega', \The)}{1-\expf{-x'}[1-f_{\mu'}]}
\end{align}
is weakly non-linear in $f_\mu$, which modulates the importance of stimulated scattering effects. For large $\mu$, one has $1-\expf{-x'}[1-f_{\mu'}]\approx 1$, while for $\mu\ll 1$, one has $1-\expf{-x'}[1-f_{\mu'}]\approx 1/x'$. The latter causes strong blackbody-induced stimulated Compton scattering, but these are mostly independent of $\mu$. We find that the $f$-dependence of the kernel to leading order can usually be neglected. 
Overall, the above expression suggests an iterative approach in the form\footnote{Using appropriate weight factors, the integral of a function $g(x)$ is rewritten as $\int g(x') \id \omega'=\sum_{i} g(x_i) \,w_i$. In our computations we use fifth order Lagrange interpolation coefficients and their integrals to discretize the solution \citep{Chluba2010}.}
\begin{align}
\label{eq:QS_final_FH_iteration}
f^{(i+1)}_\mu
= \frac{\int K(x,x',f^{(i)}_{\mu'}) \,f^{(i)}_{\mu'}\id\omega'}{\int K(x,x',f^{(i)}_{\mu'})\id\omega'+\Lambda(x)/x^3},
\end{align}
where one inserts an estimate for the solution $f^{(i)}$ on the right hand side to obtain an improved solution $f^{(i+1)}$ until convergence is reached\footnote{This is usually achieved in $\simeq 10^3-10^4$ steps. At high temperature, the number of iterations is typically significantly smaller than at low temperatures, as expected from the variation of the timescales on which photons diffuse across the spectrum.}. In addition, after each step, using the constraint in Eq.~\eqref{eq:QS_final_b}, the overall normalization of the solution is adjusted to ensure that energy is conserved. We find this approach to work quite efficiently when using $\mu^{(0)}=\muc\,\expf{-\xc/x}$, which is based on the original analytic approximations \citep{Sunyaev1970mu, Danese1982}, as the starting point. For more extensive tables, required to compute the distortion visibility function, we slowly vary the temperature and redshift using the previous solution as a starting point.
We also attempted using an explicit quadrature method to convert the integrals into a system of non-linear equations that can then be solved numerically. However, the iterative approach based on Eq.~\eqref{eq:QS_final_FH_iteration} was more stable.

\subsubsection{Adding time-dependent corrections}
Time-dependent corrections can be added to the problem using a suitable source term. This then leads to an inhomogenous Fredholm equation of second kind, which again can be solved iteratively at a given time. To obtain the source term, we rewrite the left hand side of the photon Boltzmann equation, Eq.~\eqref{eq:Dn_evol}, as
\begin{align}
\nonumber
\left.\frac{\partial \, n}{\partial \tau}\right|_x - x\!\left.\frac{\partial \, n}{\partial x}\right|_\tau \!\partial_\tau \ln a \Te
&=-n(1+n)\Big[\!\left.\partial_\tau\mu\right|_x-x(1+\!\left.\partial_x\mu\right|_\tau) \partial_\tau \ln a \Te\Big].
\end{align}
The idea is now to use the homogeneous Fredholm equation approach to compute the solution $\mu^{(0)}(x,\tau)$. This solution can then be used to compute $x\,\partial_x\mu$ and the required energy exchange integrals or photon production rates to then determine $\partial_\tau\mu$ and $\partial_\tau \ln a \Te$, perturbatively. More explicitly, we have 
\bsub
\begin{align}
\label{eq:QS_final_FH_iteration_time}
f^{(i+1)}_\mu
&=\frac{S^{(i-1)}(x)+\int K(x,x',f^{(i)}_{\mu'}) \,f^{(i)}_{\mu'}\id\omega'}{\int K(x,x',f^{(i)}_{\mu'})\id\omega'+\Lambda(x)/x^3},
\\[1mm]
S^{(i)}(x)&=\frac{\expf{-\mu^{(i)}}}{1-\expf{-x-\mu^{(i)}}}\,\left[x\left(1+\partial_x\mu^{(i)}\right) \!\partial_\tau \ln a \Te^{(i)}
-\partial_\tau\mu^{(i)} \right].
\end{align}
\esub
To fix $\partial_\tau\mu^{(i)}$ and $\partial_\tau \ln a \Te^{(i)}$, we use Eq.~\eqref{eq:evol_murho_final} but neglect the terms related to $\id \xi/\id \tau$. This means $\partial_\tau\mu^{(i)}\approx\mu^{(0)}\,\partial_\tau \ln \mu^{(i)}_\rho$ and that the photon emission term $\id \ln a^3 N_\gamma/\id \tau$ together with the integrals $\mathcal{M}_k$ and $\kappa$ fully determine the required correction.

For small distortions, the contributions due to $\propto \id \xi/\id \tau$ have been shown to be sub-dominant at high redshifts \citep{Chluba2014}. We furthermore find below that for large $\mu$, the overall leading order time-dependent corrections do not affect the general picture and also reduce for large $\mu$ (Fig.~\ref{fig:J_bb_compare_time}). We thus expect the terms $\propto \id \xi/\id \tau$ to also only change the results at a secondary level. In addition, consistently including these effect with the treatment outlined above almost amounts to solving the full problem thermalization problem, a task that will be left to future work using more efficient methods.

As shown by \citet{Chluba2014}, the dominant contribution to time-dependent effects stems from the derivative of the electron temperature. This can be seen from the strong scaling of the source terms $S$ with frequencies, which renders the term $\propto \partial_\tau \ln a \Te$ important at $x\gg 1$, where the energetics of the problem are fixed. Although the electron temperature is always very close to the Compton equilibrium temperature in the momentary radiation field, as thermalization proceeds, $\Te$ drops. This causes an asymptotic behavour $\mu\simeq {\rm C} + \ln(x) \partial_\tau \ln a\Te$ at $x\gg 1$ or $n\simeq n_0(\tau)\,x^\gamma\,\expf{-x}$ with $\gamma\simeq -\partial_\tau \ln a\Te$, which describes power-law corrections to the Wien spectrum that arise from the slower diffusive motions of photons up/downward in energy \citep{Chluba2014}. As we explain below (Sect.~\ref{sec:Klein-Nishina}), Klein-Nishina corrections to the high-frequency scattering kernel modifying the above limiting behavior noticeably.

\section{Illustration of the solutions for $\mu$}
In this section, we illustrate the solutions for the chemical potential for several case, highlighting the importance of various physical effects. In particular, we distinguish between the classical small distortion and large distortion regimes showing how the character of the solution changes quite significantly due to the diminishing effect of stimulated scattering and non-linear terms.

\subsection{Illustrations of the classical regime}
For small energy release, the spectral shape of the chemical potential distortion is independent of the overall distortion amplitude, $\mu(t, x)\approx \mu_\rho(t)\,\xi(t, x)$. The shape of the distortion is solely set by the competition between photon number changing processes and Compton scattering with the main asymptotics $\mu\rightarrow 0$ at $x\ll 1$ and $\mu \simeq {\rm const}$ at $x\gg 1$. The transition depend on the temperature and density of the particles in the plasma as well as the frequency-dependence of the emission processes. All these effects become important \citep{Chluba2014} but can be easily captured numerically.

\begin{figure}
\centering 
\includegraphics[width=\columnwidth]{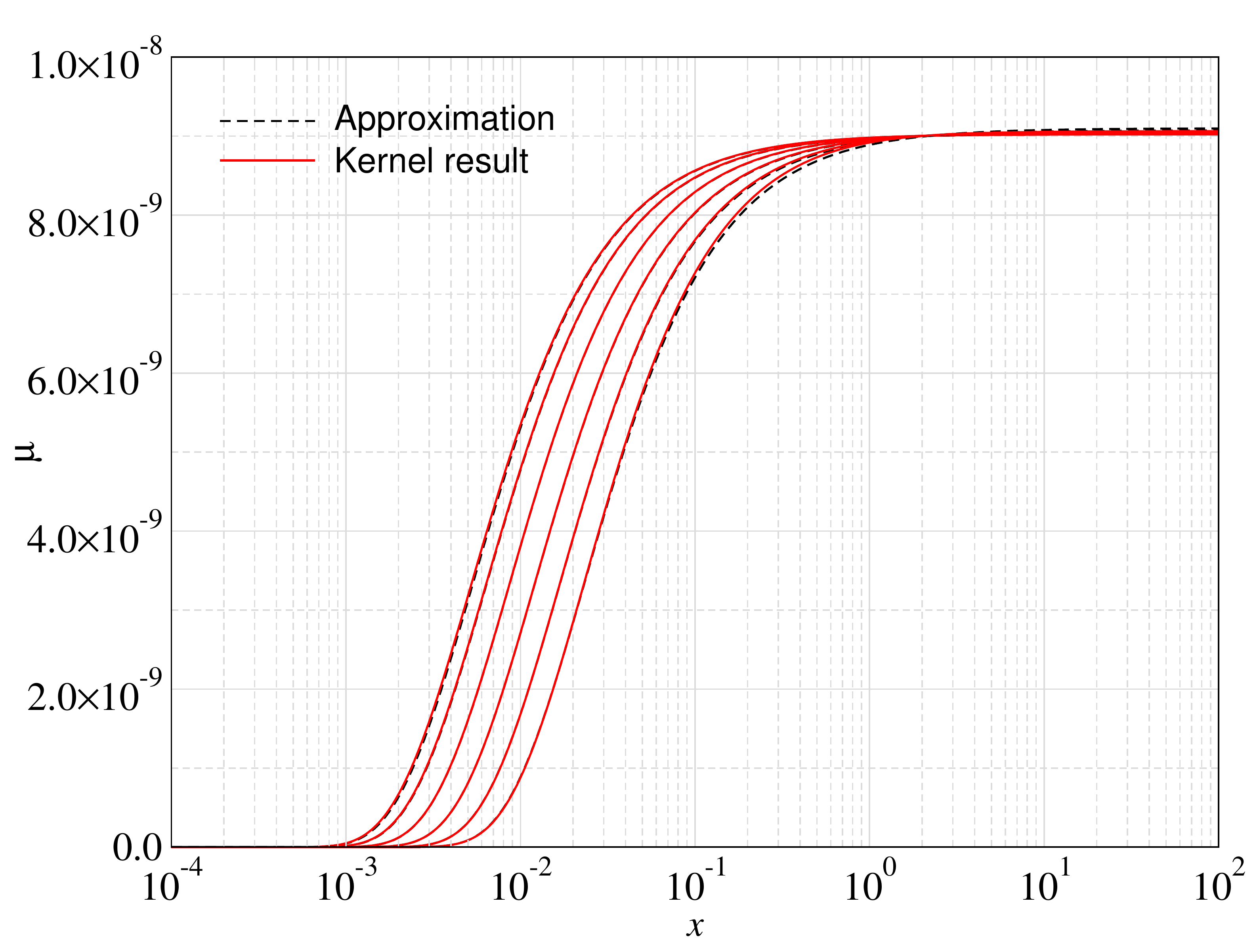}
\caption{Quasi-stationary chemical potential solution for small energy release at $\zh=\{\pot{5}{5}, \pot{1}{6}, \pot{2}{6}, \pot{4}{6}, \pot{8}{6}, \pot{1.6}{7}\}$ (left to right). The simple approximation, Eq.~\eqref{eq:QS_classical_sol} with $\xc$ given by Eq.~\eqref{eq:QS_xc}, is compared to the exact kernel result. In each case, $\Delta \Te/\Tg^{\rm eq} =\pot{2.5}{-9}$ and the photon energy density is set to $\rho_\gamma = \rho_{\rm CMB}(\zh)$ (and $\epsilon_N=0$), which implies slightly varying values for $\mu_\rho\approx 3.6\,\Delta \Te/\Tg^{\rm eq}\simeq \pot{9}{-9}$. For the considered redshifts, we have $\The\simeq 10^{-4}-10^{-2}$.}
\label{fig:small_mu}
\end{figure}
Figure~\ref{fig:small_mu} shows the quasi-stationary solution for $\mu$ after some small energy release at various redshifts. We fixed the energy density of the CMB to that of the standard blackbody at the corresponding redshift $\zh$, i.e., $\rho_\gamma =\rho_{\rm CMB}(\zh)$ (and $\epsilon_N=0$), and set the electron temperature such that $\Delta \Te/\Tg^{\rm eq} =\pot{2.5}{-9}$. We compare the simple approximation, Eq.~\eqref{eq:QS_classical_sol} with $\xc$ given by Eq.~\eqref{eq:QS_xc}, to the exact kernel result. The agreement is excellent over the shown range of redshifts with small departures becoming visible mainly at the lowest and highest redshifts. On the low redshift end, frequency-dependent corrections from BR become relevant, while at high redshifts Compton scattering relativistic corrections start playing a role. Both effect are not precisely captured by the approximation but can in principle be added analytically \citep[e.g.,][]{Chluba2014}; however, with little benefit over the numerical scheme used here.

\begin{figure}
\centering 
\includegraphics[width=\columnwidth]{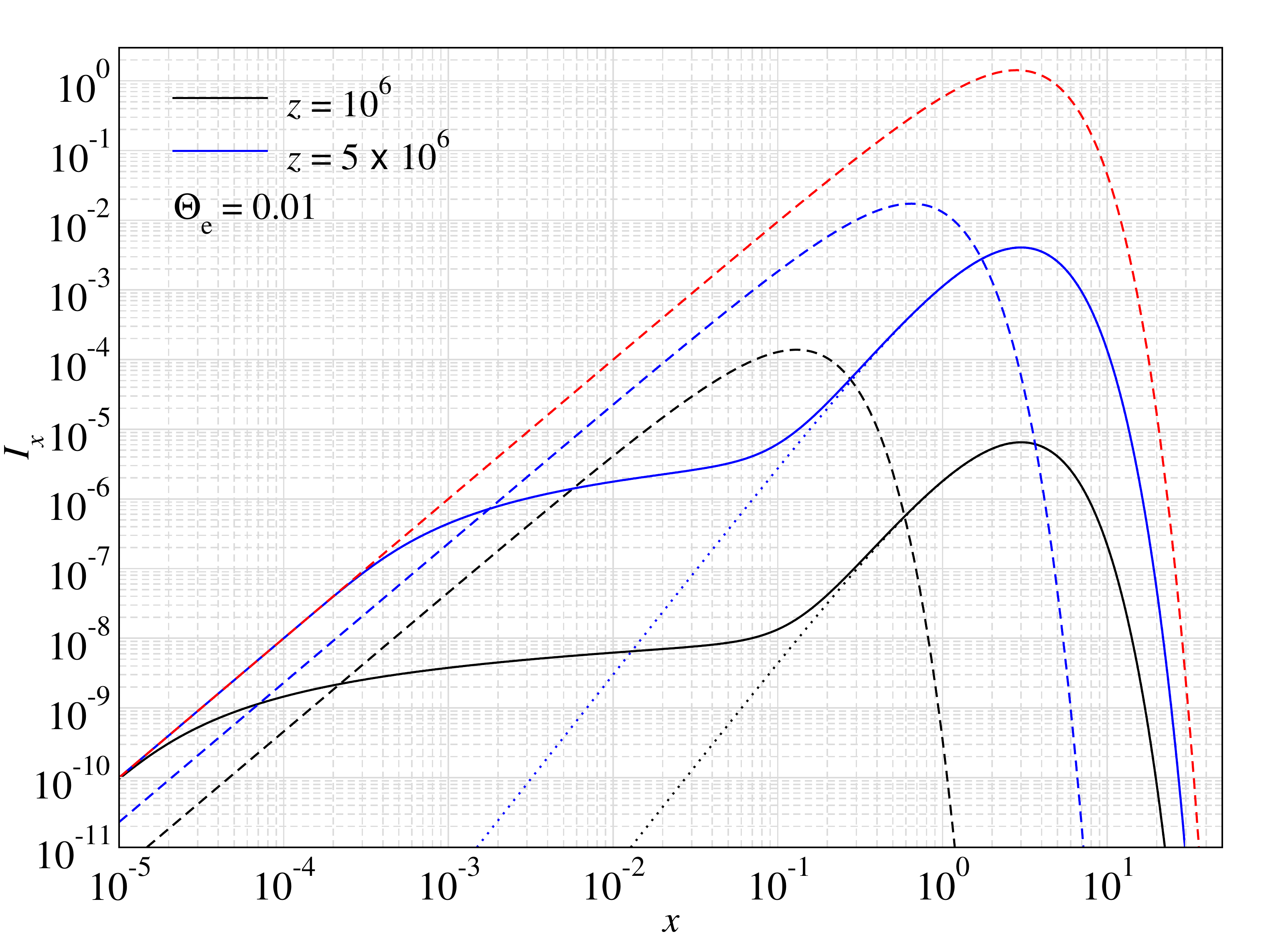}
\caption{Photon spectrum $I_x=x^3 n(x)$ for various solution for the photon occupation number $n(x)$. The dashed lines all show blackbody spectra at varying temperature. The red dashed is for a blackbody at $\The=0.01$, while the other two are for blackbodies at the CMB temperature corresponding to the annotated redshifts (i.e., $\theta_{\rm CMB}\approx \pot{4.60}{-4}$ and $\theta_{\rm CMB}\approx \pot{2.30}{-3}$). The dotted lines are the corresponding constant chemical potential solution, with $\muc\simeq 12.3$ (black) and $\muc\approx 5.80$ (blue), both in good agreement with $\muc=4\ln(0.9804\,\Te/\Tg^{\rm eq})$ following from Eq.~\eqref{eq:BE_n_sol_large}.}
\label{fig:Ix_Theta_0.01}
\end{figure}

\subsection{Effect of large injection and role of stimulated terms}
For large energy release, two main effects become important. At a given redshift, the energy density is already consistent with that of the equilibrium CMB, however, the deficit of photons in the photon spectrum implies that the average energy of the photons has to be significantly increased. This means i) a large chemical potential at high frequencies and ii) a significantly increased electron temperature, $\Te > T_{\rm CMB}(z)$. From i) it follows that in comparison to cases closer to equilibrium [$\Te\simeq  T_{\rm CMB}(z)$], stimulated scattering terms become less important at low frequencies, hence changing the shape of the distortion notably. From ii) it follows that even at rather moderate redshifts relativistic temperature corrections to the involved processes can become important, as we illustrate below.

In Fig.~\ref{fig:Ix_Theta_0.01}, we give the exact quasi-stationary solution for the spectrum at two redshifts assuming an electron temperature of $\The = 0.01$ after the energy release. The photon energy density in both cases is fixed to that of the CMB blackbody at the corresponding redshift. At low frequencies, the spectrum returns to a blackbody at the temperature of the electrons due to the combined action of DC and BR emission. At intermediate frequencies, a significant deficiency of photons with respect to the final equilibrium blackbody at temperature $\TCMB(z)$ (dashed lines) is seen, while in both cases an excess of photons is created in the distant Wien tail. It is this excess that through Compton scattering allows the electron temperature to lie significantly above $\TCMB(z)$. 

\begin{figure}
\centering 
\includegraphics[width=\columnwidth]{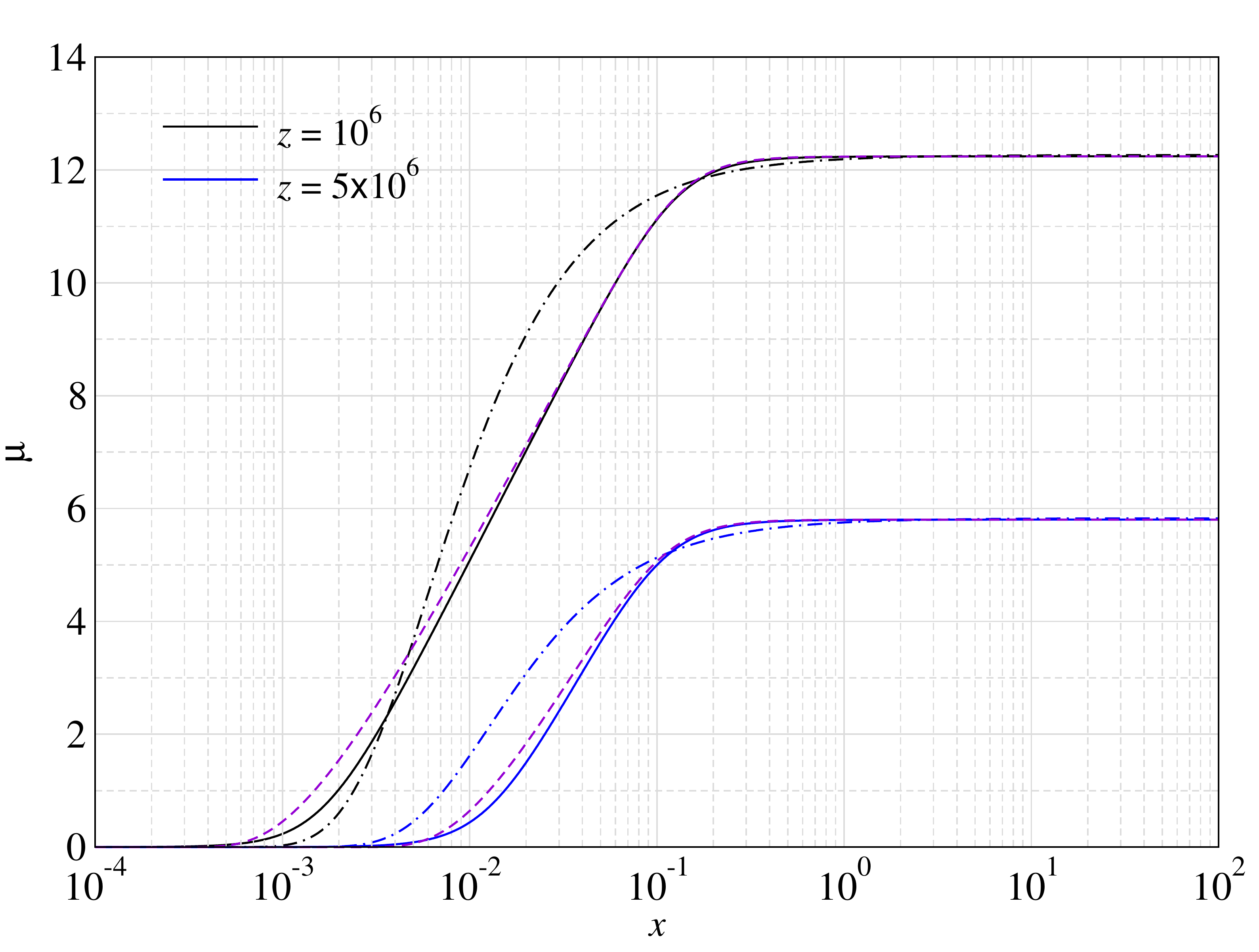}
\caption{Quasi-stationary chemical potential solution for large energy release at $z=10^6$ and $z=\pot{5}{6}$ (cases as in Fig.~\ref{fig:Ix_Theta_0.01}). The classical approximation (dash-dotted line), Eq.~\eqref{eq:QS_classical_sol} with $\xc$ given by Eq.~\eqref{eq:QS_xc}, is compared to the exact kernel result (solid line). We also show the approximation Eq.~\eqref{eq:QS_classical_sol_large_mu_full} with $\xc^*\rightarrow \sqrt{2}\xc^*$ (dashed lines). In all cases the electron temperature is $\The=0.01$. Relativistic temperature corrections and the reduction of stimulated effect and the DC emissivity are noticeable.}
\label{fig:large_mu}
\end{figure}

For further illustration, the exact solution for the chemical potential is also compared to the classical approximation in Fig.~\ref{fig:large_mu}. In both cases, relativistic temperature corrections and stimulated effects as well as a reduction of the DC emissivity become noticeable. Even by optimizing $\xc$ freely, the classical solution does not reproduce the shape of the distortion well.
In contrast the result from approximation Eq.~\eqref{eq:QS_classical_sol_large_mu_full}, which was obtained assuming large $\mu$ (or $f_\mu=1-\expf{-\mu}\approx 1$), works a lot better. By comparing with the numerical result we found that $\xc^*\rightarrow \sqrt{2}\xc^*$ gives the best match at high frequencies, which is shown. At low frequencies one has $f_\mu < 1$ such that a complete omission of stimulated terms is no longer justified. This is where we see strongest departures from the full numerical result (Fig.~\ref{fig:large_mu}), indicating photons move less efficiently upwards due to the extra 'drag' towards lower frequencies induced by stimulated terms \citep{Chluba2008d}. Nevertheless, overall the new approximation captures the main features of the solution for large $\mu$ quite well.

\begin{figure}
\centering 
\includegraphics[width=\columnwidth]{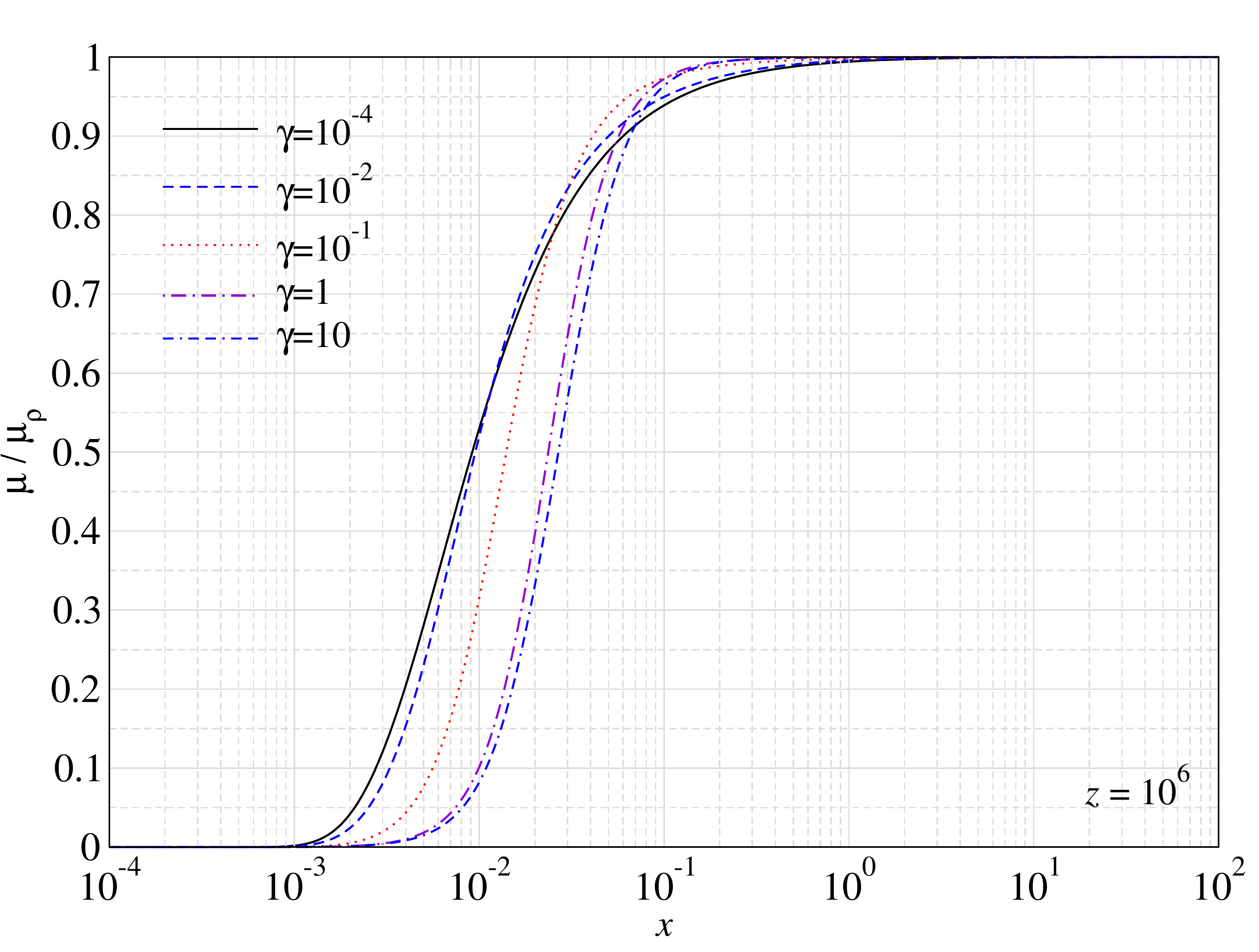}
\caption{Quasi-stationary chemical potential solution for energy release at redshift $z=10^6$ and various electron temperatures. For simplicity, we just fixed $\Te=T_{\rm CMB}(z)(1+\gamma)^{1/4}$, with the annotated values of $\gamma$. This leads to 
$\mu_\rho\approx \{ \pot{9.0}{-5}, \pot{9.0}{-3}, \pot{8.7}{-2}, \pot{6.5}{-1}, 2.3 \}$ in each of the cases, respectively. The transition regime steepens with rising $\mu_\rho$.}
\label{fig:large_mu_transition}
\vspace{1mm}
\end{figure}
To further illustrate the transition from small to large chemical potential distortions we compare the exact solutions at fixed redshift $z=10^6$ for varying amount of energy release in Fig.~\ref{fig:large_mu_transition}. For simplicity, we parametrize the energy release by varying the electron temperature. For the considered cases, we obtain $\mu_\rho\approx \{ \pot{9.0}{-5}, \pot{9.0}{-3}, \pot{8.7}{-2}, \pot{6.5}{-1}, 2.3 \}$, covering the transition from the classical to large distortion limit. As is evident, the shape of the distortion depends explicitly on $\mu_\rho$ once $\mu_\rho\gtrsim 10^{-3}-10^{-2}$. The transition regime around the critical frequency steepens, which in turn directly affects the DC emissivity, reducing the active emission region, which is essentially defined by $\simeq (\expf{\mu}-1)/x^3$ [see Eq.~\eqref{eq:col_em_mu}], for larger values of $\mu$. 

\begin{figure}
\centering 
\includegraphics[width=\columnwidth]{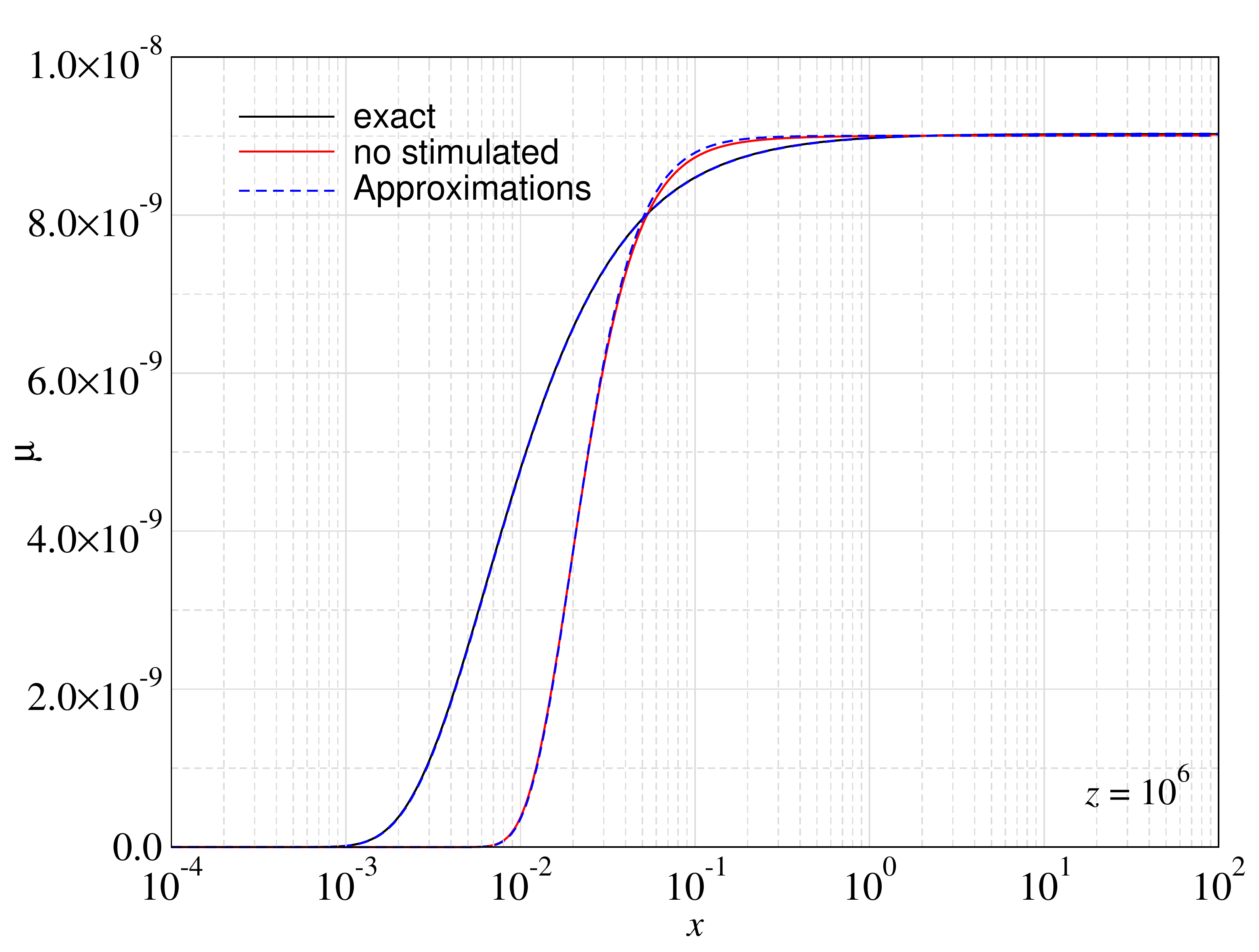}
\caption{Quasi-stationary chemical potential solution for small energy release at $z=10^6$ and $\Delta \Te/\Tg^{\rm eq} =\pot{2.5}{-9}$, leading to $\mu_\rho\approx \pot{9.0}{-9}$. The effect of stimulated electron scattering is illustrated, showing a steepening of the transition regime when omitted. This effect is captured by the analytic approximations Eq.~\eqref{eq:QS_classical_sol} and \eqref{eq:QS_classical_sol_large_mu} with $\xc$ given by Eq.~\eqref{eq:QS_xc}.}
\label{fig:mu_transition_ana}
\vspace{1mm}
\end{figure}

The cause of this steepening is mostly due to stimulated electron scattering effects. This can be further illustrated for small distortions ($\mu\ll 1$), but when omitting the factor $1/(1-\expf{-x'-\mu'})$ in the kernel. The comparison of this calculation with the full solution is given in Fig.~\ref{fig:mu_transition_ana}, exhibiting the aforementioned steepening. The analytic approximation Eq.~\eqref{eq:QS_classical_sol_large_mu} indeed captures this effect.
We note, however, that for large overall distortions, this simple approximation does not correctly capture the shape of the distortion. In this case Eq.~\eqref{eq:QS_classical_sol_large_mu_full} should be used.

\begin{figure}
\centering 
\includegraphics[width=\columnwidth]{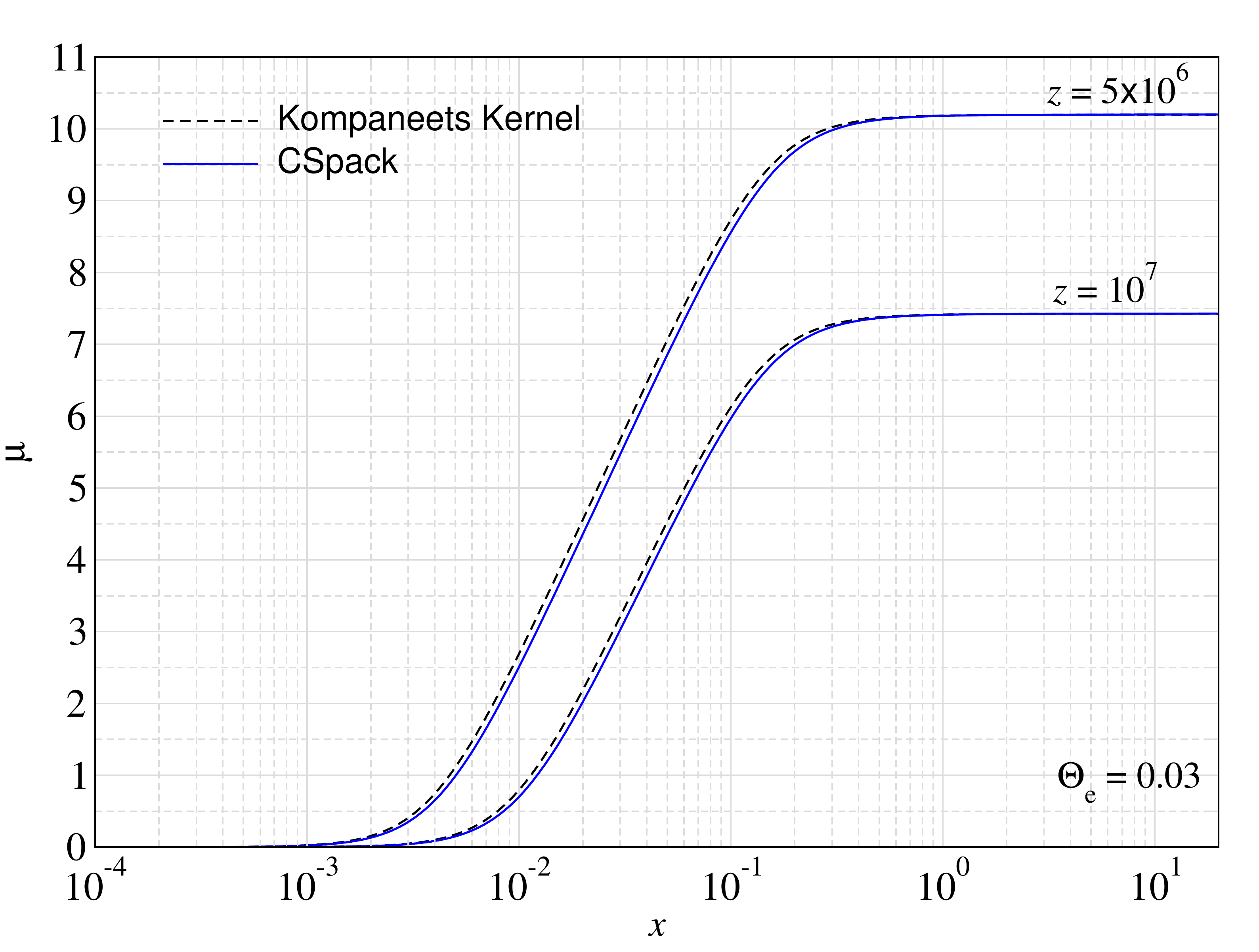}
\caption{Quasi-stationary chemical potential solution for large energy release at $z=\pot{5}{6}$ and $z=10^7$ yielding $\The=0.03$ ($\simeq 15\,\keV$). The dashed lines are obtained using the Kompaneets kernel from \citet{Sazonov2000}, while the solid lines are for the exact kernel. Compton relativistic corrections help photons move upwards more efficiently, thereby reducing the chemical potential at higher frequencies than without these corrections.}
\label{fig:large_mu_CS_rel}
\vspace{1mm}
\end{figure}
\subsection{Relativistic temperature corrections to Comptonization}
\label{sec_mu_rel_CS}
In Fig.~\ref{fig:large_mu_CS_rel} we illustrate the effect of relativistic temperature correction to the scattering process. These are not captured by the Kompaneets equation, which is usually used for the computations \citep[e.g.,][]{Burigana1991, Hu1993, Chluba2011therm}. To mimic the Kompaneets solution using the Fredholm equation approach, we apply the Kompaneets kernel, Eq.~(23) of \citet{Sazonov2000}. These kernel approximations fail at high frequencies, so here we restrict our computations to $x\lesssim 20$.
The shown examples in Fig.~\ref{fig:large_mu_CS_rel} are for large energy release, such that even at redshifts $z\simeq 10^6-10^7$ relativistic corrections can become noticeable but otherwise would be more subtle at these redshifts, when the system is already close to full equilibrium. In our computation using {\tt CSpack}, these corrections are taken into account exactly. 

Relativistic temperature corrections increase the efficiency of photon up-scattering \citep[e.g.,][]{Sazonov2000, CSpack2019} such that Compton scattering more photons reach the high-frequency part, reducing the value of the chemical potential and thereby increasing the effective critical frequency \citep{Chluba2005, Chluba2014}. We confirmed that this statement is mostly independent of whether the distortion is large or small or if we use the Lightman-Thorne DC emissivity or CS12. The relative effect remains qualitatively the same, even if the level can vary slightly.
Although $\xc$ increases, which with Eq.~\eqref{eq:DN_N_small} naively suggests an increase of the thermalization efficiency, when computing the accurate DC emissivity, Compton scattering relativistic corrections cause a net decrease of the thermalization efficiency \citep{Chluba2005, Chluba2014}, as we also see below (Fig.~\ref{fig:J_bb_compare_CS}).

\vspace{0mm}
\subsection{Time-dependent corrections}
\label{sec:Klein-Nishina}
In this section, we illustrate the effect of time-dependent corrections. For small distortions, these lead to an increase in the thermalization efficiency with time-dependent corrections due to the electron temperature being most relevant. With our computations, we can furthermore show that in the Wien tail additional Klein-Nishina corrections become relevant for the distortion shape, however, these do not affect the thermalization process as much, unless very high electron temperatures are reached. 

\begin{figure}
\centering 
\includegraphics[width=\columnwidth]{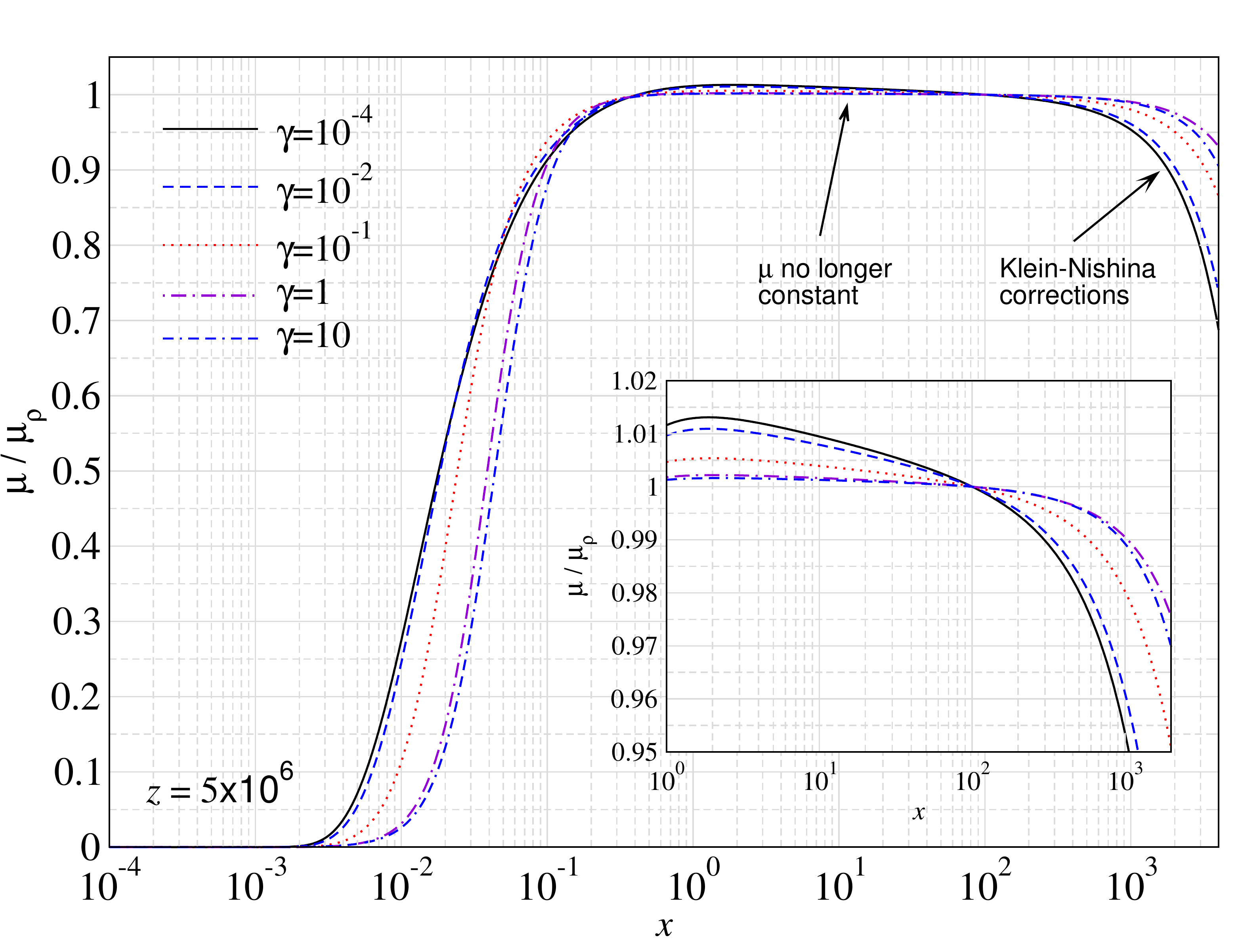}
\caption{Quasi-stationary chemical potential solution for energy release at $z=\pot{5}{6}$ and various electron temperatures. The main time-dependent corrections (neglecting terms $\propto\id \xi/\id \tau$) were included. For simplicity, we just fixed $\Te=T_{\rm CMB}(z)(1+\gamma)^{1/4}$, with the annotated values of $\gamma$. This leads to $\mu_\rho\approx \{ \pot{9.0}{-5}, \pot{9.0}{-3}, \pot{8.7}{-2}, \pot{6.5}{-1}, 2.3 \}$ in each of the cases, respectively. As without time-dependent corrections, the transition regime steepens with rising $\mu_\rho$; however, in addition a downward tilt of the spectrum at high frequencies becomes visible. At very high frequencies, Klein-Nishina corrections also become important.}
\label{fig:large_mu_time}
\end{figure}

In Fig.~\ref{fig:large_mu_time}, we compare the solutions for $\mu(x)$ at $z=\pot{5}{6}$, varying the electron temperature. As without time-dependent corrections, the low- to high-frequency transition regime steepens with rising electron temperature or $\mu_\rho$; however, we can now also observe a logarithmic downward tilt of the spectrum at high frequencies. At intermediate frequencies ($x\simeq 1-100$), this tilt is consistent with $\mu\simeq {\rm C} + \ln(x) \partial_\tau \ln a\Te$ \citep{Chluba2014}, which directly implies that $\partial_\tau \ln a\Te<0$ in all cases, as expected from Eq.~\eqref{eq:evol_murho_final}. Time-dependent effects on the spectrum at a given redshift are furthermore largest for smaller values of $\mu_\rho$, as will also be reflected in the distortion visibility function.

At $x\gtrsim 100$, additional Klein-Nishina corrections to the scattering kernel become visible, which are not captured by the Kompaneets equation but lead to an additional drop of the chemical potential. {\it How does this work?} We obtained $\mu\simeq {\rm C} + \ln(x) \partial_\tau \ln a\Te$ using the Kompaneets equation. As explained in \citet{Chluba2014}, this is due to delays in the reshaping of the spectrum in the Wien-tail: photons move up / downward in frequency on a short time-scale, which causes a drop / rise in the value of $\mu_\rho$. But given sufficient variation in the electron temperature, the spectrum has to steepen ($\partial_\tau \ln a\Te>0$) or become more shallow ($\partial_\tau \ln a\Te<0$). This requires a differential motion of the photons and causing a logarithmic drop in the chemical potential for $\partial_\tau \ln a\Te<0$. 

With the full Compton kernel, the time-scale for these changes now becomes frequency-dependent (see Fig.~\ref{fig:Kernel}) and the reshaping takes longer at high frequencies than at lower frequencies. This explains the additional steepening in the downward trend of the chemical potential, as more photons to stay behind. The effect is not present when time-dependent corrections are omitted, which our numerical computations confirm. In addition, when $\partial_\tau \ln a\Te>0$, we find that as expected the high-frequency spectrum shows an increase of the chemical potential with an additional flaring as Klein-Nishina corrections become relevant. 

While physically very interesting, Klein-Nishina corrections do not have a significant effect on the thermalization efficiency. Time-dependent corrections furthermore become less important for large chemical potential, as we will see below.
However, the effect of $\partial_\tau \ln a\Te\neq0$ at intermediate and low frequencies does change the thermalization efficiency of the plasma. This directly confirms why an accurate treatment of the thermalization problem, even for small distortions, cannot be carried out independently of the energy release history. The shape of the distortion is no longer independent of the thermal history even if for $\mu_\rho \ll 1$ it does not directly depend on the value of $\mu_\rho$.

\subsection{Precise emission rates for BR and DC}
\label{sec:DC_BR_precision}
In all of the computation presented so far, we used {\tt BRpack} to model the BR emissivity and Eq.~\eqref{eq:Lam_CS}, modified by the factor $g_{\mu}=\mathcal{I}_4/\mathcal{I}^{\rm pl}_4$ from Eq.~\eqref{eq:mu_Int}, for the DC emissivity (referred to as CS12 DC approximation). The BR process is mainly important at lower redshifts ($z\lesssim \pot{4}{5}$). Switching back to the approximations by \citet{Draine2011Book} and \citet{Itoh2000} we find no noticeable differences in the solutions for $\mu$ at early phases. Similarly, switching to the more accurate treatment in Eq.~\eqref{eq:Lam_improved} for the DC process, we find extremely small differences (at a relative level $\lesssim 10^{-3}$ around the critical frequency) in the results for $\mu(x)$. 
Given that the extra DC integrations are indeed very time-consuming, we now use the simpler approximation for computations. 
However, one can expect logarithmic corrections due to the shape of the high-frequency spectrum to enter the thermalization efficiency. These depend directly on the injection history, so that a more detailed comparison for the effects on the distortion visibility function will be carried out in our future work on continuous energy release scenarios.

\section{Results for the distortion visibility}
\label{sec:Jbb_section}
We now have all the pieces together to carry out computations for the distortion visibility function. 
To compute the distortion evolution, we need to evolve the $\mu_\rho$ using equation Eq.~\eqref{eq:evol_murho_final}. Assuming a single burst of energy release with no associated photon injection means we only need the DC and BR photon production rate $\id \ln N_\gamma/\id \tau$. Given the solutions $\mu(x, \tau)$ at each stage, these can be computed directly using Eq.~\eqref{eq:DN_N_def}. 
By starting from an initial distortion amplitude following the energy release, one could in principle solve the problem alternating between obtaining the stationary solutions for $\mu(x, \tau)$ and then advancing the solution to the next time. However, this can be avoided by simply precomputing the photon production rate, $\id \ln N_\gamma/\id \tau$, for a range of values $\mu_\rho$ at fixed redshifts $z$. The obtained tables can then be interpolated in $\mu_\rho$ and $z$ to advance the solution from redshift to the other. The benefit of this procedure is that the quasi-stationary solutions only have to be computed once for a given chemical composition of hydrogen and helium to obtain reliable results for the visibility functions and distortion limits. As mentioned above, for single energy release no changes to the Hubble expansion factor are necessary.

\subsection{Distortion visibility function}
The distortion visibility function, $\mathcal{J}_{\rm bb}(z)$, defines what fraction of the energy injected at a redshift $z$ is still present as a spectral distortion today after the thermalization process. It has two main contributions, the $y$-distortion and $\mu$-distortion visibilities, $\mathcal{J}_{y}(z)$ and $\mathcal{J}_{\mu}(z)$, respectively; however, their individual contributions can only be distinguished by looking at the specific shape of the distortion \citep[e.g.,][]{Chluba2013PCA, Chluba2016}. With this definition, the fraction of energy leading to a change of the average CMB temperature is given by $\mathcal{J}_{T}(z)=1-\mathcal{J}_{\rm bb}(z)$, making $\mathcal{J}_{\rm bb}(z)$ one of the central aspects of the thermalization calculation.

For small spectral distortions, a simple approximation of the distortion visibility function is given as \citep{Sunyaev1970mu, Danese1982, Hu1993}
\begin{align}
\label{eq:J_bb_DC}
\mathcal{J}_{\rm DC}(z)&=\expf{-(z/\zmu)^{5/2}}.
\end{align}
with $\zmu\approx \pot{1.98}{6}$ for the standard cosmological model. For estimates, this approximation is extremely useful although refinements can be included analytically \citep{Chluba2005, Khatri2012b, Chluba2014}. However, for large energy release, the distortion visibility is significantly increased, as we show now.

\begin{figure}
\centering 
\includegraphics[width=\columnwidth]{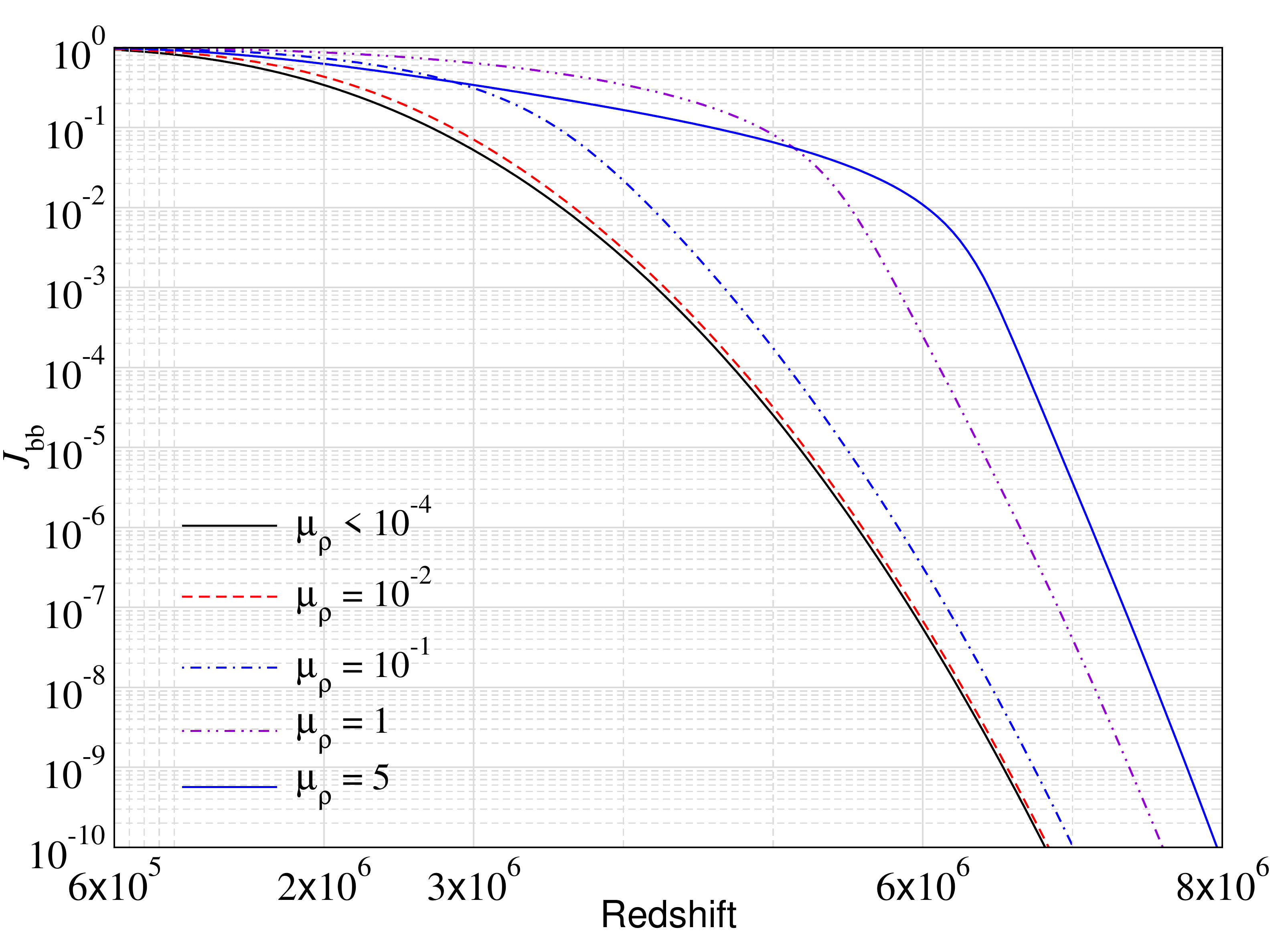}
\caption{Visibility function for single injection of energy at various redshifts and for various initial values of $\mu_\rho$. At early times, the visibility for large distortions is orders of magnitudes higher than for small distortions.}
\label{fig:J_bb}
\end{figure}

To obtain the distortion visibility function from our computation, we fix the initial value of $\mu_\rho$ at a given heating redshift $\zh$. From the solution $\mu(x, \tau)$, we can then determine the required injected energy, $\left.\Delta \rho_\gamma/\rho_\gamma \right|_{\rm in}$. We then evolve $\mu_\rho$ using Eq.~\eqref{eq:evol_murho_final} until redshift $z=\pot{2}{5}$, which was shown to define the moment when Comptonization becomes too inefficient for soft photons created by BR and DC to reach the high-frequency part of the CMB spectrum, thus freezing the high-frequency distortion that is present at $z=\pot{2}{5}$ \citep{Chluba2014}. If energy release occurred at $z>\pot{3}{5}$, the solution should correspond to a $\mu$-type distortion today. 
At the final moment, we then compute the final $\left.\Delta \rho_\gamma/\rho_\gamma\right|_{\rm f}$, and then compute the visibility function numerically as 
\begin{align}
\mathcal{J}^{\rm num}_{\rm bb}(z)&=\frac{\left.\Delta \rho_\gamma/\rho_\gamma\right|_{\rm f}\,\,}{\left.\Delta \rho_\gamma/\rho_\gamma\right|_{\rm in}}.
\end{align}
A similar procedure was used previously \citep{Chluba2011therm, Chluba2014}. It is important to mention that simply computing the ratio of the  final and initial values for $\mu_\rho$ gives a slightly different result since at the earliest times the distortion shape modifies the effective heat capacity of the CMB spectrum \citep{Chluba2014}.

Following the above procedure, we obtain the results summarized in Fig.~\ref{fig:J_bb}. For small distortions $\mu\lesssim 10^{-4}-10^{-3}$, the distortion visibility converges to a unique solution, which is reasonably well approximated by Eq.~\eqref{eq:J_bb_DC} (we will show a more detailed comparison below). At $\mu\gtrsim 10^{-3}-10^{-2}$, corrections start to become important, significantly increasing the distortion visibility.  This is simply a reflection of the fact that the photon production rate is no longer only linearly dependent on the value of $\mu_\rho$. The modifications become dramatic for large values of $\mu_\rho$, increasing the visibility by several orders of magnitudes. This implies, that distortion constraints reach to significantly higher redshifts than estimated in the limit of small distortions, as also pointed out previously \citep{Burigana1991, Hu1993}. 

For large distortions, the visibility function shows a characteristic redshift at which the slope changes significantly. For $\mu_\rho=5$, this occurs at $z\simeq \pot{6}{6}$, for $\mu_\rho=1$ at $z\simeq \pot{5}{6}$. In addition, at low redshifts the thermalization efficiency again increases (visibility reduces) once $\mu_\rho$ exceeds $\simeq 1-3$. We associate this behaviour with the significant change in the shape of the distortion for large chemical potentials (compare Fig.~\ref{fig:large_mu_CS_rel} and Fig.~\ref{fig:large_mu_transition}), which affects the active photon emission region and thus the thermalization efficiency. However, a more detailed investigation is beyond the scope of this paper, also because this only concerns cases with extremely large energy release, $\Delta \rho_\gamma/\rho_\gamma\gtrsim 0.1$.

\begin{figure}
\centering 
\includegraphics[width=\columnwidth]{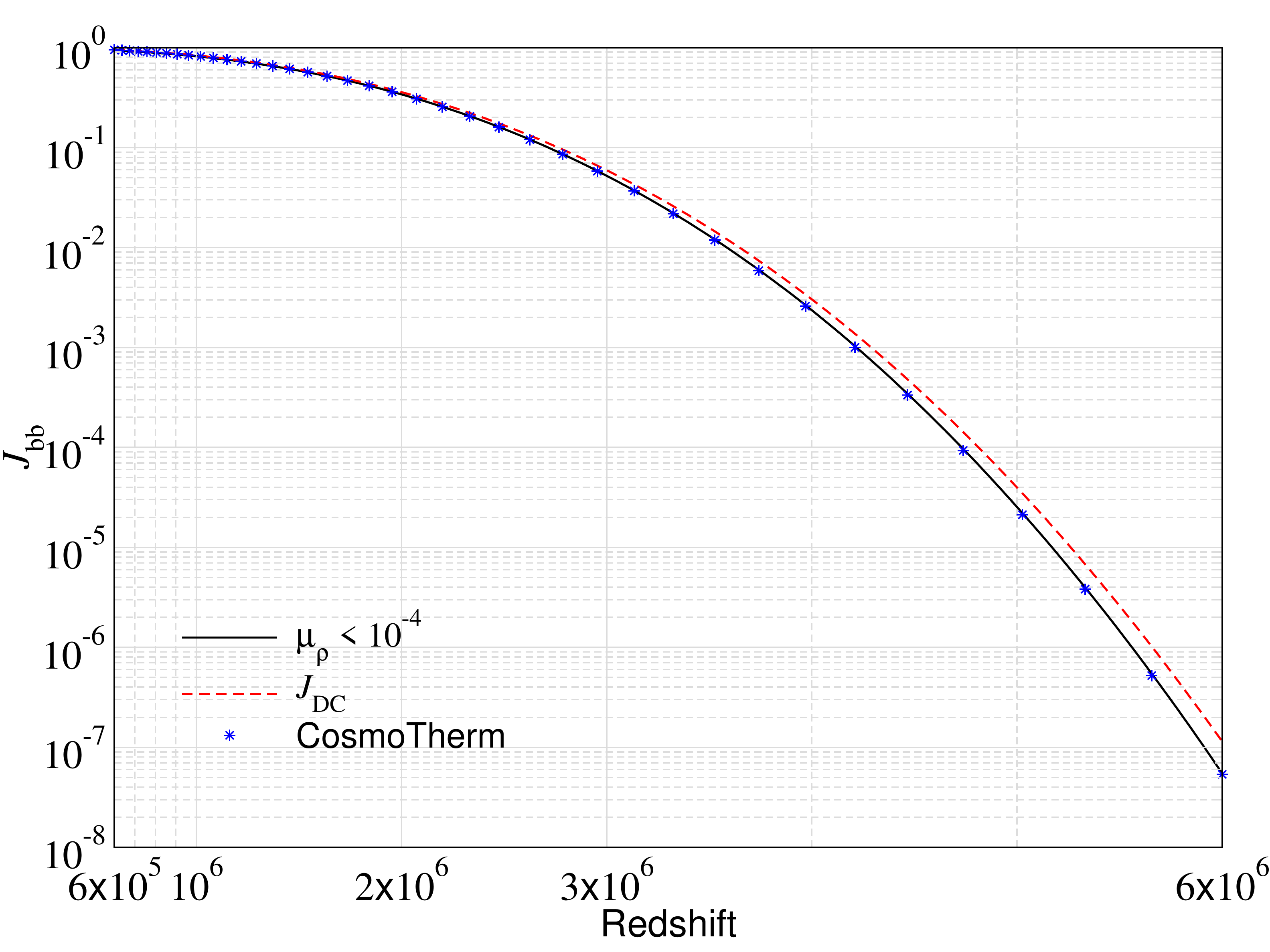}
\caption{Comparison of various approximations for the distortion visibility function for small distortions. The red dashed line gives the classical result, $\mathcal{J}_{\rm DC}(z)=\expf{-(z/\zmu)^{5/2}}$, while the stars are obtained using {\tt CosmoTherm}, which shows very good agreement in spite of neglecting Compton scattering relativistic corrections.}
\label{fig:J_bb_compare}
\end{figure}
\subsection{Importance of various corrections}
\label{sec:various_effects}
To confirm the precision of our computations, in Fig.~\ref{fig:J_bb_compare} we directly compared with the results obtained for small distortions using {\tt CosmoTherm}. The {\tt CosmoTherm} computation includes all time-dependent effects and relativistic temperature and the main DC Gaunt factor corrections; however, it does not include CS temperature and Klein-Nishina corrections. When comparing with the present computation, we thus find a small ($\simeq 10\%$) mismatch at $z\simeq \pot{6}{6}$, which we will further discuss below.

In contrast, the simple approximation $\mathcal{J}_{\rm DC}(z)=\expf{-(z/\zmu)^{5/2}}$ overestimates the visibility notably. Most of the difference is indeed related to DC and BR temperature and frequency-dependent corrections, as well as other approximations made in the derivation of $\mathcal{J}_{\rm DC}(z)$. Time-dependent terms also contribute, but play a more minor role. This was also discussed in \citet{Chluba2014} and can be further confirmed using our numerical scheme by simply switching time-dependent corrections on and off.
\begin{figure}
\centering 
\includegraphics[width=\columnwidth]{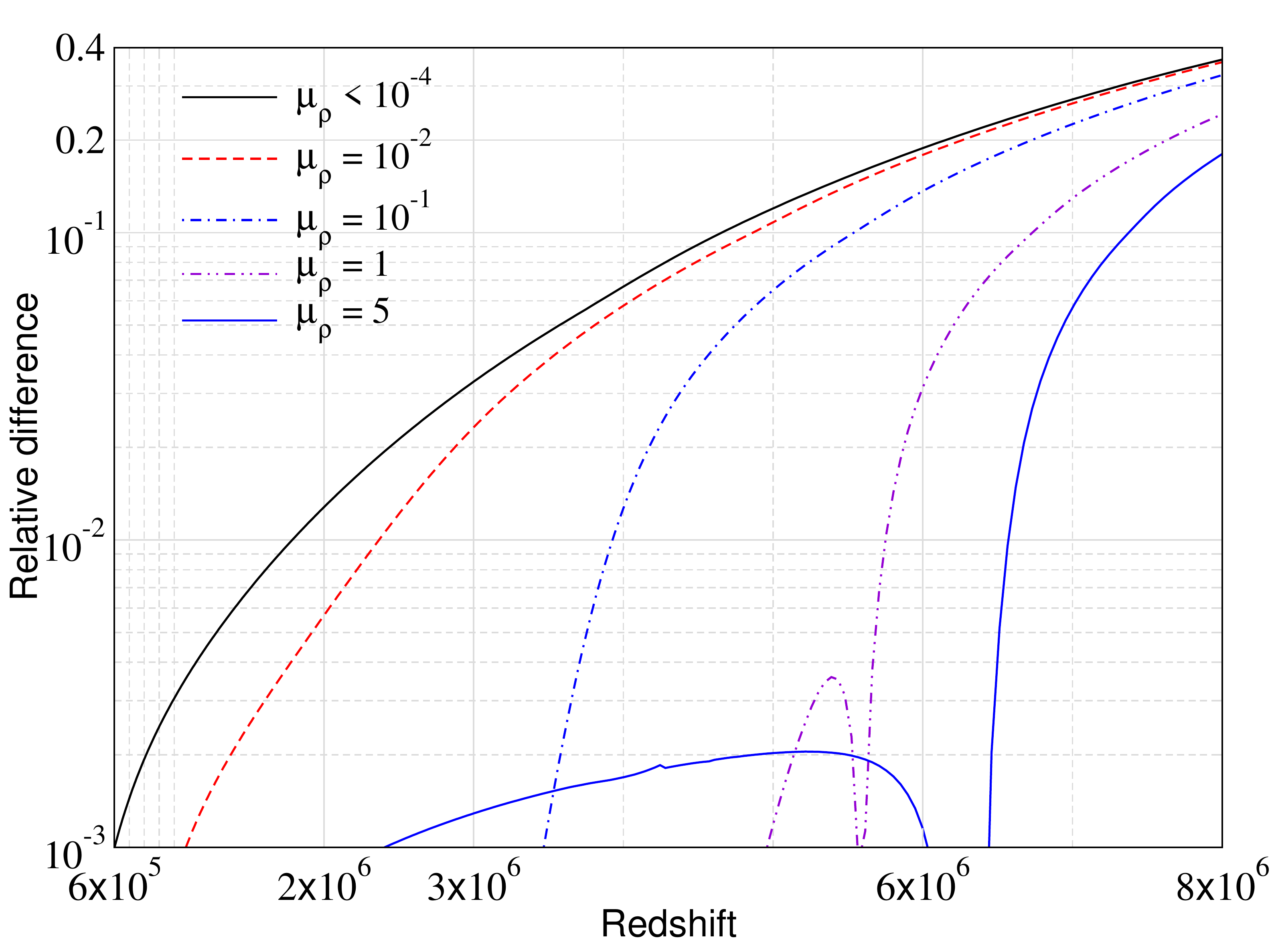}
\caption{Importance of time-dependent corrections to the distortion visibility function. We show the absolute values of the relative difference between the results obtained with and without time-dependent terms switched on. For large distortions, time-dependent corrections are generally less important than for small distortions, with the biggest effect seen early on.}
\label{fig:J_bb_compare_time}
\end{figure}
The results of this comparison are shown in Fig.~\ref{fig:J_bb_compare_time} for various values of the chemical potential. For small chemical potentials ($\mu_\rho \lesssim 10^{-4}$), the time-dependent correction reaches $\simeq 19\%$ at $z\simeq \pot{6}{6}$. Comparing this to the relative difference of the full result with respect to $\mathcal{J}_{\rm DC}(z)=\expf{-(z/\zmu)^{5/2}}$, which yields a factor of $\simeq 2.1$ difference at this redshift, shows that this makes up about $\simeq 20\%$ of the full effect. The dominant correction is due to proper evaluation of the redshift integrals, without adding new effects to the classical treatment \citep{Chluba2014}.

Our computations also show that the biggest difference due to time-dependent corrections gradually moves towards higher redshifts for increasing $\mu_\rho$ (see Fig.~\ref{fig:J_bb_compare_time}). Given that the distortion visibility function becomes more and more shallow as we increase $\mu_\rho$, this is not surprising and merely highlights that large distortions thermalize more slowly.
Here, we limited the redshift range to $z\leq\pot{8}{6}$, since the constraints from {\it COBE/FIRAS} and even future experiments similar to {\it PIXIE} become extremely weak beyond this point (Sect.~\ref{sec:Drho_limit}). However, it is worth mentioning that our numerical results show that time-dependent corrections become more important again at $z\gtrsim 10^7$. In addition, at this stage we have not included the corrections due to $\propto \id \xi/\id \tau$, which may have another noticeable effect, as will be considered in a future publication.

\begin{figure}
\centering 
\includegraphics[width=\columnwidth]{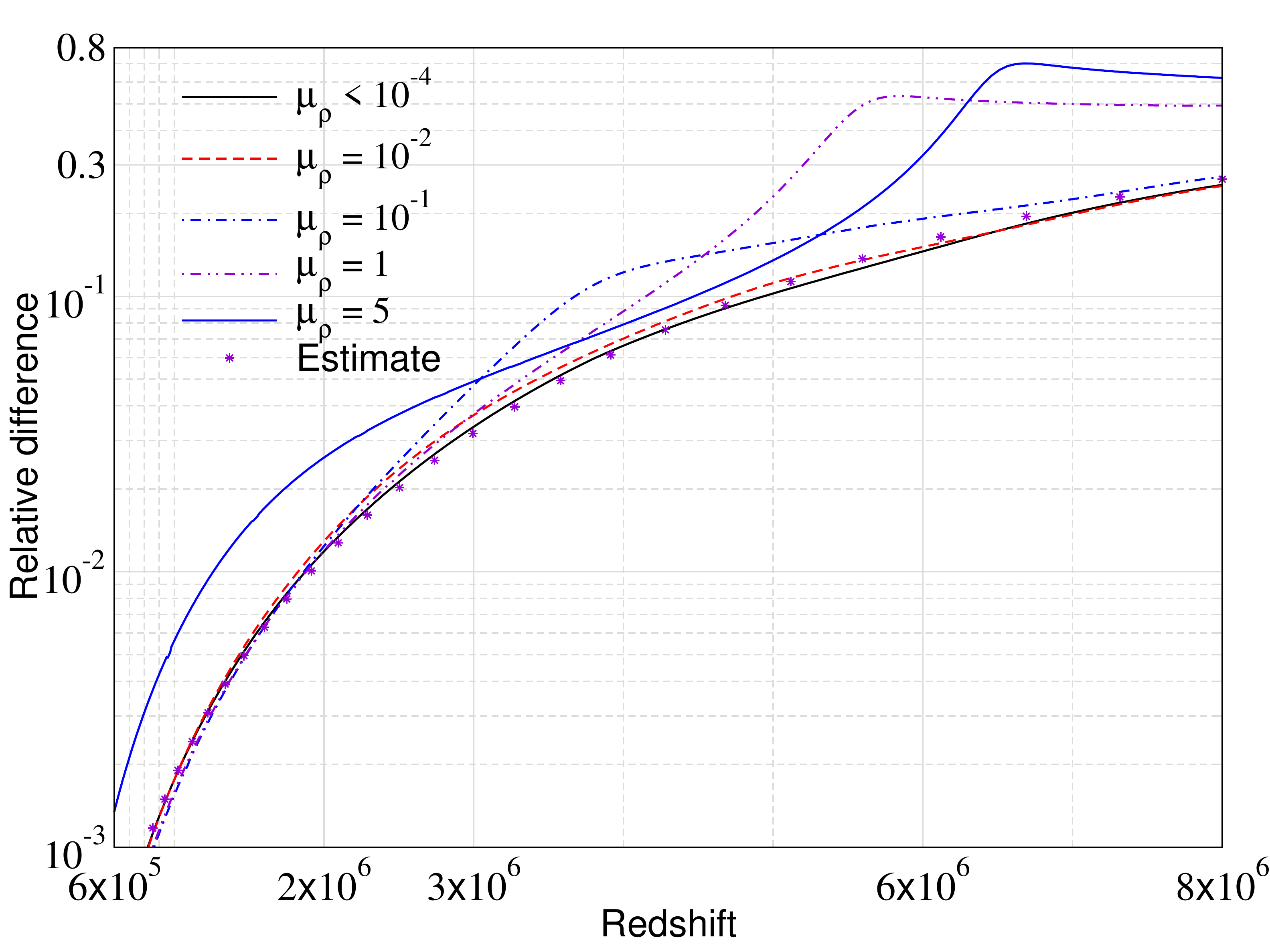}
\caption{Importance of DC relativistic corrections. We show the absolute values of the relative difference between the results obtained evaluating the CS12 DC emissivity in comparison to the Lightman-Thorne approximation (with suppression factor $g_\mu\propto \expf{-\muc}$). DC relativistic corrections lead to a net acceleration of the thermalization process. For small distortions, the effect is in good agreement with the estimate of \citet{Chluba2014}, Eq.~\eqref{eq:DC_rel_corrs}, while for large distortions the effects can become dramatic at early times.}
\label{fig:J_bb_compare_DC}
\end{figure}

In Fig.~\ref{fig:J_bb_compare_DC} we highlight the importance of DC relativistic corrections. With the Lightman-Thorne approximation [which includes the $\mu$-dependent suppression factor Eq.~\eqref{eq:mu_Int}], the thermalization efficiency is underestimated. DC temperature corrections cause a decrease of the thermalization efficiency, while frequency dependent corrections compensate for this effect, leaving a net increase or lower distortion visibility. For small distortions this was studied earlier \citep{Chluba2005, Chluba2014}, giving the correction factor
\begin{align}
\label{eq:DC_rel_corrs}
\mathcal{F}_{\rm DC}&\approx\exp\left[\left(\pot{3.7}{-3}+1.43 x^{\rm nr}_{\rm c, DC} -5.06\,\Thg\right) \,(z/\zmu)^{5/2}\right],
\end{align}
to account for the associated visibility change. Here, $\Thg=\pot{4.60}{-10}(1+z)$ is CMB blackbody temperature in units of $\me c^2$. 
At $z=\pot{6}{6}$ this expression implies $\Delta \mathcal{J}/\mathcal{J}\simeq -16\%$ and numerically we find $\Delta \mathcal{J}/\mathcal{J}\simeq -14\%$ here. However, for large distortions, the difference can become dramatic early on, significantly exceeding those for small distortions (see Fig.~\ref{fig:J_bb_compare_DC}).

\begin{figure}
\centering 
\includegraphics[width=\columnwidth]{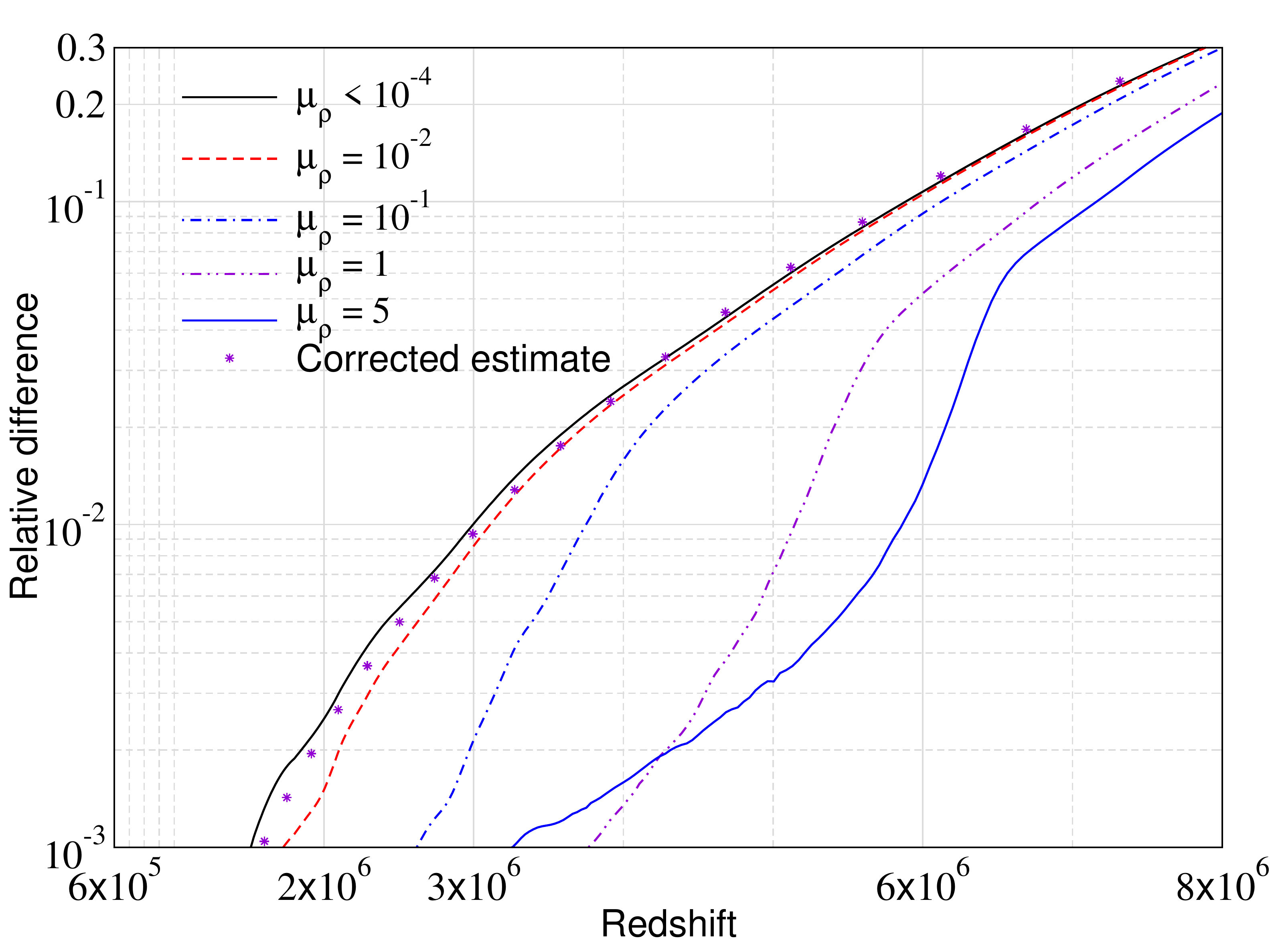}
\caption{Importance of CS relativistic corrections. We show the absolute values of the relative difference between the results with the non-relativistic case providing the reference. For small distortions, the correction is well approximated by Eq.~\eqref{eq:CS_rel_corrs} after multiplying by a factor of two.}
\label{fig:J_bb_compare_CS}
\end{figure}
We close our discussion by highlighting the differences due to our exact treatment of Compton scattering (CS). Previously, CS temperature corrections were only estimated analytically \citep{Chluba2005, Chluba2014}, yielding the correction factor
\begin{align}
\label{eq:CS_rel_corrs}
\mathcal{F}_{\rm CS}&\approx\exp\left[1.23\,\Thg \,(z/\zmu)^{5/2}\right].
\end{align}
No numerical confirmation of this estimate was yet given.  
By comparing our numerical results with those obtained using the Kompaneets kernel of \citet{Sazonov2000}, here we can isolate the effect numerically. 
The outcome of this comparison is shown in Fig.~\ref{fig:J_bb_compare_CS}. For both the exact and the approximate computations, we only included frequencies $x\leq 20$, since the Kompaneets kernel becomes inaccurate above that.
The overall effect of temperature corrections is to reduce the thermalization efficiency. The difference naturally increases with redshift but also significantly depends on the amplitude of the chemical potential. 
The result is slightly surprising, given the relativistic correction to CS increase the critical frequency (cf. Fig.~\ref{fig:large_mu_CS_rel}) such that also the photon production rate naively is expected to increase slightly, thereby suggesting the an opposite behavior. However, when carefully evaluating the emission integral, its value reduces \citep{Chluba2005}. One can anticipate this effect by realizing that the emission rate (which is set by DC) does not change; however, the position at which most emission happens increases. This naturally gives a drop in the total rate, explaining the finding.

When comparing with Eq.~\eqref{eq:CS_rel_corrs}, we find that this approximation underestimates the effect roughly by a factor of $\simeq 2$. Thus, by using $\mathcal{F}_{\rm CS}^*\approx\exp\left[2.46\,\Thg \,(z/\zmu)^{5/2}\right]$ instead we find very good agreement with the numerical result given here (see Fig.~\ref{fig:J_bb_compare_CS}). {\it What is the origin of this difference?} Repeating all the steps in \citet{Chluba2014} leading to Eq.~\eqref{eq:CS_rel_corrs} we find no obvious algebraic mistake. However, it turns out that not only is there a correction to the shape of the $\mu$-distortion spectrum but one {\it also} has to re-evaluate the DC emissivity. The origin of the latter lies in a simple temperature correction to the leading order CS terms, which yields
\begin{align}
\label{eq:QS_classical_large_mu_new}
0\approx (x^2 \partial_x^2 \mu +2x \,\partial_x\mu)\left[1+\frac{5}{2}\The\right] - \frac{\xc^2}{x^2}\,\mu.
\end{align}
Indeed, this term was mentioned in \citet{Chluba2014}, but then incorrectly omitted in the final evaluation. It essentially leads to the replacement of the critical frequency by $\xc\rightarrow \xc/(1+\frac{5}{2}\The)^{1/2}$. In addition, the frequency-dependent CS correction function also picks up another temperature-dependent term $\simeq -\frac{11}{16} \The$, which was absorbed into the normalization coefficient  \citep[see Eq.~(74) of][]{Chluba2014}. This yields the relevant contribution
\begin{align}
\label{eq:DCS_corr}
\Delta \mu\approx - \The \left[\frac{11}{16}+\frac{\zeta}{2}\left(\frac{11}{4}+\frac{21}{20}\zeta-\frac{7}{30}\right)\right]\expf{-\zeta}.
\end{align}
with $\zeta=\xc/x$. Including all these term in the evaluation of the required emission integrals given in \citet{Chluba2014}, we then find
\begin{align}
\label{eq:CS_rel_corrs_corrected}
\mathcal{F}^*_{\rm CS}&\approx\exp\left[2.55\,\Thg \,(z/\zmu)^{5/2}\right],
\end{align}
which now is very close to what was used in Fig.~\ref{fig:J_bb_compare_CS}. In summary, CS modifies the relative emission rate of DC, reducing its efficiency, an effect that was incorrectly omitted. Therefore, CS relativistic corrections lead to $\Delta \mathcal{J}/\mathcal{J}\simeq 10\%$ increase in the distortion visibility at $z\simeq \pot{6}{6}$. All DC and CS relativistic corrections together thus almost cancel each other, leaving a net correction of only $\Delta \mathcal{J}/\mathcal{J}\simeq -3\%$ at $z\simeq \pot{6}{6}$.

\section{Constraints on large energy release}
\label{sec:Drho_limit}
\begin{figure}
\centering 
\includegraphics[width=\columnwidth]{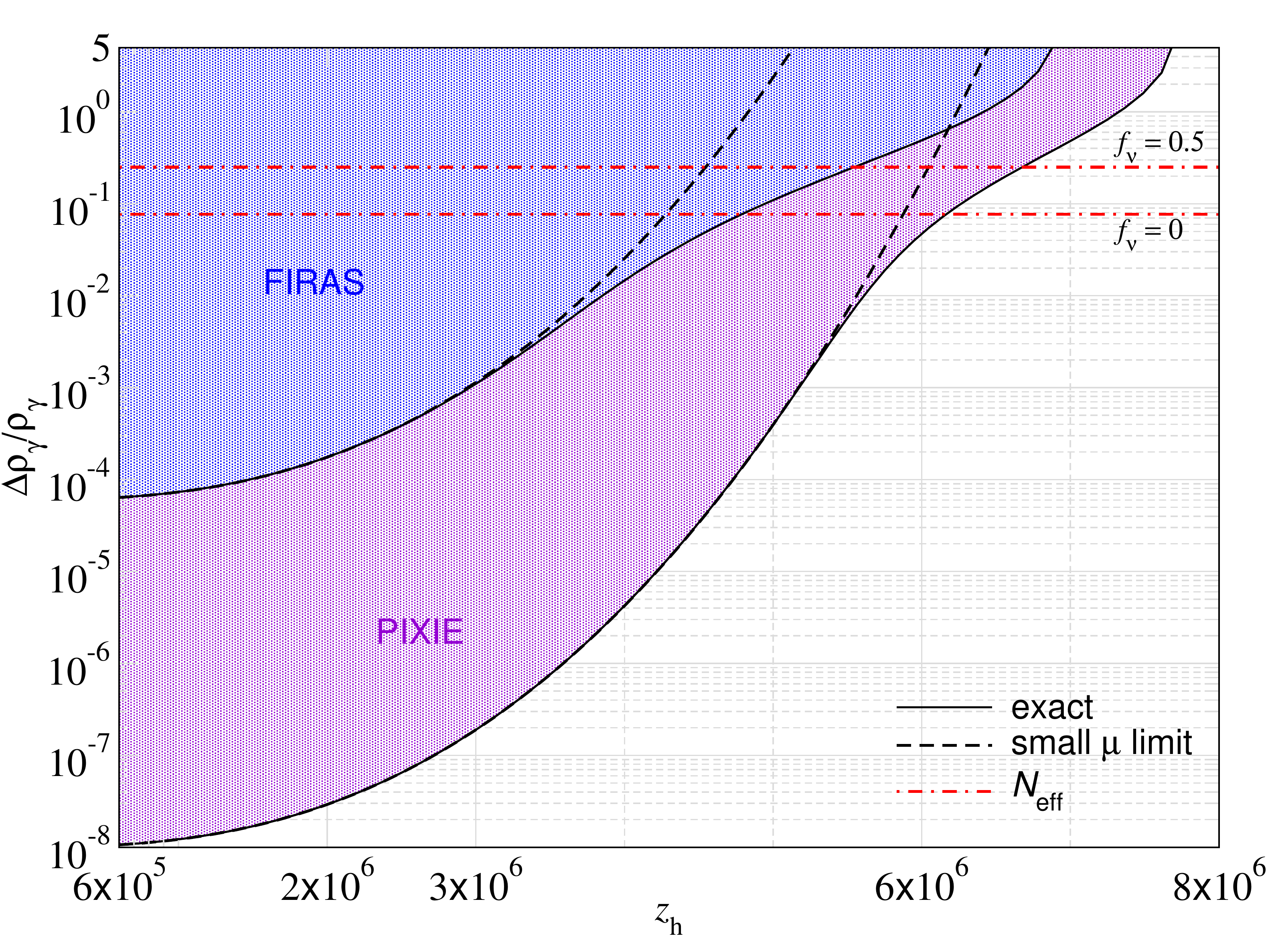}
\caption{CMB spectral distortion constraints on single energy release obtained with {\it COBE/FIRAS} [$\Delta \rho_\gamma/\rho_\gamma<\pot{6}{-5}$ ($95\%$ c.l.)] and for a future experiment similar to {\it PIXIE} [$\Delta \rho_\gamma/\rho_\gamma<10^{-8}$ ($95\%$ c.l.)]. For comparison, we show the standard estimate assuming small distortion evolution. We also give the limits derived from current CMB measurements of $\Neff$ assuming a fraction $f_\nu$ of the total injected energy goes into neutrinos (see Sect.~\ref{sec:Drho_limit_Neff} for details). This limit can become extremely weak around $f_\nu\simeq 0.4$.}
\label{fig:dlnrho_limits}
\end{figure}

We are now in the position to convert the visibility function into constraints on the energy release at various redshifts. Given an observational limit $\Delta \rho_\gamma/\rho_\gamma \big|_{\rm lim}$, for small distortions (i.e., small energy release) one can simply write
\begin{align}
\nonumber
\Delta \rho_\gamma/\rho_\gamma\lesssim \frac{\Delta \rho_\gamma/\rho_\gamma \big|_{\rm lim}}{\mathcal{J}_{\rm bb}(\zh)}.
\end{align}
For large distortions this is no longer possible because the visibility function directly depends on the amplitude of the distortion. One thus has to explicitly find the root to the inequality 
\begin{align}
\nonumber
\mathcal{J}_{\rm bb}\left(\zh, \Delta \rho_\gamma/\rho_\gamma\right)\,\Delta \rho_\gamma/\rho_\gamma\lesssim \Delta \rho_\gamma/\rho_\gamma \big|_{\rm lim},
\end{align}
which can be done numerically. The results of this exercise are presented in Fig.~\ref{fig:dlnrho_limits} for $\Delta \rho_\gamma/\rho_\gamma \big|_{\rm lim}=\pot{6}{-5}$ and $\Delta \rho_\gamma/\rho_\gamma \big|_{\rm lim}=10^{-8}$ (both at 95\% c.l.). The first limit corresponds roughly to the one obtained with {\it COBE/FIRAS} \citep{Fixsen1996, Fixsen2011}, while the second can be imagined for a future CMB spectrometer similar to {\it PIXIE} \citep{Kogut2011PIXIE, PRISM2013WPII, Kogut2016SPIE, Chluba2019Voyage}. We can see that simply using the small distortion bound (dashed red lines) yields a significantly weaker limit at early times, where large energy release can in principle still be digested by the cosmic plasma. This corresponds to epochs about $t\simeq 10^6-10^7\,{\rm s}$ after the big bang, allowing us to reach about one order of magnitude in time deeper into the cosmic fluid.

\vspace{-2mm}
\subsection{Change of $N_{\rm eff}$ after single injection}
\label{sec:Drho_limit_Neff}
By assuming that energy is injected after BBN, we can also derive a constraint from measurements of $\Neff$ \citep{SS2008}. The theoretical value for $\Neff$ from standard model neutrinos is $\Neff\simeq 3.046$ \citep{GG1998,MMPPPS2005,DS2016, Akita2020}. Measurements of the CMB anisotropies limit this value to $\Neff=2.99\pm 0.17$ [68\%~c.l.] \citep{Pl2018}. In the future, the error could improve by about one order of magnitude \citep{SOWP2018, Baumann2018}.

Assuming standard BBN physics, right after electron-positron annihilation is over we have \citep{DodelsonBook, Steigman2007, Pospelov2010}
\begin{equation}
\rho_\nu=R_\nu \,\rho_{\gamma}=\Neff\left(\frac{7}{8}\right)\left(\frac{4}{11}\right)^{4/3}\rho_{\gamma}
\approx 0.6917 \,\left[\frac{\Neff}{3.046}\right] \, \rho_{\gamma}.
\end{equation}
Defining $\rho=\rho_\nu+\rho_\gamma=\rho_\gamma\left[1+R_\nu\right]$ and assuming that $\epsilon=\Delta \rho / \rho$ of energy is injected with a fraction $f_\nu$ going into neutrinos, then after the energy release we have 
\bsub
\beal
\rho^*_\nu&=\rho_\nu+f_\nu \Delta \rho =\rho_\gamma\big[ R_\nu+f_\nu\,\epsilon (1+R_\nu)\big]
\\
\rho^*_\gamma&=\rho_\gamma+(1-f_\nu) \Delta \rho  =\rho_\gamma\big[1+(1-f_\nu)\,\epsilon (1+R_\nu) \big].
\end{align}
\esub
We do not specify how the particles are distributed in energy but only need to consider the overall energetics. In this case, $\rho^*_\gamma=\rho_{\rm CMB}$ after the release is over.
In terms of post-release $\Neff$ this means
\beal
\Neff^*&=\left(\frac{8}{7}\right)\left(\frac{11}{4}\right)^{4/3}\,\frac{R_\nu+f_\nu\,\epsilon (1+R_\nu)}{1+(1-f_\nu)\,\epsilon (1+R_\nu)}
\nonumber\\
&=\Neff\,\frac{1+f_\nu\,\epsilon (1+R_\nu)/R_\nu}{1+(1-f_\nu)\,\epsilon (1+R_\nu)}
\end{align}
Converting this to $\delta \ln \Neff\approx (\Neff^*-\Neff)/\Neff$ and then solving for $\epsilon$, we find
\begin{equation}
\frac{\Delta \rho}{\rho}=\frac{R_\nu}{1+R_\nu}\frac{\delta \ln \Neff}{f_\nu-(1-f_\nu)(1+\delta \ln \Neff)R_\nu}.
\end{equation}
Assuming that $\delta \ln \Neff\ll 1$, we can also write
\begin{equation}
\left|\frac{\Delta \rho}{\rho}\right|\lesssim \left|\frac{0.2417}{f_\nu-0.4089}\,\delta \ln \Neff\right|.
\end{equation}
For $f_\nu=1$ (i.e., everything goes into neutrinos), with $R_\nu\simeq 0.6917$ we obtain the strongest possible limit on the total energy release $|\Delta \rho/\rho|\lesssim 0.4089\,|\delta \ln \Neff|$, implying $|\Delta \rho/\rho|\lesssim 0.046$ (95\% c.l.) from current CMB anisotropy data. In this case, there is {\it no} direct distortion constraint expected.
If on the other hand we assume $f_\nu=0$, due to the changes of the photon energy density we find $|\Delta \rho/\rho|\lesssim 0.5910\,|\delta \ln \Neff/(1+\delta \ln \Neff)|\lesssim 0.066$ (95\% c.l.), understanding that in this case $\delta \ln \Neff<0$. In terms of $\Delta \rho_\gamma/\rho_\gamma$, this means $|\Delta \rho_\gamma/\rho_\gamma|\lesssim 0.077$ (95\% c.l.). This case seems to be consistent with the discussion of \citet{SS2008}.
For $f_\nu\approx R_\nu/(1+R_\nu)\approx 0.4089$ one finds $\Neff^*\approx \Neff$, leaving the energy release unconstrained by measurements of $\Neff$. Indeed, this is close to the typical value found in ultrahigh-energy particle cascade computations, where neutrinos initially can carry $\simeq 40\%-50\%$ of the total energy \citep[e.g., see Fig.~4 of][]{CCHHKPRSS2011}. Assuming $f_\nu=0.5$ as a fiducial value, this means $|\Delta \rho/\rho|\lesssim 2.7\,\delta \ln \Neff$, which with current observations of the CMB anisotropies implies
$|\Delta \rho/\rho|\lesssim 0.3$ or $|\Delta \rho_\gamma/\rho_\gamma| \lesssim 0.25$ (95\% c.l.). 

The limits on $\Delta \rho_\gamma/\rho_\gamma$ for neutrino heating efficiencies $f_\nu=0$ and $0.5$ are shown in Fig.~\ref{fig:dlnrho_limits}. These already put significant pressure on large energy release scenarios, however, with the understanding that uncertainties in the value of $f_\nu$ affect the constraint significantly. In particular for $f_\nu\approx 0.3-0.5$, the constraint can be entirely avoided. The evolution of ultrahigh-energy neutrinos in the cosmic plasma, will furthermore lead to strongly delayed energy release that can allow us to place tight constraints on neutrino heating processes at very early times, highlighting the important interplay between particle physics and CMB distortions. 
In the future, improved measurements of $\Neff$ will thus provide a powerful way for ruling out thermal histories with large energy release after BBN, complementing the tight constraints obtained at later times from measurements of CMB spectral distortions. For {\it COBE/FIRAS}, the distortion constraints supersede those from current limits on $\Neff$ at $z\lesssim \pot{4}{6}-\pot{5}{6}$, or some $\gtrsim 10^6\,{\rm s}$ after the big bang.

Here it is also important to mention that planned CMB anisotropy searches for the presence of new particles in the pre-BBN era through measurements of $\Neff$ \citep[e.g.,][]{Baumann2018} rely on robust limits on post-BBN energy injection, as these can otherwise hamper the conclusions. In particular, if the energy release mainly involves low-energy processes, CMB spectral distortions may provide one of the only ways of learning about these scenarios, highlighting the necessity for careful distortion calculations in obtaining robust limits for large energy release.

\subsection{Revised distortion constraints on PBHs}
\label{sec:Drho_limit_PBHs}
PBHs are formed from density spikes in the radiation-dominated era when the radiation pressure is unable to resist the gravitational collapse \citep{ZN1967,H1971,CH1974,NPSZ1979} with the mass of the black hole being of the order of the horizon mass \citep{CH1974,CKSY2010}. Due to formation at different epochs, the mass of PBHs can vary from Planck mass relics to $\simeq 10^{10}$ times heavier than the mass of the Sun \citep{CKS2016}. Black holes radiate particles at a temperature\footnote{Here we use $c=\hbar=k_{\rm B}$=1} \citep{H1974,CKSY2010},
\begin{equation}
T_{\rm BH}=\frac{1}{8\pi GM_{\rm BH}}\approx 0.11 \left[ \frac{M_{\rm BH}}{10^{11}\,{\rm g}}\right]^{-1} {\rm TeV}.
\end{equation}
The lifetime of evaporating black holes is approximately given by \citep{MW1991,CKSY2010},
\begin{equation}
 t_{\rm BH}\simeq \pot{4.1}{5}\left[\frac{f(M_{\rm BH})}{15.35}\right]^{-1} \left[ \frac{M_{\rm BH}}{10^{11}\,{\rm g}}\right]^{3}\,{\rm s},
 \end{equation}
where $f(M_{\rm BH})$ carries the information about the emitted particles, which depends on the accessible particle energies, the spin degrees of freedom and other particle properties \citep[e.g., see][for details]{MW1991,AK2020}. Since $f(M_{\rm BH})$ is a slowly varying function, we quote all values for $f(M_{\rm BH})\approx 15.35$ relevant to the mass range of interest here. 
%
  \begin{figure}
    \includegraphics[width=\columnwidth]{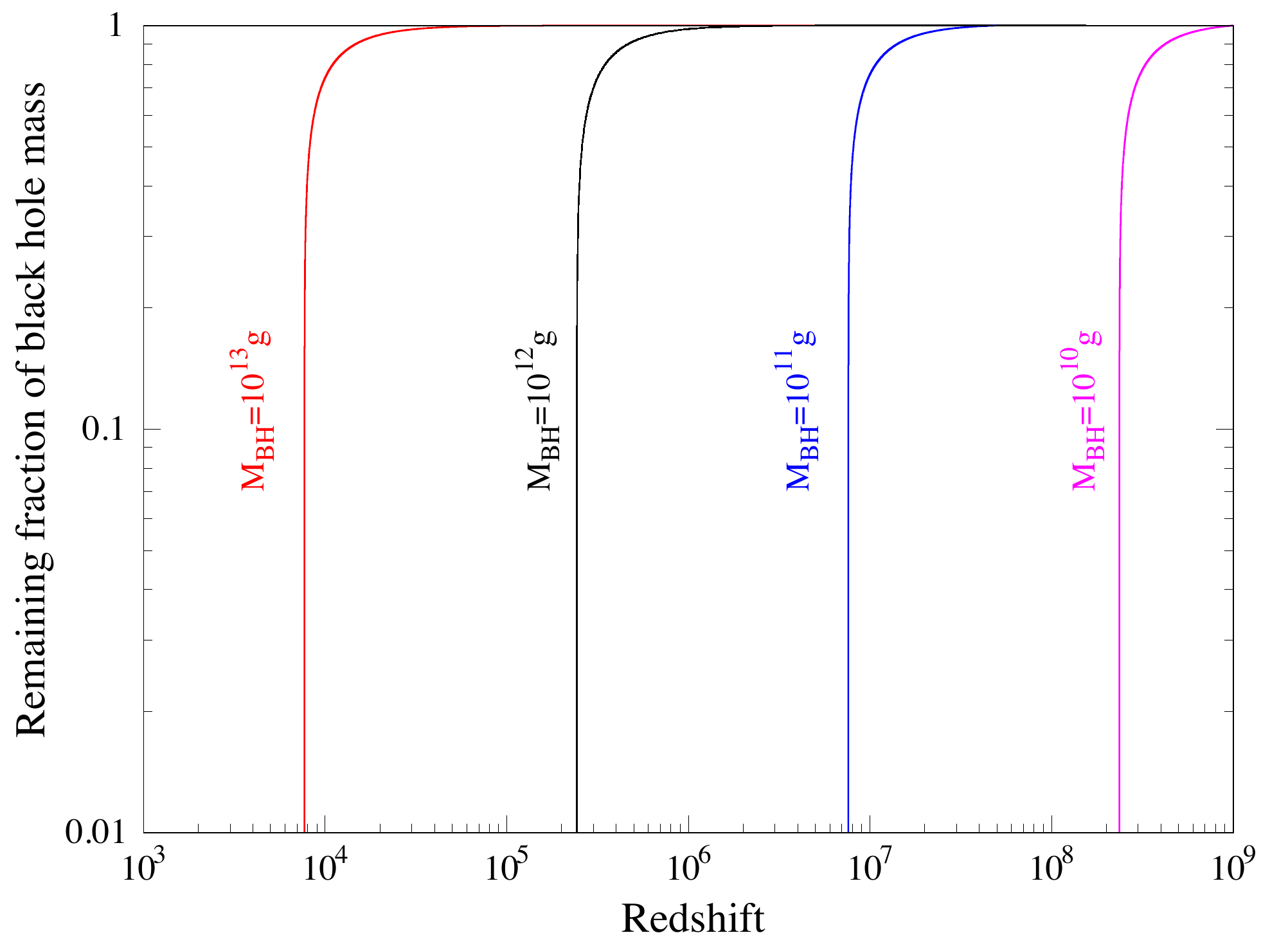}
    \caption{Fraction of black hole mass yet to be evaporated as a function of redshift. Close to the end of the evolution a runaway process starts, leading single burst of energy release around $(1+z_{\rm BH})\simeq M_{\rm BH}^{-3/2}$.}
    \label{fig:blackholefraction1}
  \end{figure}
\begin{figure}
\centering
\includegraphics[width=\columnwidth]{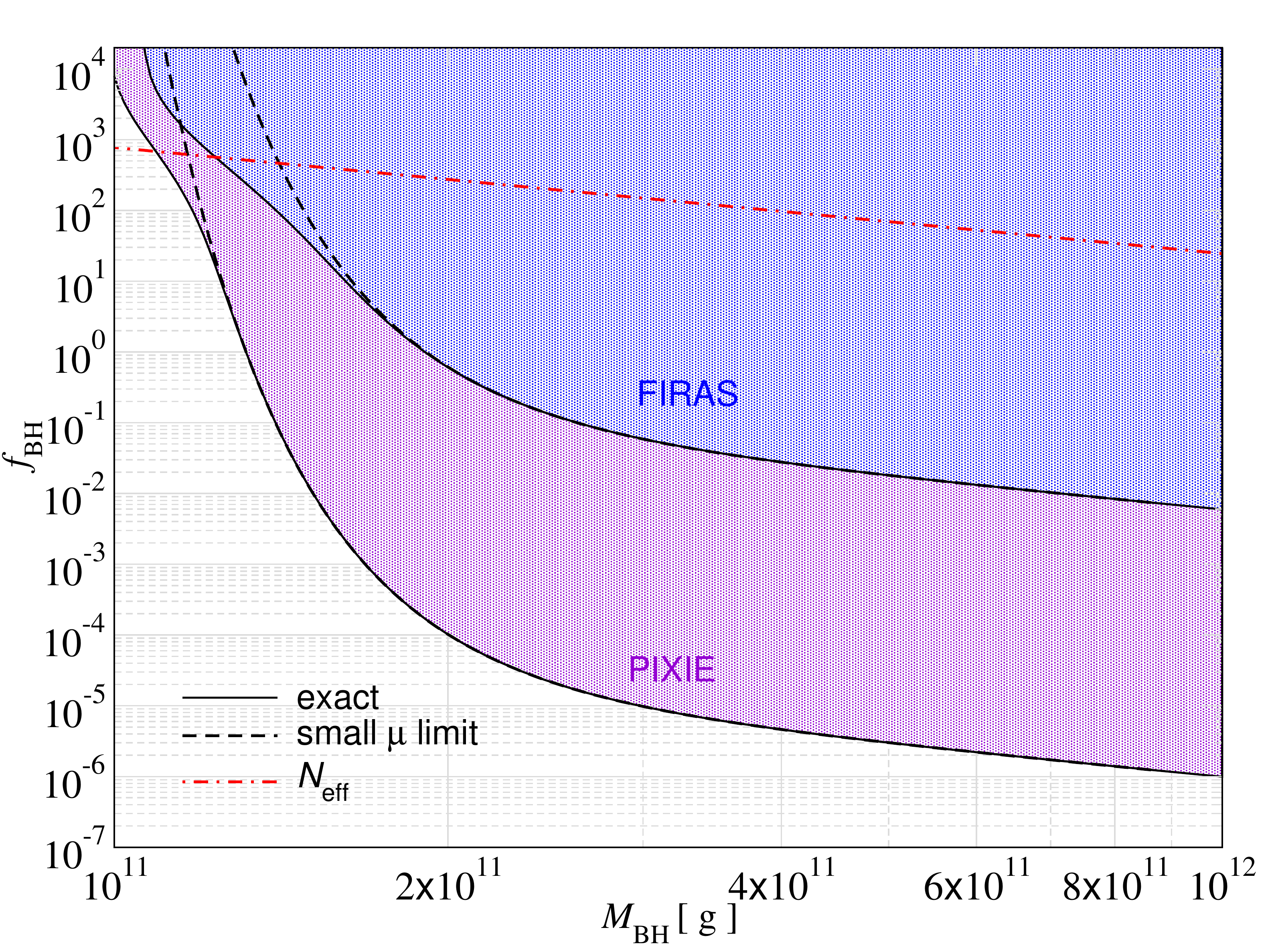}
\caption{Constraints on abundance of evaporating primordial black holes as a function of their mass. We compare the CMB spectral distortions constraint from {\it COBE/FIRAS} and a {\it PIXIE}-like experiment. In addition we give the constraints derived from measurements of the effective number of neutrinos species for energy injection fraction $f_\nu=0.5$.}
\label{fig:bhconst}
\end{figure}
%
With this, the PBH mass-loss rate can be written as \citep{MW1991}, 
\begin{equation}
\frac{\id M_{\rm BH}}{\id t}\simeq -\pot{8.2}{4} \left[\frac{f(M_{\rm BH})}{15.35}\right] \left[ \frac{M_{\rm BH}}{10^{11}\,{\rm g}}\right]^{-2}\,{\rm  g \,s^{-1}}.
\end{equation} 
In Fig.~\ref{fig:blackholefraction1}, we show the fraction of mass of black hole that is yet to be evaporated for different black hole mass. Most of the black hole mass evaporates in a short redshift interval at the end of the evolution. We can therefore approximate black hole evaporation as a quasi-instantaneous energy release at a redshift 
\begin{equation}
(1+z_{\rm BH})\simeq  \pot{7.6}{6}\left[\frac{f(M_{\rm BH})}{15.35}\right]^{1/2} \left[ \frac{M_{\rm BH}}{10^{11}\,{\rm g}}\right]^{-3/2},
\end{equation} 
showing that higher mass black holes evaporate later. In this case, the limits obtained from Fig.~\ref{fig:dlnrho_limits} directly apply.
The total PBH energy injection at $z_{\rm BH}$ can be written as,
\begin{equation}
\nonumber
\left.\frac{\Delta \rho_\gamma}{\rho_\gamma}\right|_{\rm BH}
\!=\frac{f_\gamma f_{\rm BH}\,\Omega_{\rm cdm}\rho_{\rm c,0}}{\rho_{\gamma, 0}(1+z_{\rm BH})}
\approx \pot{4.9}{-3} f_\gamma f_{\rm BH}
\left[ \frac{1+z_{\rm BH}}{10^6}\right]^{-1} \left[ \frac{\Omega_{\rm cdm} h^2}{0.12}\right],
\end{equation}  
where we parametrized the energy density of PBHs relative to the dark matter density, $\rho_{\rm PBH}=f_{\rm BH}\,\Omega_{\rm cdm}\,\rho_{\rm c,0}(1+z_{\rm BH})^3$ and defined the fraction of energy going into the photon field, $f_\gamma=1-f_\nu$. This then yields the constraints shown in Fig.~\ref{fig:bhconst}, where we compare the 2$\sigma$ limits on abundance of evaporating black holes from CMB spectral distortions and from $\Neff$ (for $f_\nu=1/2$).
For CMB spectral distortions, we show the difference between small energy release and large energy release case. As already claimed, the two results start to differ at redshifts $z_{\rm BH} \gtrsim \pot{3}{6}$. On the high-mass end, one can also observe the slight tilt with mass caused by the mass scaling of the total energy release, $\Delta \rho_\gamma/\rho_\gamma\big|_{\rm BH}\propto (1+z_{\rm BH})^{-1}\propto M_{\rm BH}^{3/2}$. 

While we only considered non-rotating PBHs in this work, extending the calculations to rotating black holes should be straightforward. In that case, the lifetime of black holes and the particle emission spectra will be functions of both mass and black hole angular momentum. Another related scenario for CMB spectral distortion is due to black hole superradiance \citep{Z1971,PT1972}. Photons in the ionized Universe have an effective mass (which is a function of redshift) due to efficient scattering with the background electrons. These photons can extract rotational energy from Kerr black holes. The efficiency of energy extraction is a function of effective mass of photon or redshift. The energy extracted will show up as a spectral distortion signal. Previous estimates \citep{PA2013} assumed the spectral distortion to be of thermal nature. However, the photon spectrum can be non-thermal and thus differ from thermal distortions for $z_{\rm BH}\lesssim \pot{{\rm few}}{4}$ \citep[e.g.,][]{Acharya2018}. Finally, spectral distortions may allow shedding light on the precise mechanisms of BH evaporation, which within loop-quantum gravity occurs after a lifetime $t\propto M_{\rm BH}^2$ instead of $t\propto M_{\rm BH}^3$ \citep{Martin2019CQG}.

\subsection{Revised distortion constraints on decaying particles}
\label{sec:Drho_limit_Decay}
Ever since the early works on distortion and the spectroscopic measurements by {\it COBE/FIRAS}, spectral distortions have been used to constrain the decay of long-lived particles in the early Universe, being most sensitive to lifetimes $t_X\simeq 10^7\,{\rm s}-10^{12}\,{\rm s}$ \citep{Sarkar1984, Ellis1985, Ellis1992, Hu1993b, Dimastrogiovanni2015}. 
The constraints are usually based on simple estimates integrating the total amount of energy released in the process weighted by the distortion visibility function
\begin{align}
\label{eq:dlnrho_lim_tot}
\frac{\Delta \rho_\gamma}{\rho_\gamma}\Bigg|_{\rm tot} = \int \frac{\mathcal{J}_{\rm bb}(z)}{\rho_\gamma}\,\frac{\id Q}{\id z}\,\id z.
\end{align}
Here $\id Q/\id z$ describes the energy release and $\mathcal{J}_{\rm bb}(z)$ is commonly approximated using the classical approximation, Eq.~\eqref{eq:J_bb_DC}. In addition, the integral is often approximated as single injection at an effective redshift where most of the injection happens. 

\begin{figure}
\centering 
\includegraphics[width=\columnwidth]{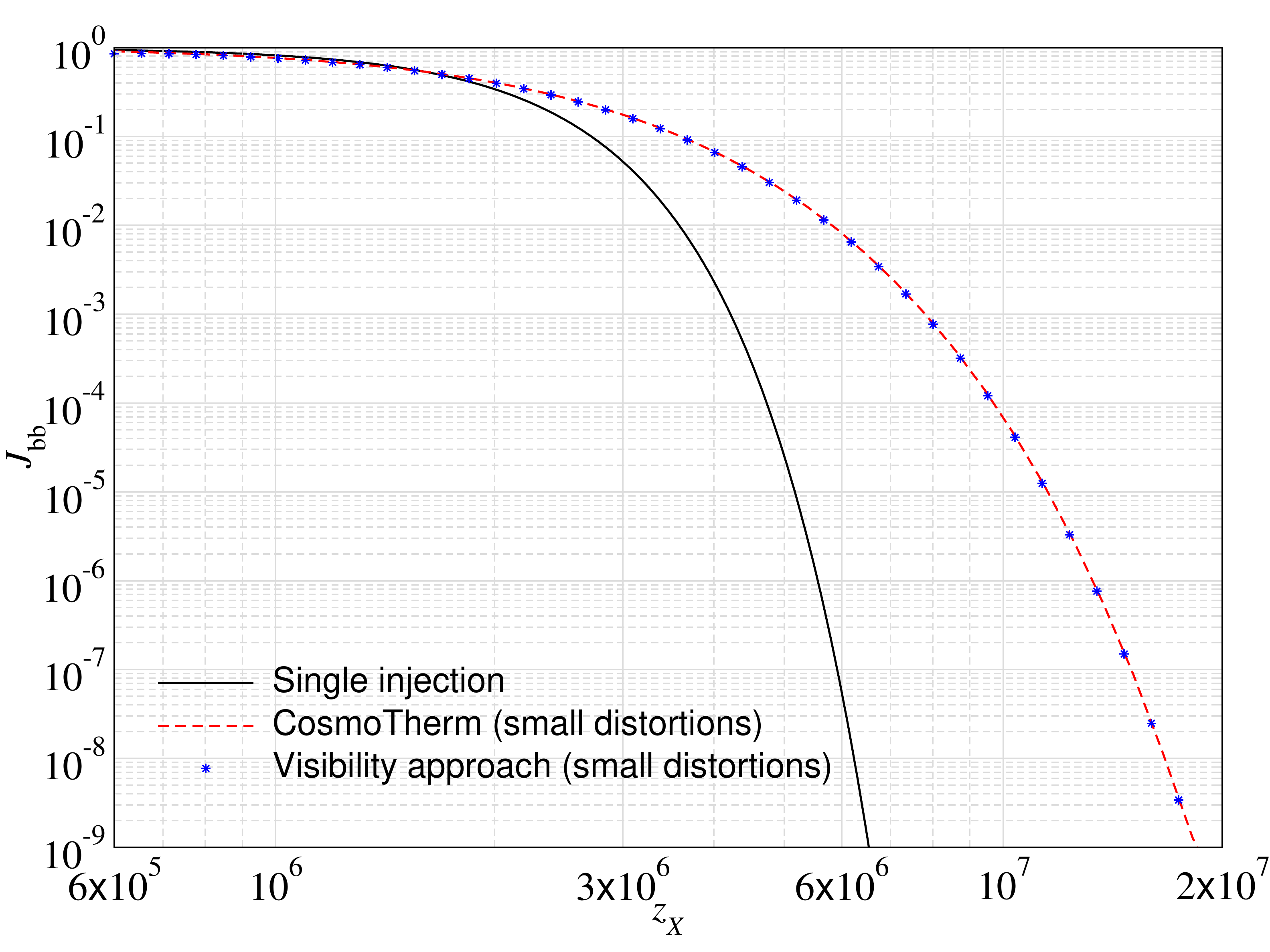}
\caption{Fraction of energy that would still be visible as CMB spectral distortions today and was injected by a long-lived decaying particle with lifetime $t_X=\pot{2.4}{7}[(1+z_X)/10^6]\,{\rm s}$. The full numerical result from {\tt CosmoTherm} agrees well with the estimate based on the distortion visibility approach, departing at the few percent level.}
\label{fig:dlnrho_limits_decay}
\end{figure}
For decaying particles, \citet{Chluba2011therm} obtained the distortions by explicitly solving the time-dependent problem using {\tt CosmoTherm}. This showed that the constraints for early lifetimes tighten significantly when carefully carrying out the energy release integral \citep[see Fig.~16 of][]{Chluba2011therm}; however, the importance of time-dependent corrections to the distortion visibility was not separately highlighted, but according to our discussion may become inaccurate even for small distortion. 

In Fig.~\ref{fig:dlnrho_limits_decay}, we show the constraint on the energy release by decaying particles as a function of lifetime. The estimate obtained by inserting the exact small-distortion visibility function into Eq.~\eqref{eq:dlnrho_lim_tot} is compared to the explicit {\tt CosmoTherm} computation, as was also done previously \citep{Chluba2013fore, Chluba2013PCA}. 
We can see that the two results are very close to each other, with departures at the percent level, and that the omitted time-dependent effects largely average out. 
Since in the small distortion limit the constraint weakens significantly for lifetimes $t_X\lesssim 10^{7}\,{\rm s}$, we expect the constraint to significantly tighten in this regime when considering the evolution of large distortions. In the case, several aspects have to be treated more carefully (see Sect.~\ref{sec:procedure_cont_cosmo}), such that we leave a more detailed computation to the future. 

\begin{figure}
\centering 
\includegraphics[width=\columnwidth]{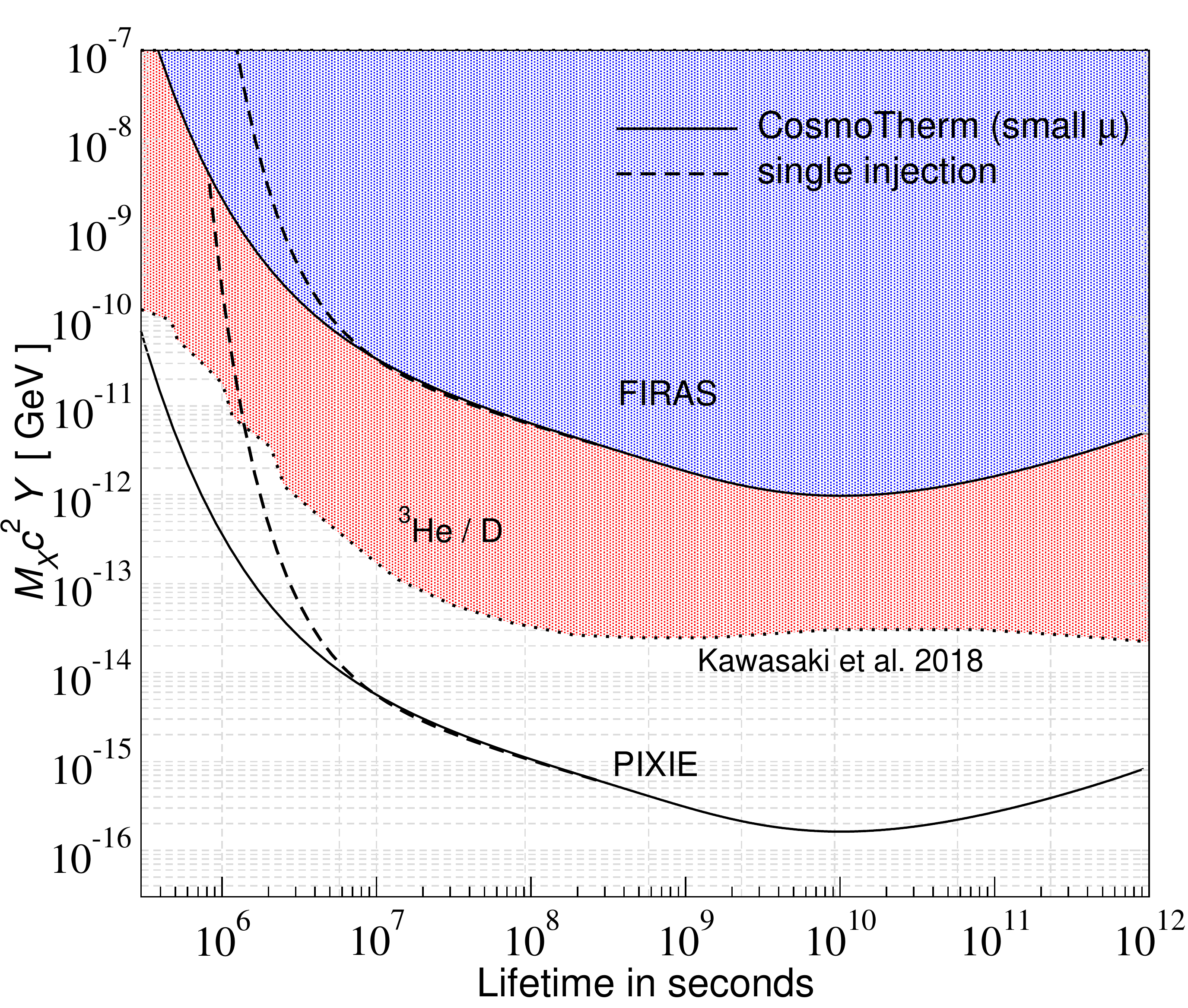}
\caption{Limits on decaying particle yields for various lifetimes from {\it COBE/FIRAS} and expected from a {\it PIXIE}-like spectrometer. As an example, we also compare with the limits derived from measurements of the He$^3$/D abundance ratio for the $b\bar{b}$ channel \citep{Kawasaki2017}.}
\label{fig:dlnrho_limits_decay_Yield}
\vspace{-3mm}
\end{figure}
For completeness, we also converted Fig.~\ref{fig:dlnrho_limits_decay} into a distortion limit on the commonly-used particle yield variable, $Y_X=N_X/s$ \citep[e.g.,][]{Kawasaki2005, Kawasaki2017}. This allows us to compare the distortion constraint with existing limits from measurements of light element abundances. To obtain the figure, we need to compute the total energy release into to the CMB, which relates to
\begin{align}
\nonumber
\frac{\id Q}{\id z}&= \frac{M_X c^2}{H (1+z)}\, \frac{N_{0, X}(1+z)^3}{t_X} \, \expf{-t/t_X}
\nonumber \\
&\approx M_X c^2 Y_X\,s\, \frac{(1+z_X)^2}{(1+z)^3} \, \expf{-\left[\frac{1+z_X}{1+z}\right]^2}
\nonumber \\
&=\pot{1.1}{13}\left[\frac{M_X c^2 Y_X}{1\,\GeV}\right]\,\frac{(1+z_X)^2}{(1+z)^4} \, \expf{-\left[\frac{1+z_X}{1+z}\right]^2}\,\rho_\gamma
\end{align}
where $s=(4/3)[g^*_s/2]\,\rho_\gamma/k\TCMB$ is the entropy density post-BBN with $g^*_s=43/11\approx 3.909$. 
The total energy release thus is
\begin{align}
\nonumber
\left.\frac{\Delta\rho_\gamma}{\rho_\gamma}\right|_{\rm dec}
&
\approx \frac{\pot{4.9}{12}}{1+z_X}\,\left[\frac{M_X c^2 Y_X}{1\,\GeV}\right],
\end{align}
assuming that all the energy goes into the CMB. Multiplying this by the $J_{\rm bb}$'s given in Fig.~\ref{fig:dlnrho_limits_decay} and comparing to the distortion limits, we obtain the curves given in Fig.~\ref{fig:dlnrho_limits_decay_Yield}. The distortion constraints are consistent with those presented in \citet{Chluba2013PCA} and demonstrate that using the simple single injection approximation underestimates the constraints significantly at lifetimes $t_X\lesssim 10^7\,{\rm s}$. However, even the limits presented here in the small distortion regime from {\tt CosmoTherm} are expected to tighten significantly at lifetimes $t_X\lesssim 10^6\,{\rm s}$, warranting a more detailed future analysis.

As an example, we also compare with the limits obtained with measurements of the He$^3$/D abundance ratio from \citet{Kawasaki2017} \citep[see also][]{Kawasaki2005}. For a wide survey of BBN limits on various decay channels and masses in comparison with the commonly quoted CMB distortion limits (similar to the dashed lines in Fig.~\ref{fig:dlnrho_limits_decay}), we refer the reader to \citet{Kawasaki2020}.
While light element abundance limits supersede those obtained from {\it COBE/FIRAS} in many scenarios, a {\it PIXIE}-like spectrometer could significantly improve the constraints on long-lived decaying particles over a wide range of lifetimes. However, significant uncertainties in the fractions of energies going into photons, neutrinos and destruction of light elements exist, so that a more careful reanalysis should be undertaken. In addition, light element limits usually require energetic particles above the nuclear dissociation thresholds to apply, if the injection occurs mainly in the post-BBN era. However, if large energy release indeed occurs at lower energies, but then is ingested by the CMB, also the Wien-tail photon distribution of the non-blackbody CMB will be affected, potentially leading to some extra destruction of light elements in excess of the standard blackbody radiation. This problem is currently not treated explicitly, but could cause an interesting interaction of CMB distortion and BBN calculations at late times. This again highlights the great potential of combining CMB distortion measurements and calculations with other probes of the thermal history.

\subsection{Distortion constraints on small-scale acoustic modes}
\label{sec:Drho_limit_Dissipation}
The dissipation of small-scale acoustic modes is also known to cause distortions of the CMB \citep{Sunyaev1970diss, Daly1991, Hu1994, Hu1994isocurv}. It now has become possible to accurately treat the dissipation process using direct computations of the photon transfer functions \citep{Chluba2012, Pajer2012b, Chluba2013iso}. Assuming linear perturbations, the energy carried by a single mode at wavenumber $k$ is dissipated in a bursty manner over a range of redshifts peaking around \citep{Chluba2012inflaton}
\begin{align}
(1+z_{\rm diss})\approx \pot{4.5}{5}\left[\frac{k}{10^3\,\Mpc^{-1}}\right]^{2/3}.
\end{align}
In the small distortion approximation, distortion limits weaken significantly around $k\simeq 10^4\,\Mpc^{-1}$ \citep[e.g.,][]{Chluba2012inflaton, Chluba2013PCA}. Even without including the effects of non-linear evolution of the acoustic modes\footnote{This will change the relation between maximal heating and the wavenumber of the mode.}, our results imply that single modes with $10^4\,\Mpc^{-1}\lesssim k \lesssim 10^6\,\Mpc^{-1}$ will see tightened constraint when including the effect of large distortions on the thermalization process. However, time-dependent corrections have to be treated carefully in this regime, and we leave detailed computations to another paper.

Improved calculations will complement and allow us to refine constraints obtained from changes to $N_{\rm eff}$ caused by acoustic heating, which limit the integrated power of curvature perturbations at $10^4\,\Mpc^{-1}\lesssim k \lesssim  10^5\,\Mpc^{-1}$ to $\Delta^2_{\rm R}\lesssim 0.007$ \citep{Jeong2014}. A refined study is thus justified, with expected implications for the formation of supermassive PBHs \citep{Kohri2014} or features in the small-scale power spectrum \citep{Chluba2015IJMPD, Byrnes:2018txb}, further shrinking the allowed parameter space.

\section{Conclusion}
\label{sec:conclusion}
In this paper, we carefully considered the thermalization of distortions in the $\mu$-distortion era, deep into the pre-recombination Universe at redshift $z\gtrsim 10^5$. We consistently included all relativistic corrections to the main thermalization processes, using state-of-the-art descriptions for Compton, double Compton and Bremsstrahlung event with {\tt CSpack}, {\tt DCpack} and {\tt BRpack}. Most importantly, we systematically studied the evolution of large distortions in the earliest phases, which in principle are still possible at $z\gtrsim 10^6$, given current constraints from {\it COBE/FIRAS}.

We first provided a rigorous discussion of the main evolution equations for macroscopic quantities like the electron temperature and chemical potential amplitude, without assuming $\mu\ll 1$ (Sect.~\ref{sec:formulation}). This led to a clear picture of how to evolve distortion during the quasi-stationary evolution phase across time, showing that due to time-dependent corrections limits on continuous energy release scenarios generally cannot be obtained independently of the energy release history, even in small distortion limit (Sect.~\ref{sec:procedure_cont_cosmo}). In simple words: the time-dependence of the energy release history directly affects the microscopic reactions of the plasma (e.g., effective heat capacity) and thus its thermalization efficiency, very much like the density of the free electrons and ions do. 
In general this invalidates simple distortion visibility approaches taken in previous works to include various effect \citep[e.g.,][]{Chluba2005, Khatri2012b, Chluba2014} when computing spectral distortion constraints on different energy release scenarios.

To obtain the momentary spectrum under quasi-stationary conditions we developed a Fredholm equation approach. This allows one to obtain accurate solutions for the frequency-dependent chemical potential in a numerically stable and efficient manner. For the first time, we were able to consistently include the effect of Compton scattering relativistic corrections and Klein-Nishina terms, showing their effect on the solution at high frequencies (Sect.~\ref{sec_mu_rel_CS} and Sect.~\ref{sec:Klein-Nishina}). Klein-Nishina corrections become in particularly important when time-dependent terms are included (see Fig.~\ref{fig:large_mu_time}), highlighting that differential motion of photons across frequencies always happens on a finite time-scale.

With our computations, we show that for large distortions, the transition between the low- and high-frequency limits steepens with increasing value of the high-frequency chemical potential (Fig.~\ref{fig:large_mu_transition}). This greatly modifies the active region for photon emission and thus the thermalization efficiency.  We also derived two new analytic approximations for the frequency-dependent chemical potential applicable to cases with large distortions [Eq.~\eqref{eq:QS_classical_sol_large_mu} and \eqref{eq:QS_classical_sol_large_mu_full}]. These solutions highlight the importance of stimulated Compton scattering terms for the shape of the distortion and the transition between small and large energy release. 

Prepared with these tools, we then studied the distortion visibility function as a function of the overall chemical potential amplitude. This revealed that the thermalization efficiency is significantly reduced for large energy injection (Fig.~\ref{fig:J_bb}). Consequently, limits on energy release are significantly tightened at early times (Fig.~\ref{fig:dlnrho_limits}), implying that future distortion missions have a better grasp at constraining various early-universe processes.

We in detail explained the contributions from several physical processes to changes in the distortion visibility function (Sect.~\ref{sec:various_effects}). In particular, we identified that for small distortions, CS relativistic corrections were underestimated by roughly a factor of $\simeq 2$ (see Fig.~\ref{fig:J_bb_compare_CS}). The origin of the mismatch with \citet{Chluba2014} was traced back to a incorrect omissions of first order temperature corrections to the scattering efficiency, and an improved estimate for the effect in the limit of small distortions was given [Eq.~\eqref{eq:CS_rel_corrs_corrected}]. Interestingly, the net effect of relativistic corrections, even if individually more significant, is greatly reduced. However, for large distortions, these can become more pronounced given that in this case significantly higher electron temperatures can be reached (Fig.~\ref{fig:J_bb_compare_DC}). Time-dependent corrections are found to play a secondary role (Fig.~\ref{fig:J_bb_compare_time}), but a complete and self-consistent treatment is left to a forthcoming publication, in which we also will consider constraints on continuous energy release scenarios.

Using our results, we also gave revised CMB spectral distortion constraints on the allowed energy release at high redshift (Fig.~\ref{fig:dlnrho_limits}). The limits tighten significantly at redshift $z\gtrsim \pot{3}{6}$, complementing limits derived from measurements of $\Neff$ and light element abundances. This allowed us to obtain improved distortion limits on the PBHs abundance for masses, $M_{\rm BH} \simeq 10^{11}-10^{12} \, {\rm g}$. At $M_{\rm BH} \lesssim 10^{11}\,{\rm g}$, limits from $\Neff$ dominate, while being clearly superseded by spectral distortion constrains on the high-mass end. 

For decaying particle scenarios, we explicitly confirmed that a small distortion treatment gives consistent estimates for the fraction of energy that is still visible as a distortion today (Fig.~\ref{fig:dlnrho_limits_decay}). Converting these results into limits on the particle yield (Fig.~\ref{fig:dlnrho_limits_decay_Yield}) showed that a more careful forecast for particles with lifetimes $t_X\lesssim 10^7\,{\rm s}$ would be important. Estimates for the distortion constraints obtained using the single-injection visibility furthermore greatly underestimate the limits, as previously pointed out \citep{Chluba2011therm, Chluba2013PCA}.
An extension to large distortion scenarios from decaying particles and a derivation of revised limits on the small-scale power spectrum (briefly discussed in Sect.~\ref{sec:Drho_limit_Dissipation}) are left to a forthcoming publication, but can be expected to tighten for distortions created early on. Our work again highlights the complementarity between particle physics and early Universe cosmology with CMB spectral distortions as a novel probe.

Finally we mention that our analysis focuses mainly on {\it CMB distortion limits} to early energy release. Other limits from measurements of light element abundances rule out many large energy scenarios, and often supersede current distortion limits \citep{Ellis1992, Kawasaki2005, Poulin2015, Kawasaki2017, Keith2020, Kawasaki2020}. However, the BBN limits are mainly applicable if particles with typical energies above the nuclear dissociation energies ($\gtrsim {\rm few}\times \,\MeV$) are involved. Both the considered CMB bounds and discussion of $\Neff$ presented here apply also in less violent scenarios, e.g., from the dissipation of acoustic modes or decay of excited low-energy states of particles. 
In addition, the physics going into computations of the CMB constraints is extremely simple and the quasi-exact limits derived here provide a robust reference for comparison. This again highlights the relevance of CMB spectral distortion measurements as a probe of the thermal history and a sensitive {\it cosmic calorimeter}.

{\small
\vspace{-3mm}
\section*{Acknowledgments:}
We thank the referee for their comments on the manuscript.
This work was supported by the ERC Consolidator Grant {\it CMBSPEC} (No.~725456) as part of the European Union's Horizon 2020 research and innovation program.
JC was also supported by the Royal Society as a Royal Society URF at the University of Manchester, UK. In addition, SKA is grateful for financial support from the Royal Society (No. URF{\textbackslash}R{\textbackslash}191023) during his extended visit to Manchester.

\vspace{-5mm}
\section*{Data availability:} For data see {\tt www.Chluba.de/large-release}.
}

{\small
\bibliographystyle{mn2e}
\bibliography{Lit,pbh}
}

\end{document}